\documentclass[12pt, a4paper, twoside, openright]{book}

\usepackage[top=3cm,bottom=3cm,left=3.5cm,right=3cm]{geometry}
\usepackage{mathtools}
\usepackage{amssymb}
\usepackage{amsmath}
\usepackage{hyperref}
\usepackage{bbm}
\usepackage{enumitem}
\usepackage{ragged2e}
\usepackage{environ}
\usepackage{empheq}
\usepackage{emptypage}
\usepackage{epigraph}
\usepackage{siunitx}
\usepackage[nottoc]{tocbibind}
\usepackage{tocloft}
\usepackage{csquotes}
\usepackage[markcase=noupper]{scrlayer-scrpage}
\usepackage{tensor}
\usepackage{graphicx}
\graphicspath{{./Images/}}
\usepackage[compat=1.1.0]{tikz-feynman}
\newcommand{\Xmatrix}{\mathbb{X}}
\newcommand{\stuckelberg}{St\"uckelberg }
\newcommand{\HRule}{\rule{\linewidth}{0.6mm}}

\newcommand{\mfp}{m_{\text{fp}}}
\newcommand{\munu}{{\mu\nu}}
\newcommand{\numu}{{\nu\mu}}
\newcommand{\gmunu}{g_\munu}
\newcommand{\del}{\partial}
\newcommand{\half}{\frac{1}{2}}
\newcommand{\eqnref}[1]{eq\eqref{#1}}
\newcommand{\detsqrt}[1]{\sqrt{|#1|}}
\newcommand*{\fullref}[1]{\hyperref[{#1}]{\autoref*{#1}}}
\NewEnviron{equationsplit}{%
	\begin{equation}
	\begin{split}
	\BODY
	\end{split}
	\end{equation}
}
\begin{document}
\frontmatter
\label{frontmatterstart}

\begin{titlepage} 

\center 
\HRule
\\[0.4cm]
{\Large\bfseries Theories of Massive Gravity in 2+1 Dimensions}
\\[0.4cm] 
\HRule
\\[1.5cm]

	\begin{flushright}
	\large
	Master's Thesis in Theoretical Physics
	\end{flushright}
%
	\begin{flushright}
		\large
		\textit{Author}\\[-0.3 em]
		Lokesh \textsc{Mishra}
	\end{flushright}



	

\vfill\vfill 

	


	
\vfill 
	
\end{titlepage}
\begin{titlepage} 

\center 
\HRule
\\[0.4cm]
{\Large\bfseries Theories of Massive Gravity in 2+1 Dimensions}
\\[0.4cm] 
\HRule
\\[1.5cm]

	\begin{flushright}
	\large
	Master's Thesis in Theoretical Physics
	\end{flushright}
%
\begin{flushright}
	\large
	\textit{Author}\\[-0.3 em]
	Lokesh \textsc{Mishra}
\end{flushright}
\begin{flushright}
	\large
	\textit{Supervisor}\\[-0.3 em]
	Prof. Dr. Jochum Johan \textsc{van der Bij} 
\end{flushright}


\vfill\vfill
\begin{figure}[h]
	\centering
	\includegraphics[width=0.5\textwidth]{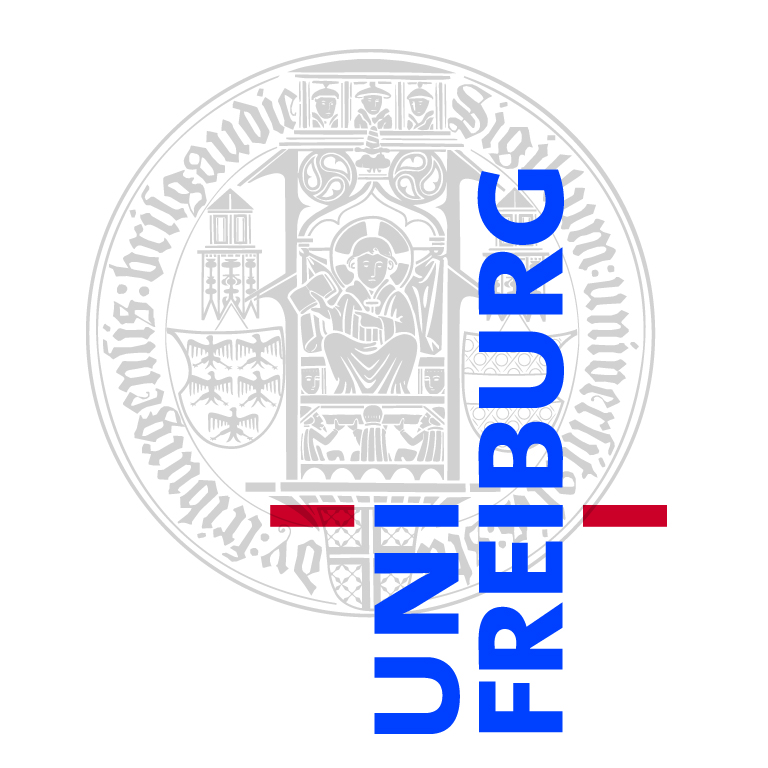}
	\\[1cm] 
\end{figure}

	

\vfill\vfill 

{\large $10^{th}$ of July, 2018} 
\center
{\large 
Institute of Physics
\\
Faculty of Mathematics and Physics
\\
Albert Ludwigs University of Freiburg
}
	


	
\vfill 
	
\end{titlepage}
\thispagestyle{empty}
\begingroup
\footnotesize

\vspace*{\fill}
\begin{flushleft}
	Front Matter - pages \pageref{frontmatterstart} to \pageref{frontmatterend} 
	\\
	Main Matter - pages \pageref{mainmatterstart} to \pageref{mainmatterend}
	\\
	Back Matter - pages \pageref{backmatterstart} to \pageref{backmatterend}
	\\[4em]
	Typeset in \LaTeX 
	\\
	July 2018, Lokesh Mishra
\end{flushleft}

\endgroup
\clearpage
\cleardoublepage
\thispagestyle{empty}
\begin{flushright}
	To my family,
\end{flushright}
\vspace*{\stretch{1}}
\begin{center}
	\textbf{It is, indeed, an incredible fact that what the human mind,\\ at its deepest and most profound,\\ perceives as beautiful finds its realization in external nature.}
	\\[3 em]
	{\hfill Subrahmanyan Chandrasekhar \cite{chandrabeauty}}
\end{center}
\vspace{\stretch{3}}
\cleardoublepage
\chapter{Abstract}

This thesis is dedicated to the study of theories of massive gravity. The formulation of higher spin gauge field theories, along with a Chern-Simons (CS) like term for fields of higher spins is presented. Through this setup, general features of theories describing massless/massive fields of higher spin in arbitrary dimensions is discussed.

In 2+1 dimensions, the existence of multiple mass-generating mechanisms i.e. metric-based masses and topological masses, offers the possibility for gauge bosons to acquire mixed masses. This scenario is first introduced in a theory of photons where both Proca mass term and CS mass term is simultaneously present. The mass-mixing which occurs in this theory is further analysed through the \stuckelberg formalism. Motivated by these results, a theory of massive gravity where gravitons obtain masses from both Fierz-Pauli mass term and CS mass term is studied. This theory allows for 3 propagating massive graviton modes, and their masses have undergone considerable mixing. This mass-mixing is made explicit through a \stuckelberg analysis. 

The bimetric theory of gravity is a non-trivial generalization of the theory of General Relativity. This theory provides a consistent non-linear theory of massive gravity which is studied in detail. In this thesis, the bimetric theory of gravity is extended in 2+1 dimensiosn to the theory of Topologically Massive Bimetric Gravity (TMBG). In the theory of TMBG, the mass-mixing which arises from the interaction of the two metrics and their corresponding CS terms occurs at the non-linear stage itself. For the version of TMBG studied in this thesis, the linearized theory shows that there are 3 massive graviton modes, which corresponds to the same 3 modes found earlier. Finally, quantum loop corrections to the graviton propagator from massive photons (which acquire mass from both CS and Proca mechanism) is calculated. 
\chapter{Acknowledgements}

I am thoroughly grateful to my supervisor Prof. Dr. Jochum Johan van der Bij for having given me an opportunity to study and work in this exciting and active field of research. This thesis could not have reached completion without his scientific guidance and endless patience. Additionally, his stories, humour, vast experience and characteristic support have only enhanced my working experience.

I am also indebted to my colleagues Dr. Christian Steinwachs and Matthijs van der Wild for countless illuminating discussions and showing me the `nitty-gritty' side of theoretical physics. 

Furthermore, I would like to thank my friends with whom I have enjoyed provoking discussions on diverse topics and shared the vicissitudes of everyday life as a student in Freiburg. I especially thank my friend Vladislavs P\c{l}e\v{s}anovs for the delights we have experienced together during our struggles, in and outside of Physics. 

Finally, my heartfelt thanks to my family for their unending support, earnest love and for always encouraging me to pursue my passions.

\chapter{Preface}

Gravitational interactions, omnipresent as they are, really require no introduction. Their theoretical description, however, necessitates the following comments. The theory of General Relativity, as proposed by Einstein in 1916, provides the current standard and well-tested description of gravitational phenomena. The present thesis is dedicated to the theoretical-study of a sub-class of theories which are collectively referred to by the name \emph{massive gravity}. Theories of massive gravity are naturally branched under a class of theories known as modified gravity. As the name suggests, Einstein's description of gravitation is modified in these theories which has its own motivations, advantages as well as disadvantages. 

Regarding the present thesis, it is primarily noted that the title of this thesis, ``Theories of Massive Gravity in 2+1 Dimensions", is a juxtaposition of sorts. On the one hand this thesis pertains to Lorentz invariant theories of massive gravity; theories in which the field that is responsible for gravitational interactions (i.e. the gravitational field) has a non-zero mass. In recent years, perturbations in this gravitational field or gravitational waves have been directly observed \cite{gravwavemass}. These observations place an extremely strong constraint on the mass of the gravitational field \cite{gravitonmasslimit}: 
$$m_g \le \SI{7.7e-23}{\tfrac{eV}{c^2}} \approx \SI{e-58}{kg} $$

Foregoing this enormous constraint, on the other hand many of the theoretical studies undertaken in this thesis (although done for spacetimes of arbitrary dimensions) have been specialized for a spacetime of 2+1 dimensions i.e. two dimensions of space and one dimension of time. Ubiquitous experience of everyday life suggests that the world we dwell in, is 3+1 dimensional with three dimensions of space and one dimension of time. 

Therein lies the aforementioned juxtaposition. Clearly, there is a need to strongly motivate a study in which two overbearing facts, if only for a moment, are boldly ignored. After providing a brief overview of the General Theory of Relativity in \fullref{ch:intro}, the reasons for studying massive gravity will be elucidated in \fullref{sec:whymassive}. Studying these theories in 2+1 dimensions has its own sincere intrinsic motivations which will be discussed in \fullref{sec:why2+1}. 

In order to provide a logically coherent connection between seemingly disparate ideas within different chapters, all chapters begin with a section which is dedicated to providing a contextually relevant perspective. The rest of the thesis is organized in the following manner:
\begin{enumerate}
	\item \hyperref[ch:notation]{Notations and Conventions}\newline
	Notations and conventions utilized throughout this document are listed for later convenience.
	
	\item \hyperref[ch:intro]{Chapter 1: Introduction}\newline
	A brief perspective on the theory of General Relativity is provided. Experimental and theoretical observations which motivates the modifications of the standard theory of gravity are then presented. A case is made for studying the theories of massive gravity and, that too, in 2+1 dimensions. 

	\item \hyperref[ch:highspin]{Chapter 2: Higher Spin Gauge Field Theory}\newline
	The formulation and setup of higher spin gauge field theories is presented. Studying the irreducible representations of the Poincar\'e group in 2+1 dimensions leads to the realization of some important distinctions between physics in 2+1 and 3+1 dimensions. Fronsdal formulation and a Chern-Simons like term for fields of higher spins is presented in detail. This formulation will be used in later studies. 

	\item \hyperref[ch:spin1]{Chapter 3: Fields of Spin-1}\newline
	This chapter serves an introductory setup for gaining an understanding into the physics of massless and massive spin-1 fields. The mass-mixing occurring for photons which gain mass from multiple mechanisms is presented and it is further analysed through the \stuckelberg formulation.

	\item \hyperref[ch:spin2]{Chapter 4: Fields of Spin-2}\newline
	From the lessons gained from previous chapters, theories of massless and massive spin-2 fields are presented. The problems prevalent in linear massive gravity or the Fierz-Pauli theory are detailed. A theory of massive gravity where gravitons acquire masses through multiple mechanisms is possible in 2+1 dimensions. Giving a broad overview into the \stuckelberg formulation, this analysis is used to understand how the different degrees of freedom excited by the two mechanisms lead to physically propagating massive graviton modes.
	
	\item \hyperref[ch:bimetric]{Chapter 5: Topologically Massive Bimetric Gravity}\newline
	The bimetric theory of gravity is studied in detail. This theory is extended to develop the theory of topologically massive bimetric gravity. This extensions allows for the mass-mixing to occur at the non-linear stage. Linear perturbations which develop into propagating massive graviton modes are calculated.

	\item \hyperref[ch:quantumloopcorrections]{Chapter 6: Quantum Loop Corrections}\newline
	Quantum corrections to the graviton propagator coming from minimally coupled multiply-massive photons is calculated.
	
	\item \hyperref[ch:final]{Chapter 7: Discussion and Outlook}\newline
	The work done in this thesis is summarized and directions for future work is discussed.
\end{enumerate}

\chapter{Notations and Conventions}\label{ch:notation}
\setcounter{footnote}{0}
A comprehensive list of notations and conventions frequently employed in this thesis are listed here:
\begin{enumerate}
	\item Often, the number of dimensions $d$ of the spacetime manifold, will be kept arbitrary. In certain explicit calculations, $d$ will be set to $3$ and the reader will be made aware of this change (unless explicit from context).
	
	\item The metric signature associated with flat minkowski metric $\eta_{\mu\nu}$ and general metric $g_{\mu\nu}$ is mostly minus, i.e. 
	$$ \eta_{\mu\nu}\  = diag\ (+,-,-)$$
	
	Due to this choice, the Levi-Civita symbol satisfies: 
	$ \epsilon_{012} = \epsilon^{012} = +1 $
	
	\item As is standard in high energy physics, the system of natural units will be used throughout this thesis. This means setting,
	$$ \hbar = c = 1 $$ 
	where, $\hbar$ is the reduced Planck's constant and $c$ is the speed of light. These units imply that the reduced Planck mass, $M_{pl}$ is given by:
	$$ M_{pl}^2 = \frac{1}{8 \pi G} $$
	where $G$ stands for Newton's gravitational constant. Note that this definition remains the same for both 2+1 and 3+1 dimensional physics. 
	
	\item For a rank two tensor $\phi_{\munu}$\footnote{The choice of constant $\frac{1}{2}$ in the expression for trace reversal is related to the relation between the trace of $\overline{\phi}$ and $\phi$, as well as, dimension $d$ of manifold. In the present work, it is fixed to $\frac{1}{2}$ for two reasons: (a) simplicity in computations, and (b) in keeping with standard literature\cite{carrollbook}. This implies for $d=3: \bar{\phi}' = - \frac{1}{2} \phi' $.} the following conventions are used:
	$$\text{Trace : } \phi'\ =\ \phi_{\munu}\ \eta^{\munu}$$
	$$\text{Trace Reversal : } \bar{\phi}\ =\ \phi_{\munu}\ -\ \frac{1}{2}\ \eta_{\mu\nu}\ \phi' $$ 
	$$\text{Symmetrization : } \phi_{(\munu)}\ = \frac{1}{2} \big(\phi_{\munu}\ +\ \phi_{\numu} \big) $$
	Note: These definitions will be properly extended when dealing with tensors of higher rank in \fullref{ch:highspin} on Higher Spin Field Theory.
	
	\item The following notation is used:
	$$\Box = \del_\mu \del^\mu  \quad \quad \del^\mu \phi_\munu = \del\cdot\phi_\nu$$
	
	\item When dealing with perturbations, objects with a bar such as $\overline{g}_{\munu}$ or $\overline{\nabla}$ denote objects that are defined with respect to the background metric\footnote{Since trace-reversals and perturbations will not appear simultaneously in this document, the meaning of an object with a bar on top will be clear from context.}.
	
	\item Conventions for geometric quantities such as curvature tensor, covariant derivatives etc are those of Carroll\cite{carrollbook}. Also, Einstein summation convention is always in force.

	\item In \fullref{ch:bimetric}, the bimetric theory of gravitation will be introduced. In this theory a second independent metric $f_{\munu}$ is present and it is, then, important to distinguish between geometrical quantities defined for each metric $g_{\munu}$ and $f_{\munu}$ separately. A tilde `{\textasciitilde{}}', is called upon for such a service:
	$$\text{Ricci Scalar for }g_{\munu} : R$$
	$$\text{Ricci Scalar for }f_{\munu} : \tilde{R}$$

	\item An important class of solutions in bimetric theory are derived from the Proportional Background Ansatz. The constant of proportionality used in this ansatz is denoted by $\rho$,\footnote{Although, standard literature on bimetric theory usually employs $c$ for this job, this trend is not followed here since $c$ is canonically reserved for the speed of light.} 
	$$ f_{\munu}\ = \ \rho^2\ g_{\munu} $$
	
	\item For any field of spin-s $\phi_{\mu_1\mu_2 \dots \mu_s}$, in general, there will be two kinds of masses:
	\begin{itemize}
	\item $m_{\phi} $ denotes conventional mass of field $\phi_{\mu_1\mu_2 \dots \mu_s}$ (also called St\"uckelberg mass or Proca mass for spin-1 or Fierz-Pauli mass for spin-2 etc)
	\item $\mu_{\phi}$ denotes topological mass of field $\phi_{\mu_1\mu_2 \dots \mu_s}$ coming from a Chern-Simons term in the Lagrangian.
	\end{itemize}
	
	\item Finally, a note on labelling of fields. Whenever a theory will be derived for a particular value of spin-$s$ from the Fronsdal formulation of higher-spin gauge field theory, which is presented in \fullref{ch:highspin}, then a generic Greek letter such as $\phi$ or $\psi$ will be used to denote this field. The spin of the field in question should be clear from the number of indices on the tensor. 
\end{enumerate}

\tableofcontents
\label{frontmatterend}
\mainmatter
\label{mainmatterstart}
\chapter{Introduction}\label{ch:intro}
\epigraph{``The measure in which science falls short of art is the measure in which it is incomplete as science."}{J. W. N. Sullivan\cite{chandrabeauty}}
\section{Perspective: Theory of General Relativity}\label{sec:gravityreview}
Eintstein's theory of General Relativity (GR) along with the Standard Model (SM) of particle physics provides a well established frame-work upon which modern fundamental physics firmly stands. Together these theories build up a concrete and venerable picture of nature. On one side, SM deals with questions regarding the fundamental building blocks of nature and provides a unified framework describing three of the four known fundamental interactions viz strong interactions, weak interactions and electro-magnetic interactions.  On the other side, GR exclusively deals with gravitational interactions. Amongst a multitude of ways in which these two descriptions differ from each other, arguably, there are some common aspects: (a) symmetries play a central role in each description (gauge symmetry associated with Lie groups for SM and general coordinate invariance or diffeomorphism for GR), (b) both are field-theoretic descriptions, (c) both feature boson mediated interactions\footnote{Bosons are fields with integer spins. Fields are characterized by their mass and spin; this arises naturally through Wigner's classification of irreducible representation. These ideas will be briefly reviewed for the case of 2+1 dimensions in section \ref{sec:irrep}.}, and (d) both are very well vindicated by experimental evidence. 

GR has to its credit a unique and intuitive geometrical interpretation, which has invited many to claim it as the ``most beautiful of all existing physical theories"\cite{chandrabeauty}. Einstein's field equations describe an interplay between matter and geometry\footnote{This has been inscribed by J. A. Wheeler in these famous words ``Spacetime tells matter how to move; matter tells spacetime how to curve''. Interestingly, among many contributions Wheeler is also known for the phrases `black holes', `worm holes', and `it from bit'. }. To its success, GR reproduces Newton's universal law of gravitation in the weak-gravity limit, predicts the perihelion shift of Mercury, bending of light by massive objects, gravitational lensing, gravitational time dilation, gravitational waves, black holes, among many others. After more than a 100 years of its proposal, GR firmly withstands tests within the solar-system to a high precision\footnote{In recent years, gravitational waves have been directly observed from black hole mergers. These observations test GR in the strong field limit and have not observed any deviations from the theory\cite{gravwavemass}.}. 

The field equations of GR which are known as Einstein's field equations, are equations of motion for the gravitational field or the metric tensor $\gmunu$ describing the coupling between sources of the gravitational field (stress-energy tensor $T_\munu$) and the field $\gmunu$ itself. These equations can be succinctly derived from the Einstein-Hilbert action which will be detailed in \fullref{sec:gravity2+1}. The famous equations are:
\begin{equation}\label{eq:einsteinfieldeq}
R_\munu\ -\ \frac{1}{2}\ \gmunu\ R = \frac{1}{M_{pl}^2} T_{\munu} -\ \Lambda_{eff}\ \gmunu
\end{equation} 
Here, on the left hand side are objects built up from the geometry of the underlying manifold, Ricci Tensor $R_\munu$ and the curvature/Ricci scalar $R$. The left hand side is characteristic of curvature and describes the dynamics of spacetime geometry. The right hand side is made up of two terms. $T_\munu$ is the stress-energy tensor of matter sources, which is scaled by $\tfrac{1}{M_{pl}^2}$. This determines the strength of the coupling between matter sources and the gravitational field. This coupling strength is in turn fixed by demanding correct Newtonian limits, which gives:
\begin{equation}\label{eq:reducedplanckmass}
	\frac{1}{M_{pl}^2} = 8 \pi G \approx \SI{1.693e-37}{GeV^{-2}}
\end{equation}
A well-known fact that gravitational interaction is quite a weak one, is captured in the smallness of the above number. The other term on the right hand side is called an effective cosmological constant. Shortly, it will be described in further detail. 

The Einstein field equations are highly non-trivial equations, describing non-linear interactions including back-reactions of the gravitational field upon itself, which are very hard to tame. Therefore, carefully made assumptions based upon the nature of problem at hand can simplify the involving computational efforts immensely. For a given background, say of a massive source of mass $M$ such as the Sun, solutions of GR can be classified into three distinct regimes based upon their region of validity\cite{reviewhinterbichler}. Firstly, there is the classical linear regime, for distances $ r > r_S $, where $r_S \approx M/M_{pl}^2 $ is the Schwarzschild radius. In this regime, both the non-linear effects and quantum corrections can be ignored. For the Sun with mass $M = M_{\odot} \approx \SI{e+30}{kg}$, the Schwazschild radius is $r_s \approx \SI{1}{km}$. Thus, the linear classical approximation of GR is practically valid almost everywhere in the solar system. Secondly, there is a classical non-linear regime, for distances $r_{pl} < r <r_S$, where $r_{pl} \approx \SI{e-35}{m}$ is the length scale associated with Planck scale. In this regime, non-linear effects become important and need to be summed up, but quantum corrections can still be ignored. This is the regime which is used to describe dynamics and physics inside a black hole. Finally, there is the regime of quantum gravity for distance $r < r_{pl}$. In this regime quantum effects play a necessarily significant role and can not be ignored any more. This regime becomes important when one is describing gravitational effects very close to the singularity itself. 
\section{Why Modify General Relativity?}\label{sec:whymodifygr}
Having noted these preliminary remarks upon the impressive successes of GR, its field equations, and solution regimes, some comments are in order to motivate a study which aims to modify the beautiful theory of GR. In spite of its impressive records, GR may not be the final word as a theory of gravitational interactions. This is based on sensible theoretical grounds. As a starter, GR is not a UV complete theory and hence, at most, can be regarded as an effective field theory valid up to a cut off scale at the Planck mass $M_{pl}$. This means that calculating higher order corrections in GR results in ever increasing number of infinities which requires an ever increasing number of free-parameters to cancel those infinities, consequently making the theory lose its predictability. 

On the other side there are certain definite cosmological observations which to this date do not have a satisfactory explanation. These include the so-called (a) Cosmological Constant Problem associated with Dark Energy, and (b) Existence of Dark Matter. 

\subsection{The Cosmological Constant Problem}\label{subsec:cosmologicalconstantproblem}
The cosmological constant problem has been ascribed many labels such as `Vacuum Catastrophe', `largest discrepancy between theory and experiment in all of science', or even `worst theoretical prediction in the history of physics' \cite{cosmologicalproblem}. Exciting as these labels may be, it is worthwhile to look at this problem from its humble origins. The effective cosmological constant term as already seen in \eqnref{eq:einsteinfieldeq}, can be thought to be made up of two contributions \cite{cosmologicalconstantPADMA}. 
\begin{equation}\label{eq:effectivecosmologicalconstant}
	\Lambda_{eff} = \Lambda_b\ +\ 8\pi G\ V(\phi_{min}) 
\end{equation} 
These two terms are believed to arise due to two disparate mechanisms. On one hand, adding a constant term to the geometrical side of Einstein's field equations \eqnref{eq:einsteinfieldeq} posits a fundamental constant of Nature $\Lambda_b$, the bare cosmological constant. In this interpretation, describing the theory of gravitation requires two constants $G$ and $\Lambda_b$, and spacetime has to be treated as curved even in the absence of matter source (i.e. $T_\munu = 0$). Conversely, treating this constant as a shift in the matter side, results in a corresponding shift in the matter Hamiltonian. Although dynamics of matter fields are not affected by such shifts in the zero-point energy of the matter field configuration, for gravity there is a remarkable difference. Due to the universal nature of gravitational interactions, these shifts in the zero-point energy of matter fields have a considerable impact on gravity and it should produce a response to this constant vacuum/dark energy density. Having gained this context, the following should now be noted:
\begin{itemize}
	\item Observations of the late accelerated expansion of universe have constrained the effective cosmological constant to $|\Lambda_{eff}| < \SI{e-47}{GeV^4}$.
	
	\item In pure classical GR smallness of bare $\Lambda_b$ is suggestive that it is perhaps zero. Yet, there is no known invariance principle or symmetry argument which requires the bare cosmological constant to be zero. 
	
	\item When quantum effects are included, quantum corrections to the vacuum energy density can be explicitly calculated for all the fields present in the universe \cite{cosmologicalconstantcalculations}. These calculation, although dependent on model specific information, generally have $V(\phi_{min}) > \SI{e+48}{GeV^4}$ or is the corrections are as large as Planck scale, $V(\phi_{min}) \approx \SI{e+72}{Gev^4}$.   
\end{itemize}

As the facts stand, to exactly satisfy \eqnref{eq:effectivecosmologicalconstant}, the bare cosmological constant must have a value, which seems to be fine tuned to, an unprecedented, $60\sim120$ orders of magnitude. This is the crux of the entire \emph{cosmological constant problem}. A fine-tuning of this scale and with apparently no reason is more than hard to digest. It warrants further investigation.

\subsection{Existence of Dark Matter}\label{subsec:existenceofdarkmatter}
The existence of a Dark Matter (DM) component in the Universe has been firmly established through various experimental observations. DM has been deduced from its gravitational effects on the rotational curves and velocity dispersions in galaxies, dynamics of stars in disk environment, through direct gravitational lensing, within the Local Group of galaxies to explain the inevitable collision of Milky Way with Andromeda(M31) and through other means \cite{darkmatterlivingreview}. There is an industry of dedicated researchers who investigate hopeful candidates for DM which are postulated to follow properties similar to those of other SM particles. Yet, since gravitation is the only confirmed interaction in which DM seems to participate, this is suggestive that a DM candidate may come from a modification to the theory of gravitation\cite{darkmatterbimetric}.

\section{Why Massive Gravity?}\label{sec:whymassive}

The issues raised in the previous section lead one to realize that modifying GR might not be that bad of an idea after all\footnote{The present scenario seems like a complete apostasy from the accepted dogma which ascertains that extraordinary claims require extraordinary evidence. The extraordinary evidence, in the case of DM and $\Lambda_{eff}$ is well-pronounced and awaits theoretical conformity.}. Apart from its' possibility to address the issues raised previously, modifications to gravity are definitely interesting in their own right. 
\begin{displayquote}
\textit{``There are few better ways to learn about a structure, whether it is a car, a computer program, or a theory, than to attempt to modify it."}
\begin{flushright}
K. Hinterbichler \cite{reviewhinterbichler}
\end{flushright} 
\end{displayquote}

There are many modifications to GR which have been studied throughout the history of this theory. For this thesis, the focus is on a certain class of theories in which the modification comes in the form of giving the gravitational field a mass. Classical GR can be uniquely summarized in the statement that ``GR is the theory of a non-trivially interacting massless helicity 2 particle''\cite{reviewhinterbichler}. It is interesting to note that this statement says nothing about geometry, general coordinate invariance or the equivalence principle. The route followed by Einstein himself using such guiding principles does not necessarily lead to a unique theory of gravitation. In theories of massive gravity, the massless helicity 2 particle of GR (quanta of the gravitation field, also called graviton) is made massive. This is, what is meant by massive gravity.

With a theory of massive gravity, there is  hope that the problem of fine-tuning encountered in the cosmological constant problem  \fullref{subsec:cosmologicalconstantproblem} might receive a technically natural explanation. The argument of technical naturalness goes back to 't Hooft \cite{naturalnesstHooft}. The general idea is that a small parameter in a theory is called technically natural, if there exists a symmetry which appears when the value of the said parameter is set to zero. In other words, the \emph{principle of naturalness} states that if an underlying theory becomes more symmetric when a parameter involved is set to zero, only then should this quantity be small in nature. For example, small masses of fermions (such as electrons) are technically natural because if they were put to zero, say in the theory of Quantum Electro-Dynamics (QED), then chiral symmetry appears \cite{naturalnessdine}. In regard to the extremely small seemingly fine-tuned value of the bare cosmological constant $\Lambda_b$, no such symmetry is know and hence their low values does not conform to 't Hooft's principle of naturalness. On the other hand, in a theory of massive gravity with graviton mass $m_g$ the fine-tuning problem in $\Lambda_b$ can be redressed into the fine-tuning issue of $\frac{m_g}{M_{pl}}$. The punchline comes when one notes that when $m_g$ in such a theory is set to zero, this theory will regain its symmetry under general coordinate invariance. This provides a hope and a sincere motivation that the cosmological constant problem can be solved with a massive graviton.


To modify a theory as successful as GR, the first consistency condition on any new modified theory is that it must reproduce all the known successes of GR. Additionally, for a theory as rich and rigid in its structure as GR, any tinkering with the theoretical-structures does not occur for free. The penalties come in various forms such as propagation of unphysical ghost modes (Boulaware-Deser Ghost), discontinuity in parameter space (vDVZ discontinuity) etc, rendering such pathology-ridden theories unsuitable. It is, therefore, difficult to construct a healthy consistent theory of massive gravitons. 

The last point is rather general and, technically, is worth pausing upon. The highly successful SM of particle physics frequently and extensively employs massive and massless fields of spin $0, \frac{1}{2}, \text{and}\ 1$ to describe enormous amounts (if, not all) of physics. Thus, it is safe to say, that constructing a field theoretic description of massive as well as massless fields of spin $0, \frac{1}{2}, \text{and}\ 1$ is certainly under control \footnote{Note: The construction of a gauge invariant scheme for massive spin $1$ `vector' fields leads to the Higgs Mechanism.}. The same, however, cannot be said for spin-2 particles. Hence, naturally one is lead to ask: why shouldn't massive spin-2 fields not have a theoretically sound description? The problem of constructing a healthy non-linear theory for general spin-2 fields turned out to be more challenging than expected. This program, for non-linear massive gravity, started in the year 1939 when Fierz and Pauli wrote down the theory of linear massive spin-2 fields\cite{fierzpauli}. Since then, the pursuit of this research has lead to many deep insights into the theory; finally leading to the construction of the dRGT theory of massive gravity in 2010 \cite{dRGTtheory}  (after de Rham, Gabadadze and Tolley). This dRGT theory was further generalized into the bimetric theory of gravity by Hassan, Rosen and May in 2012 \cite{bimetrichassanrosenmay}. Between the years 1939 and 2012, needless to say, quite a lot of important and necessary work was done. A detailed description of this journey can be found in the interesting reviews \cite{reviewhinterbichler, reviewdeRham, reviewMay}. It is noted that important physical concepts such as vDVZ discontinuity and the Boulaware-Deser ghost which arised in this journey will be given due attention when they arise in \fullref{ch:spin2}.

The next fruit of studying massive gravity comes in the form of identifying suitable candidates for DM. The bimetric theory of gravity which will be discussed in detail in \fullref{ch:bimetric}, involves two independent metrics interacting in a highly non-trivial manner. This theory when expressed in terms of its massive eigenstates leads to a description of the theory in terms of massive spin-2 field and a massless spin-2 field. With the massless spin-2 field all predictions of GR can be satisfactorily recovered. This leaves the other massive spin-2 field, which can be tuned, as a rather ideal candidate for DM (see \cite{darkmatterbimetric} for an example).

This section is now satisfactorily concluded with the satisfactory convinction that a theory of massive gravity may provide interesting solutions to the problems mentioned earlier in \fullref{sec:whymodifygr}. Who knows, studying such a theory may result in more than what is being asked for?

\section{Why 2+1 Dimensions?}\label{sec:why2+1}

It was mentioned at the beginning of this chapter that there are many differences between GR and SM, two of the most established theories of fundamental physics. Out of those many differences, the most peculiar one is that the SM is based upon the framework of quantum field theory, whereas GR as a poor old cousin is only a classical field theory. Since unification of all fundamental interactions, is one of the long-cherished goals in theoretical physics, naturally, many people have tried to put forward a quantum theory of gravity, or quantum gravity. The issue of quantum gravity is an unusually demanding and involving difficulty which can not be given a concise introduction in this short overview. Their exists a vast sea of literature for the interested reader, and as a gentle starting point the reader is referred to \cite{qgBurgess, qgVeltman:1975vx} and the references therein. 

The serious nature of significant difficulties faced when dealing with the computational and conceptual challenges of quantum gravity desperately motivates one to look for simple/toy models; preferably those which retain some of original conceptual difficulties but simplify the computational effort \cite{qgCarlipbook:1998uc}. GR in 2+1 dimensions is just such a model. As a classical theory of spacetime geometry, GR in 2+1 dimensions is riddled with many of the foundational issues that exist for the 3+1 dimensional theory\footnote{Pure GR in 2+1 dimensions does not have any propagating degrees of freedom. While this initially gave the impression that 2+1 dimensional GR is too trivial, it turned out to not be the case. These points will be clarified and detailed in \fullref{ch:spin2} when dealing with spin-2 fields}. Nevertheless, it has proved as an important testing ground for many theoretical approaches and provides an excellent theoretical-laboratory for understanding the quantum nature of gravity.

When 2+1 dimensional GR was in its nascent stage, the arguments mentioned above may have formed the basis for most motivations in studying 2+1 dimensional gravity. According to the author, this has significantly changed in the present day. It turns out that fundamental physics, especially field theories in 2+1 dimensions enjoy special properties which invites dedicated studies on its own. Some, and empathetically not all, of the curious features in 2+1 dimensional physics are: 
\begin{itemize}
	\item Schonfeld and Deser, Jackiw, and Templeton discovered, in the early 1980's, a mechanism through which gauge fields acquire mass in a gauge invariant way \cite{tmgSchonfeld:1980kb, tmgravityDeserJackiwTempleton:1981wh}. This mechanism and such theories go under the name \emph{Topologically Massive Gauge Theories}\footnote{The importance of this work lies in part to its ability to provide a gauge invariant mass generating mechanism. Another alternative is the Higgs mechanism. In \fullref{sec:irrep}, group theoretical arguments will be presented for the curiosities observed in such theories.}. A major part of this thesis is dedicated to the study of such theories. There have been attempts to extend such mass terms to 3+1 dimensions\cite{tmg4dmassJackiw}.
	
	\item Induced Masses: It was observed, quite early on, that quantum loop corrections may induce a mass term for gauge bosons \cite{tmgravityDeserJackiwTempleton:1981wh}. In fact, even if one begins with a theory of massless gauge boson coupled to fermions, radiative corrections give such massless bosons a mass term! Furthermore, this is even true for the case of gravity: massless gravitons becomes massive when coupled to massive vector bosons or fermions \cite{tmgVANDERBIJ198687}. 
	
	\item Parity Violation: Weak interactions in SM are the only known interactions which violate parity. Parity transformations are those transformation under which spatial dimensions are inverted. In 2+1 dimensions, mass terms for fermions and topologically massive gauge bosons break parity and time reversal symmetry\footnote{Parity transformations in 2+1 dimensions are defined by a transformation of the type: $\vec{r}=(x,y)\to\vec{r'}=(-x,y).$}.
	
	\item Cosmic Strings: These hypothetical objects are 1-dimensional topological defects which may have been created during a symmetry breaking phase transition in early universe \cite{cosmicstringCopeland623, cosmicstringVachaspati:2015cma}. It turns out that GR in 2+1 dimensions is particularly suitable to treat these objects and study their phenomenology. Unfortunately, the strongest experimental constraints on these objects are based upon the lack of their detection through gravitational waves\cite{cosmicstringexpBLANCOPILLADO2018392}.
	
	\item Quantum Hall effect: When the topological mass mechanism is used for spin-1 vector bosons, as in $QED_{3}$,the theory so-obtained is effectively applicable for the study of quantum hall effect in condensed matter physics\cite{qhalleffectDunne:1998qy, qhalleffectTong:2016kpv}. 
	
	\item Cosmic Topology: The study of overall structure and topology of the Universe is called Cosmic Topology (see \cite{cosmictopologyuniverse2010001} for present status with Planck data). There exists an interesting argument, due to van der Bij, based upon cosmic topology which has been used to explain the number of fermion generations in the SM. Such an argument assumes that the early universe may have been 2+1 dimensional (the third space dimension grows large at later times) and hence makes exquisite usage not only of GR in 2+1 dimensions, but also other features mentioned above\cite{cosmictopologyvanderbijPhysRevD.76.121702, cosmictopologyvanderBij2011}.
	
	\item AdS/CFT, BTZ, Holography and all that: Although far removed from the purview of the present thesis, there exists a gargantuan amount of literature (including several books) rich in novel, creative, and insightful results dedicated to the study of `empty space quantum cosmology' in 2+1 dimensions. GR in 2+1 dimensions is endowed with rich structures such as a black hole solution, the BTZ black hole (valid for AdS spaces only). These black hole solutions have in turn been utilized as theoretical playgrounds to test interesting new directions such as black hole thermodynamics, AdS/CFT duality, Knot theory among many others. Since most of these topics lie outside the scope of this thesis, some general and interesting references are cited for interested readers \cite{qgCarlipbook:1998uc, carlipbtzblackhole, Witten:2007kt, btzblackholePhysRevLett.69.1849}. 
\end{itemize}

These remarks clearly demonstrate that physics in 2+1 dimensions is special, spectacular even. This, in itself, constitutes a strong motivation to dedicatedly study what effects arise when a graviton is made massive in 2+1 dimensions.

\chapter{Higher Spin Gauge Field Theory}\label{ch:highspin}
\epigraph{``It will be impossible to answer any one question completely without at the same time answering them all."}{P. A. M. Dirac\cite{diracquote}}
Although, Fierz and Pauli are often cited for originating the program of massive gravity, their original paper had a lot more to offer than that. Indeed, they had set out to systematically study and extend Dirac's earlier work on relativistic wave equations to particles of \emph{arbitrary high spin} \cite{fierzpauli}. Their approach was based upon demanding Lorentz invariance and positivity of energy after quantization. They correctly pointed out that consistent interactions between fields of spin 2 or more is non-trivial, and derived an equation for linear massive spin-2 fields. This, the Fierz-Pauli equation along with its theory will be studied in detail in \fullref{sec:fierzpaulitheory}. The present chapter, however, is devoted to their original goal: theories of higher spin fields which possess high spin gauge symmetry.

Quantum field theory is a framework, coherently based upon the principles of Quantum Mechanics (QM) and the theory of Special Relativity (SR), which describes the propagation and interaction amongst quantum fields. For flat spacetimes or Minkowski spaces, this framework associates particles with unitary, irreducible representations of the Poincar\'e group. The keywords used here describe fundamental principles: Unitarity, coming as a basic postulates from QM demands conservation of probabilities or bounded Hamiltonians; irreducible representations reflecting the elementary nature of the corresponding field/particle; and transformations under the Poincar\'e group which allows for principles of SR to be incorporated. All of these far-reaching and deep ideas are logically connected and firmly established in Wigner's theorem. Additionally, this theorem's tantamount importance lies in its ability to provide a scheme for classification of elementary particles/fields. The contents of this theorem which are relevant for physics in 2+1 dimensions will be shortly reviewed in \fullref{sec:irrep}. According to this theorem quantum fields can be massive or massless, and their spins can take up values $0, 1, 2 \dots$ for bosons and $\frac{1}{2}, \frac{3}{2}, \frac{5}{2} \dots$ for fermions. As was already pointed out in \fullref{sec:whymassive}, SM only uses fields of spin $0, \frac{1}{2}$ and $1$ to describe all known physics except gravity, which uses fields of spin $2$. 

\noindent
Naturally, this leads one to the curious question: What about Higher Spins (HS)?\footnote{Following literature, HS is used as a generic term to refer fields with spin $s>2$ \cite{highspinRahmanprimer2015}.} 

Even before the field theoretic description for HS is studied, one might wonder whether HS theories are interesting or  where do they arise in the present context of gravity. Or as Rabi might have said: Higher Spins, \emph{``Who ordered that?"} 
\section{Perspective: Why Higher Spins?}\label{whyhigherspins}

Thirty-five years after Fierz-Pauli's work, in 1974, Singh and Hagen constructed the lagrangian for fields of arbitrary spins, both bosons and fermions \cite{highspinsinghhagenboson, highspinsinghhagenfermion}\footnote{In fact, their two papers appeared back-to-back.}. Their motivation for this work was due to the technical challenge of their work, and also in part, perhaps, due to Dirac's remark: ``the underlying theory is of considerable mathematical interest". Later in 1978, Fronsdal extracts the equations governing massless HS fields from Singh and Hagen's work \cite{highspinfronsdal}. He explicitly cites twofold reasons for his interest: (1) supersymmetry predicts HS counterparts to known fields and therefore their consistent field theoretic description is needed, and (2) a thorough understanding of gauge symmetry associated with HS fields may allow one to construct a gauge principle for neutrinos (which were presumed as massless in those days). 

\noindent
It is a delight to contrast these historic motivations with the present day reasons for pursuing higher-spin gauge field theories. The following points are neither completely independent of each other, nor presented in any specific order of significance.
\begin{enumerate}
	\item \textbf{Tower of infinite spins:}
	Although theories of field with spin 2 or lesser are unique because they actually correspond to known natural phenomena, it seems these theories are unique in more ways. Any theory which contains a HS field, i.e. spin greater than 2, necessarily contains an infinite number of fields of all spins \cite{highspinvasilievelements}. Two apt-quoted examples of such theories are String theory and Vasiliev theory. Vasiliev theory, which provides a consistently-interacting theory for the tower of infinite spins in backgrounds of constant curvature for all dimensions, is a minimal theory whose spectrum consists of each massless HS fields occurring once, much simpler than the massive excitations in string theory. Interestingly, this theory does not have an associated energy scale and is, therefore, seen as a toy model for a fundamental theory beyond the Planck scale. Such a feature is general for theories with HS gauge symmetries, and noting that effects of quantum gravity become prevalent at this scale makes these theories important \cite{highspinRahmanprimer2015}. 
	
	\item \textbf{Peculiarities in 2+1 dimensions:}
	Although HS gauge theory may shed lights on trans-planckian physics, dealing with HS gauge theories can be quite complicated. It is here that 2+1 dimensions come to rescue. A feature of the theories mentioned above in three dimensions is that to obtain consistent interactions, there is no need for considering the infinite tower. Additionally, for theories which satisfy the AdS/CFT correspondence, GR in 2+1 dimensions has been considered as a simpler case of a wider class of theories with HS gauge fields  \cite{highspinasymptoticsymmetries}. This makes the case for studying HS theories necessary, as well as, natural for gaining a better understanding of quantum gravity in 2+1 dimensions.

	\item \textbf{String theory's tensionless limit:}
	Strings in string theory have an important property called string tension, which arises as a coefficient in the Nambu-Goto action. In the limits of string tensions being zero, all massive HS excitations in this theory become massless. This is because in the tensionless limit, there appears to be a large enhancement of string theory symmetry to that of a massless HS gauge symmetry. Motivated by this enhancement of symmetry, it is even conjectured that string theory is a spontaneously broken phase of an underlying HS gauge theory. Originally, the chance to better understand quantum nature of string theory provided an impetus for the development of theories of HS gauge fields with consistent interactions \cite{highspinclassicalintro}. 
	 
	\item \textbf{No-go theorems:}
	Although, Singh and Hagen's work and Fronsdal's construction was available since the late 70s, there existed major road-blocks in the development of HS gauge theories. These come in the form of two, powerful and classic, no-go theorems. These theorems describe the nature of difficulties encountered while constructing HS gauge field theories with consistent interactions in flat space. The Coleman-Mandula theorem strongly prohibits conserved charges (or symmetry generators) associated with a HS gauge group algebra; thereby severely restricting the symmetries associated with the S-matrix of an interacting QFT in four-dimensional Minkowski spacetimes. The Weinberg-Witten theorem disallows the energy momentum tensor associated with a particle of spin 2 or more to be both gauge invariant and Lorentz covariant in a flat background. This does not prevent gravitons from interacting with matter or itself, merely restating that a gravitational field's energy cannot be localized. These statements stem from the fact that interactions in HS gauge field theory seemingly contradict certain basic assumptions which are an input in the setup of canonical QFT, necessarily implying that such interactions must be unconventional. In spite of these no-go theorems, considerable success and breakthroughs has allowed the construction and analysis of HS gauge field theories featuring consistent interactions. Among many other ways, the dual use of $\Lambda$ as both a coupling constant and a cosmological constant,  allows one to evade these no-go theorems (which work in flat spaces only). The literature on these issues is wide and interested readers are referred to the reviews \cite{highspinRahmanprimer2015, highspinnogo}. Due to some exciting yes-go results obtained in the last few decades this field has seen a resurgence of interest.
\end{enumerate}

The above mentioned in-exhaustive points indicate that there is a deep theoretical connection between HS gauge field theories, String theory and possibly even quantum gravity. This makes pure HS gauge field theories almost an imperative for further work in those theories. There is a more down-to-earth application: many HS hadronic resonances have been observed in nature. HS field theory could in principle describe the dynamics of these composite particles, which are currently described using complicated form-factors. Finally, HS gauge field theory represents a generalization of physics based on lower spin-fields. Noting that these lower spin fields comprise most of known fundamental physics, to the author, a generalization to HS is worthy of attention in its own right.

\section{Irreducible Representations: Poincar\'e Group in 2+1 dimensions}\label{sec:irrep}

Invariance under the transformations of the Poincar\'e group (which includes rotations, boosts and translations) is a natural, as well as, technical demand which is imposed upon physical theories. These transformations form a mathematical group and hence ideas from group theory become quite relevant in field theories. Wigner's theorem, mentioned earlier, is a group theoretical result and provides, in some sense, a connection between the mathematical description of fields with physical particle in nature. Since these ideas are textbook-old, only a brief overview is given. Apart from mentioning some details pertinent to the case of 2+1 dimensions, the reader is referred to the original paper by Binegar \cite{groupBinegar} and an excellent review valid for $d \geq 3$ \cite{groupbekaert}. 

The subgroup of linear inhomogeneous proper orthochronous Lorentz transformations is called the Poincar\'e group, $ISO(d-1,1)^{\uparrow}$. Being a Lie group, the commutative properties of all transformation generators of the Poincar\'e group constitute, what is called, a Lie algebra. The importance of studying Lie algebra lies in the fact that these are properties of the group which are independent of the group representations. Groups are studied via their representations, which is an operation that assign a linear operator to an abstract elements of the group. The Lie algebra for $ISO(d-1,1)^{\uparrow}$ is given below. These equations describe the difference between performing two subsequent transformations on a $d$ dimensional vector (rotations, boosts or translations) in one order and then the other way around.
\begin{equationsplit}
	i [M_\munu, M_{\rho\sigma}] &= \eta\indices{_\nu_\rho} M\indices{_\mu_\sigma} - \eta\indices{_\mu_\rho} M\indices{_\nu_\sigma} - \eta\indices{_\sigma_\mu} M\indices{_\rho_\nu} + \eta\indices{_\sigma_\nu} M\indices{_\rho_\mu}
	\\
	i[P_\mu, M\indices{_\rho_\sigma}]  &= \eta\indices{_\mu_\rho} P_\sigma - \eta\indices{_\mu_\sigma} P_\rho
	\\
	i[P_\mu, P_\nu] &= 0
\end{equationsplit}

While studying representations, an important role is played by Casimir operators. These operators, having the special property of commuting with all generators, are proportional to the identity. The constant of proportionality provides a label for the representations of the group. There are two Casimir operators for the Poincar\'e group: (a) the square of momentum $P_\mu P^\mu$, and (b) the Pauli-Lubanski vector $W_\mu = \frac{1}{2} \epsilon_{\mu\nu\rho\sigma} M^{\nu\rho} P^\sigma$, giving the labels of mass $m$ and spin $s$ respectively. \footnote{There is an interesting group-theoretical result due to Beltrametti and Blasi, to find the number of Casimir operators for any Lie algebra \cite{numberofcasimir}.}

Wigner's celebrated theorem re-expressed the demand of positivity of energy from Fierz-Pauli to the condition that one particle states carry a Unitary Irreducible Representation (UIR) of the Poincar\'e group. The next task in this logical scheme is to classify different representations of the Poincar\'e group based upon the UIR's they carry. Wigner introduced the method of induced representation, based upon the representations of the stability sub-group called little group (transformations which leave the momenta invariant). All possible states of different momenta's which can be connected to any chosen momentum state using boosts forms an orbit. There are 6 classes of orbits, whose classification leads to the following UIRs\footnote{Only relevant UIR's for three dimensions are discussed. There are also UIR's corresponding to tachyons, anyons etc.}:
\begin{enumerate}
	\item \textbf{Massive particles with spin $s$}\\
	The orbit of such states satisfies $p^2 = m^2$ and forms a hyperboloid mass-shell. A momenta state can be picked as: 
	$$\text{Particle's rest frame: } p_\mu = (m, 0, 0) $$
	
	As can be seen the little group is $SO(2)$, with complex dimension one (real: two), this implies that massive states in 2+1 dimensions should have 2 propagating degrees of freedom (dof), \emph{irrespective} of their spin.
	
	\item \textbf{Massless particles with discrete spin}\\ 
	The orbit of such states satisfies $p^2 = 0$ and forms a light-cone. An exemplary momentum state, for energy $E$ is:
	$$\text{Motion along y-axis: } p_\mu = (E, 0, E) $$
	
	Evidently, the little group is: $ Z\otimes \mathbb{R}$, where $Z = \{1,-1\}$ is simply the multiplicative group and $\mathbb{R}$ is the group of real numbers. There are two UIRs (upto equivalence) of $Z$. Due to the degeneracy of $\mathbb{R}$, this orbit corresponds to excitation with no polarizations. Equivalently, massless UIRs in 2+1 dimensions of any spin correspond to \emph{scalar fields}\footnote{Although, not transparent from the presented remarks, scalar and spinors are the only kinds of massless particles that can propagate in 2+1 or lower dimensions\cite{groupBinegar, groupbekaert, highspinRahmanprimer2015}.}. 
\end{enumerate}

Interestingly, the group of massless UIRs is degenerate: implying that, group-theoretically, for massless UIRs there is only one spin \cite{groupBinegar}. This implies: Firstly, GR in 2+1 dimensions will have no propagating dof \footnote{This will also be seen from the vanishing of the Weyl tensor in \fullref{sec:gravity2+1}, and from a comprehensive dof counting in \fullref{sec:countinghighspindof}}. Additionally, even HS gauge field theories will suffer the same end in 2+1 dimensions. Being gauge theories they necessarily deal with massless excitations, and for 2+1 dimensions this already implies that there will be no local dof in such theories \cite{highspin2+1Campoleoni:2011tn, highspinasymptoticsymmetries}. This supports the idea that GR in 2+1 dimensions can be viewed as a specific example of HS theories. Finally, the absence of dof for fields of spin $s>1$ will also arise in \fullref{sec:countinghighspindof}, in which the dof for massive and massless fields of arbitrary spin and in arbitrary dimensions will be comprehensively calculated. 

It is important to note that such a group theoretical analysis cannot probe topological theories, such as pure Chern-Simon theories\footnote{By pure Chern-Simons theory it is meant that the theory contains only a Chern-Simons term.}. This is due to the lack of physical dof in such theories, which corresponds to identically vanishing UIRs of their little groups. In this sense, there is an equivalence between pure GR/HS theories and topological theories \cite{Witten:2007kt}. Nevertheless, local dof can come into existence by deforming a pure HS theory with topological terms, resulting in a rather, curious interplay between these two kinds of theories (example - topologically massive gauge theories). 

Finally, a comment upon how these physical UIRs are incorporated in QFT. Since the transformation properties of individual UIRs are diverse and complicated, constructing a theory describing interactions between multiple UIRs in a covariant way is quite involving. Instead in QFT, one repacks these UIRs neatly into tensorial fields with definite transformation properties which guarantees covariance right from the start\footnote{Due to Poincar\'e's duality, only symmetric tensors are required for $d=3$ or $4$. In higher dimensions, tensors of mixed symmetry also need to be considered.}. The price for replacing UIRs with covariant fields, which are generally not irreducible, is the propagation of many unwanted dofs. These unphysical dofs, then, have to be removed using subsidiary conditions, preferably coming from a well-chosen Lagrangian. 

\section{Massless Fields: Fronsdal Formulation}\label{sec:fronsdalformulation}
Since individual UIRs are packaged into covariant tensorial fields which may not be irreducible representations themselves, a correct description for HS usually requires multiple lower-spin auxiliary fields which vanish on-shell. In the successful approach of Singh and Hagen, the description of massive HS field of spin-$s$ requires, the following symmetric traceless tensorial fields\footnote{\label{footnote:extradimensions}An interesting way of obtaining a massive spin-$s$ lagrangian in $d$ dimensions is to perform a Kaluza-Klein reduction of a massless spin-$s$ field in $(d+1)$ dimensional manifold, and compactify the extra dimension as a circle with radius $\frac{1}{m}$\cite{highspinRahmanprimer2015}. This approach is rather general given that consistent theories of massive gravity have also been obtained in a similar manner (see part-1 of \cite{reviewdeRham}). Additionally, in \fullref{sec:countinghighspindof} this will be explicitly seen from a general counting of the number of dof for both massive and massless fields.}:
$$ \psi_{\mu_1\mu_2\dots\mu_s},\  \underbrace{\psi_{\mu_1\mu_2\dots\mu_{s-2}}, \psi_{\mu_1\mu_2\dots\mu_{s-3}}, \dots ,\psi_{\mu_1}, \psi}_\text{auxiliary fields} $$

The Singh-Hagen Lagrangian can be considerably simplified for describing massless spin-$s$ fields. Indeed, as worked out by Fronsdal, only two traceless symmetric tensorial fields are required. These are:
$$ \psi_{\mu_1\mu_2\dots\mu_s}\ \text{and}\ \psi_{\mu_1\mu_2\dots\mu_{s-2}}$$

These two traceless fields are combined into one single field, referred to as the Fronsdal field\footnote{Writing explicit dependence of these tensorial fields upon spacetime has been supressed for brevity, compactness and readability.}:
\begin{equation}\label{eq:fronsdalfield}
	\phi_{\mu_1\mu_2\dots\mu_s}\ = \psi_{\mu_1\mu_2\dots\mu_s} + \eta_{(\mu_1\mu_2}\psi_{\mu_3\mu_4\dots\mu_s)}
\end{equation}
Notably, a \emph{double-tracelessness} condition for the Fronsdal field $\phi_{\mu_1\mu_2\dots\mu_s}$ follows immediately.
\begin{equationsplit}\label{eq:doubletracelessness}
	\eta^{\mu_i\mu_j}\ \eta^{\mu_k\mu_l}\ \phi_{\mu_1\mu_2\dots\mu_s} = 0 
	\quad \quad
	\forall \ i,j,k,l = \{1,2,3,\dots s\}
\end{equationsplit}
This is a good starting point to sketch out some details of the Fronsdal theory of massless HS fields. The field equation for Fronsdal theory is neatly stated in terms of the Frondal tensor ($F_{\mu_1\dots\mu_s}$), given by:
	\begin{equation}\label{eq:fronsdaltesnor}
	F_{\mu_1\dots\mu_s}(\phi) = \Box \phi_{\mu_1\dots\mu_s} - s\  \partial_{(\mu_1}\partial^\sigma\phi_{\mu_2 \dots\mu_s)\sigma} + \frac{s(s-1)}{2}\ \partial_{(\mu_1}\partial_{\mu_2} \phi_{\mu_3 \dots\mu_s)}{_\sigma}{^\sigma}
	\end{equation}
The fields equations in Fronsdal theory compactly becomes:
\begin{equation}\label{eq:fronsdalequation}
	F_{\mu_1\dots\mu_s}(\phi(x)) = 0
\end{equation}
As it should, the Fronsdal equation reduces to well known equations of motion: 
\begin{equationsplit}
	&\textbf{spin-1:} \quad \quad \partial^\mu F_\munu = 0
	\\
	&\textbf{spin-2:} \quad \quad R_\munu = 0
\end{equationsplit}
The above equations can be readily recognized as the Maxwell's equation for electromagnetism in terms of the electromagnetic field strength tensor $F_{\munu}$, and the linearized vacuum Einstein equations for the Ricci tensor $R_\munu$.

There are four key aspects about Fronsdal theory which are of immediate relevance. These are: (a) Gauge symmetry in the Fronsdal theory, (b) Explicit verification of massless excitations, (c) Lagrangian formulation, and (d) Counting the number of dof. The first three are dealt in the following sub-sections. The last will be dealt in a more general way in \fullref{sec:countinghighspindof}
\subsection{Gauge symmetry in Fronsdal Theory}
The well known examples, of spin-1 and spin-2 mentioned above, satisfy gauge symmetries. Gauge symmetries, in a  very broad sense, can be thought of as a redundancy in the chosen description of nature. The UIR's coming from the representations of Poincar\'e group are packed in a covariant field which has many more independent components. This brings the redundancy and gauge symmetry can be thought of as the freedom to use any of those independent components in the tensorial field as the UIR. Additionally, since the Fronsdal theory is for massless fields, it should be expected that \eqnref{eq:fronsdalequation} is invariant under a gauge transformation. The gauge transformations for the Fronsdal field $\phi_{\mu_1\dots\mu_s}$ are defined as\footnote{Prime $'$ here does not indicate a trace!}:
\begin{equationsplit}\label{eq:fronsdalfieldgauge}
	&\phi_{\mu_1\dots\mu_s} \rightarrow \phi'_{\mu_1\dots\mu_s} = \phi_{\mu_1\dots\mu_s} + \delta \phi_{\mu_1\dots\mu_s}
	\\
	& \delta \phi_{\mu_1\dots\mu_s} = \partial_{(\mu_1} \xi_{\mu_2\dots\mu_s)}
\end{equationsplit} 
Here, the gauge parameter $\xi_{\mu_1\dots\mu_{s-1}}$ is an arbitrary symmetric tensor. This results in the following gauge transformation for the Fronsdal tensor:
\begin{equation}
	\delta F_{\mu_1\dots\mu_s} \propto \partial_{(\mu_1}\partial_{\mu_2} \partial_{\mu_3} \xi_{\mu_4 \dots\mu_s)}{_\sigma}{^\sigma}
\end{equation}
Thus, gauge invariance of the Fronsdal equation \eqnref{eq:fronsdalequation}, forces the gauge parameter $\xi_{\mu_1\dots\mu_{s-1}}$ to be a traceless tensor. Note, the tracelessness of the gauge parameter necessarily implies gauge invariance of the double-tracelessness condition of the Fronsdal field \eqnref{eq:doubletracelessness}, since any gauge-variation in this equation will necessarily involve tracing the traceless gauge-parameter. The double-tracelessness condition will only arise when dealing with fields of spin$s\geq4$. For fields with spin 1 and spin 2, the gauge-symmetry for this lagrangian will be verified individually when those cases are discussed in later chapters. 

\subsection{Massless excitations in Fronsdal Theory}
Since, the gauge invariance of Fronsdal theory is established, that it indeed propagates massless excitations can also be seen. To this end, denoting the trace of Fronsdal field by $\phi'_{\mu_3\dots\mu_s}$, its gauge variation is:
\begin{equation}
	\phi'_{\mu_3\dots\mu_s} \propto \partial^{\mu_1} \xi_{\mu_1\dots\mu_{s-1}}
\end{equation}
Thus, a partial gauge choice can be made for the trace part of the Fronsdal field (by setting $\partial^{\mu_1} \xi_{\mu_1\dots\mu_{s-1}}$ to zero):
\begin{equation}
	\phi'_{\mu_3\dots\mu_s} = 0
\end{equation}
Thus, the Fronsdal equation reduces to:
\begin{equationsplit}
	F_{\mu_1\dots\mu_s} &= \Box \phi_{\mu_1\dots\mu_s} - s\  \partial_{(\mu_1}\partial^\sigma\phi_{\mu_2 \dots\mu_s)\sigma}
	\\
	&= \partial^{\mu_1} \partial^{\mu_2} \phi_{\mu_1\dots\mu_s}  = 0	
\end{equationsplit}
Note, the Einstein summation convention was used to get rid of the symmetrization present above. There is still some residual gauge symmetry in the Fronsdal equation which can be seen from:
\begin{equationsplit}
	\del^{\sigma} \delta \phi_{\mu_2\dots\mu_s\sigma}  &= \del^{\sigma} \partial_{(\mu_2}\xi_{\mu_3\dots\mu_s\sigma)} 
	\\
	&= \frac{1}{s}\ \Box \xi_{\mu_2\dots\mu_{s}}
\end{equationsplit}
Note, additional terms which would have arised above were already fixed with the first gauge choice. This can be used to further fix an additional gauge as:
\begin{equation}
	\partial^{\mu_1} \phi_{\mu_1\dots\mu_s} = 0
\end{equation}
With the above choices, the remaining Fronsdal equation simply becomes (in addition to the gauge conditions chosen above):
\begin{equationsplit}
	F_{\mu_1\dots\mu_s} = \Box \phi_{\mu_1\dots\mu_s} &= 0
	\\
	\phi'_{\mu_3\dots\mu_s} &= 0
	\\
	\partial^{\mu_1} \phi_{\mu_1\dots\mu_s} &= 0
\end{equationsplit}

The first equation, in the above three, is a Klein-Gordon type equation for a massless particle. This demonstrates that the Fronsdal theory, indeed, propagates massless excitations. Additionally, as mentioned in the beginning of \fullref{sec:fronsdalformulation}, the Fronsdal theory required an auxiliary field of spin-$(s-2)$, which was used to make the Fronsdal field in \eqnref{eq:fronsdalfield}. Its unphysical pure-gauge nature has emerged as the second equation above, whereby using gauge freedom this component of the Fronsdal field is set to zero. 

This section is closed by mentioning that the residual gauge symmetry obeys:
\begin{equationsplit}
	\Box \xi_{(\mu_1\dots\mu_{(s-1)})} &= 0
	\\
	\partial^{\mu_1} \xi_{\mu_1\dots\mu_{s-1}} &= 0
	\\
	\xi'_{\mu_3\dots\mu_{s-1}} &=0
\end{equationsplit}

\subsection{Fronsdal Action}\label{subsec:fronsdalaction}
An important quantity to neatly define the Fronsdal lagrangian is the trace-reversed Fronsdal tensor $\overline{F}_{\mu_1\dots\mu_s}$. This is defined as:
\begin{equation}\label{eq:tracereversedfronsdal}
	\overline{F}_{\mu_1\dots\mu_s} = F_{\mu_1\dots\mu_s}\ - \frac{1}{2} \eta_{(\mu_1\mu_2} F_{\mu_3\dots\mu_s)\sigma}{^\sigma}
\end{equation}
At spin-1, this object is equivalent to the divergence of the electromagnetic field strength tensor $\partial^\mu F_\munu$, and at spin-2 this object is equivalent to the linearized Einstein tensor $G_\munu$. 

The lagrangian which leads to the Fronsdal equation, now has a compact form:
\begin{equationsplit}\label{eq:lagrangianfronsdal}
	\mathcal{L}_{Fronsdal} &= \frac{(-1)^{s+1}}{2} \phi^{\mu_1 \mu_2...\mu_s} \overline{F}_{\mu_1 \mu_2...\mu_s}
	\\
	\sigma_s\ \mathcal{K}_{(\phi_{\mu_1\dots\mu_s})}&=\sigma_s\ \phi^{\mu_1 \mu_2...\mu_s} \overline{F}_{\mu_1 \mu_2...\mu_s}
\end{equationsplit}
The overall sign and the pesky factor of $\tfrac{1}{2}$  is needed to ensure standard normalization of the kinetic term and unitary positive-definite Hamiltonian, for the `mostly-minus' metric signature employed in this thesis. In the second equation, written for clarifying future notational conventions, this lagrangian is labelled as the kinetic-term $\mathcal{K}_{(\phi_{\mu_1\dots\mu_s})}$ for a spin-$s$ field. Since this lagrangian will be used again-and-again in this thesis, an overall constant factor $\sigma_s$ is defined which contains both the factor $\tfrac{1}{2}$ and an overall spin-dependent sign. For the author, this greatly simplified keeping overall factors separate from the dealing of tensorial structures. There should be no confusion between the usage of $\sigma$ as an index or as an overall constant. For later reference, 
\begin{equation}\label{eq:factorsigmas}
	\sigma_s = \frac{(-1)^{s+1}}{2} 
\end{equation}
In $d$-dimensions, the above lagrangian lead to the Fronsdal Action:
\begin{equation}\label{eq:actionfronsdal}
	S = \int d^dx \quad \phi^{\mu_1\dots\mu_s} \overline{F}_{\mu_1\dots\mu_s}
\end{equation}

This leads to the equations of motion for the trace-reversed Fronsdal tensor, which can be explicitly checked to be equivalent to \eqnref{eq:fronsdalequation}. Also, the gauge-invariance of this action can be firmly established \cite{highspinvasilievnotes, highspinKessel:2017mxa}. In the next section, a general degree-of-freedom count for massive and massless fields of arbitrary spins for $d$-dimensions is presented. The massless case, \fullref{subsec:countinghighspindofmassless}, establishes that the Fronsdal equation does indeed propagate the correct number of degrees-of-freedom.

\section{Degrees of freedom in $d$-dimensions}\label{sec:countinghighspindof}

In this section, a general calculation of the number of propagating physical dof for a general field of spin-$s$ for both massive and massless fields in arbitrary dimensions is presented. Such a calculation is not only a mathematical curiosity, but also wields the power to bring out certain very general theory-independent physical conclusions. The calculations have been divided into two eponymous subsections. 

\subsection{Massless fields of spin-$s$}\label{subsec:countinghighspindofmassless}
A massless spin-$s$ field is denoted with a tensor of rank $s$, such as the Fronsdal field \eqnref{eq:fronsdalfield}. Due to the symmetric nature of the tensor, the order of indices is unimportant.

A symmetric rank-$s$ tensor in $d$ dimensions requires $c_1$ independent components, where $c_1$ is:
\begin{equation}\label{eq:c1}
	c_1 = \binom{d-1+s}{s} = \frac{(d-1+s)!}{(s)!\ (d-1)!}
\end{equation}
For $s=2$, $d=4$ the above formula gives $c_1 = 10$ and $s=2$, $d=3$ implies $c_1 = 6$. This matches with the usual counting done with a matrix representation of spin-2 tensors. Next, the double-traceless condition of  \eqnref{eq:doubletracelessness} will remove $c_2$ components. The gauge condition is described by a rank-$(s-1)$ symmetric traceless tensor. Fixing a partial gauge removes $c_3$ components and fixing the residual gauge removes another $c_4$ components. These components are given by:
\begin{equationsplit}
	c_2 &= \binom{d-1+s-4}{s-4}
	\\
	c_3 &= \underbrace{\binom{d-1+s-1}{s-1}}_\text{symmetric} - \underbrace{\binom{d-1+s-3}{s-3}}_\text{traceless}
	\\
	c_4 &= \binom{d-1+s-1}{s-1} - \binom{d-1+s-3}{s-3}
\end{equationsplit}
Finally, the number of dof associated with a massless spin-$s$ field becomes:
\begin{equationsplit}\label{eq:dofmasslessspins}
	\text{dof spin-s, massless} &= c_1 - c_2 - c_3 - c_4
	\\
	&= \frac{(d-4+2s)\ \Gamma(d-4+s)}{\Gamma(d-3)\ \Gamma(s+1)}
\end{equationsplit}
This formulae is equivalent to eq(3.15) of reference \cite{highspinKessel:2017mxa} and eq(2.29) of reference \cite{highspinRahmanprimer2015}\footnote{The two equations cited do not have the same representation of the formula given here. Yet, these two formulas including the one presented here are equal.}. Putting numbers, this formula gives:
\begin{equationsplit}
	\text{For d=6, arbitrary spin-s} \quad &= \quad (1+s)^2
	\\
	\text{For d=5, arbitrary spin-s} \quad &= \quad 2s+1
	\\
	\text{For d=4, arbitrary spin-s} \quad &= \quad 2
	\\
	\text{For d=3, arbitrary spin-s} \quad &= \quad0
	\\
	\text{For d=2, arbitrary spin-s} \quad &= \quad0
	\\
	\text{For d=1, arbitrary spin-s} \quad &= \quad0
	\\
	\text{For d=0, arbitrary spin-s} \quad &= \quad0	
\end{equationsplit}

Some comments are now in order. First, the formula derived in \eqnref{eq:dofmasslessspins} gives the correct number of dof for massless fields in 4-dimensions, which is a well known result. Second, for 3-dimensions, this formula correctly predicts that the number of dof for HS fields should be zero, as was seen from group-theory arguments in \fullref{sec:irrep}. Third, the case of spin-1 in 3-dimensions is special since it is equivalent to a scalar field. The derivation presented above made explicit use of the double-traceless condition which cannot be defined for a spin-1 field. Fourth, the presence of $\Gamma(n) = (n-1)!$ should not cause much worry, since this formula should not be applied to cases where it is definitely not valid! 

Additionally, for 5-dimensions, this formula predicts $dof=2s+1$. As will be seen later, this is exactly the same number of dof for a massive spin-s field in 4-dimensions. This is not a mere coincidence. As was remarked in \fullref{footnote:extradimensions}, this is a very peculiar and general feature, which motivates the study of higher-dimensions/extra-dimensions and their compactifications. It is also interesting to note, that in 2-dimensions (1-space and 1-time), there are no massless dof's. Further, for even lower dimensions of 1 or even 0, (where counting space and time dimensions separately does not have a well defined meaning) - this formula neatly produces 0.

Clearly, the above formulae should not work if a theory uses a mass generating mechanism which has not been accounted for. There is a ready example - topologically massive gauge theories, with a Chern-Simons mass terms. 

\subsection{Massive fields of spin-$s$}\label{subsec:countinghighspindofmassive}
In the case of massive fields, there will be no complication due to gauge symmetry and unphysical modes. A massive spin-s field will in general follow Fierz-Pauli conditions which can be derived from the Singh-Hagen lagrangian. Thus, the number of dof should be equal to the number of independent components in a symmetric, traceless, divergence-free tensor of rank-$s$. A symmetric tensor of rank-$s$ in $d$-dimensions has $c_1$ independent coefficients. The traceless condition will remove $c_5$ components. To eliminate all lower spin auxiliary fields, a condition of the type: 
$$ \partial^{\mu_1} \psi_{\mu_1\dots\mu_s} = 0 $$
is used. This makes the traceless symmetric tensor $\psi_{\mu_1\dots\mu_s}$ divergenceless as well. Using this condition, another $c_6$ components can be removed. However, the trace part of the divergenceless condition, $c_7$, was already removed by $c_5$. 
\begin{equationsplit}
	c_5 &= \binom{d-1+s-2}{s-2}
	\\
	c_6 &= \binom{d-1+s-1}{s-1}
	\\
	c_7 &= \binom{d-1+s-3}{s-3}
\end{equationsplit}
Together, the number of dof in a massive spin-s field becomes:
\begin{equationsplit}\label{eq:dofmassivespins}
		\text{dof spin-s, massive} &= c_1 - c_5 - c_6 + c_7
		\\
		& = \frac{(d-3+2s)\ \Gamma(d-3+s)}{\Gamma(d-2)\ \Gamma(s+1)}
\end{equationsplit}
Plugging in the numbers, 
\begin{equationsplit}
	\text{For d=5, arbitrary spin-s} \quad &= \quad (1+s)^2
	\\
	\text{For d=4, arbitrary spin-s} \quad &= \quad 2s+1
	\\
	\text{For d=3, arbitrary spin-s} \quad &= \quad 2
	\\
	\text{For d=2, arbitrary spin-s} \quad &= \quad0
	\\
	\text{For d=1, arbitrary spin-s} \quad &= \quad0
	\\
	\text{For d=0, arbitrary spin-s} \quad &= \quad0
\end{equationsplit}

One immediately recognizes that this formula predicts the correct number of dof for massive spin-$s$ in 4 dimensions, the familiar $2s+1$. In 2+1 or 3 dimensions, for a particle with arbitrary spin there are 2 massive dofs. Thus, the number of dof for massive gravity or massive electrodynamics is 2, in 2+1 dimensions. These dof's will be carefully studied when they arrive in their respective chapters. 

Additionally, the number of massive dofs in 5-dimensions is the same as the number of massless dof in 6-dimensions. In fact, using the explicit formulas given above, the following can be asserted, for arbitrary spin-s:
\begin{equation}
\binom{\text{Number of massless dof}}{\text{in d+1 dimensions}} = 	\binom{\text{Number of massive dof}}{\text{in d dimensions}} 
\end{equation}

Finally, note that for a field of spin-$s$, its massive version always has more dofs than the massless counterpart (except ofcourse for very low dimensions or spin-0 fields). The reasons for this are discussed in detail in \fullref{subsec:stuckelbergspin2}, where the \stuckelberg analysis is used to understand the theory of massive spin-2 fields. 

\section{Topologically Massive Higher Spin Fields}\label{sec:highspintopologicallymassive}
As has been noted on multiple occasions, HS gauge fields do not propagate any local physical dof in 2+1 dimensions. This is a rather general statement which includes GR, Vasiliev theory and other HS gauge field theories \cite{highspinvasilievnotes}.  Additionally, a theory for massive fields can be obtained from a corresponding massless version set in higher-dimensions after compactification of the extra dimension \cite{reviewdeRham, highspinRahmanprimer2015}. In this sense, massless fields are like building blocks of their massive cousins. Nevertheless, there are other mass generating mechanisms which can deform the massless theory in other interesting ways. Conversely, there are topological field theories: such as pure Chern-Simon theories which are described with a Chern-Simons (CS) term in the Lagrangian. Such a theory in 2+1 is same in terms of its physical content with conventional GR in 2+1 dimensions for spin-2, and perhaps also for pure HS gauge field theories \cite{Witten:2007kt}. Hoping to learn lessons about quantum gravity from Chern-Simon theories was one of the motivations for studying such topological field theories.

An interesting mass generating mechanism is by having both these kinds of theories, together simultaneously in the same lagrangian. Such theories are called Topologically Massive Gauge Theories. Such theories for spin-1 (topologically massive electrodynamics(TME or QED3)) and spin-2  (topologically massive gravity (TMG)), exciting as they are, have been studied extensively for a long time. Additionally, theories with interactions between HS fields and gravity have also been looked into \cite{highspin2+1Campoleoni:2011tn, highspinasymptoticsymmetries}. It would be interesting, imperative even, to have and study theories in which pure HS gauge fields were made massive, in a manner analogous to the TME/TMG\footnote{The author is currently unaware regarding the availability of any such work in the literature.}. To this end, in this section, a linearized Chern-Simons terms based on the Fronsdal formulation is proposed. The CS terms for spin-1 and spin-2 have beautiful origins in the topology of the underlying manifold. The author has not worked out such a derivation for the extended spin-s CS term. In \fullref{ch:spin1} and \fullref{ch:spin2}, it will be seen how a Lagrangian based on this term gives rise to the theories of TME and TMG respectively.

Among other reasons, the CS term is said to be topological because it does not involve any metric. A metric is intimately tied to the geometric properties of a manifold and therefore leaving it out of the description, perhaps, leads to a description based only on topological properties. The CS term uses the epsilon symbol $\epsilon_{\alpha\beta\gamma}$ for contracting the various fields with derivatives. A CS-type term which is conjectured to be valid for all HS fields is proposed below. The term is:
\begin{equationsplit}\label{eq:highspinCSterm}
	\mathcal{L}_{\text{cs spin-1}} &= \sigma_1\ \phi^\mu \partial^\alpha \epsilon_{\alpha\beta\mu} \phi^\beta
	\\	
	\mathcal{L}_{\text{cs spin-s}\geq 2} &= \sigma_s\ \phi^{\mu_1\dots\mu_s}\ \partial^\alpha \epsilon\indices{_\alpha_\beta_{(\mu_1}}\overline{F}_{\mu_2\dots\mu_s)}{^\beta}
\end{equationsplit}
In these definitions, $\phi_{\mu_1\dots\mu_s}$ is the Fronsdal field and $\overline{F}_{\mu_1\dots\mu_s}$ is the trace-reversed Fronsdal tensor. It is now worth seeing how this term functions for some special case. 

\textbf{Caveat:} The CS term for spin-1 needs some qualifying remarks. Since, the case for spin-1 in 2+1 dimensions is special, it has to be dealt separately. If one plugs $s=1$ in the second definition above, one finds that the right hand side is identically zero. Moreover, the CS term for HS is third-order in derivatives whereas there is only one derivative for spin-1 CS term. This leads to an important difference, described hereafter. There are two demands placed upon a CS term: (a) On its own, it should describe a theory that is physically equivalent to a corresponding metric based field theory such as electrodynamics, GR etc and (b) When taken together with a metric-based term it should generate a mass for the conventional field. The definition given here for spin-1 provides only the latter of the two purposes\footnote{This definition is chosen here since it leads to TME and because in this thesis topologically massive gauge theories will play a central role.}. Nevertheless, this has an easy cure, which is now given.
\subsection{Spin-1}\label{subsec:spin1csonly}
There are two aspects to be checked out. Firstly, a theory with only CS term, and secondly a theory with both CS term and a metric-based term. To deal with the first issue, a modified definition (marked with a prime: $'$) of CS term for spin-1 is necessary in line with the caveat mentioned above. 
\begin{equationsplit}\label{eq:csspin1modifiedlagrangian}
	\mathcal{L'}_{\text{cs spin-1}} = \sigma_1\ \phi^\mu \Box \partial^\alpha \epsilon_{\alpha\beta\mu} \phi^\beta
\end{equationsplit}
Adding, a box/d'Alembert operator makes the above term third-order in derivative. The equations of motion from such a term are simply:
\begin{equationsplit}\label{eq:spin1dualseom}
	\Box \tensor[^*]{F}{_\mu} &= 0
	\\
	\partial^\mu \tensor[^*]{F}{_\mu} &= 0
\end{equationsplit}
Here, $\tensor[^*]{F}{_\mu}$ is a dual-vector to the electromagnetic field strength tensor. It is defined as:
\begin{equation}\label{eq:dualvector}
	\tensor[^*]{F}{_\mu} = \half \epsilon_{\alpha\beta\mu} F^{\alpha\beta} = \epsilon_{\alpha\beta\mu} \partial^\alpha \phi^\beta
\end{equation}
The repeated usage of the letter $F$ to denote: Fronsdal tensor $F_{\mu_1\dots\mu_s}$, trace-reversed Fronsdal tensor $\overline{F}_{\mu_1\dots\mu_s}$, electromagnetic field strength tensor $F_\munu$, and now it's dual $\tensor[^*]{F}{_\mu}$ - is an unfortunate coincidence trickling here from history. Confusion arising from such abusive notation is deeply regretted. 

The first equation in \eqnref{eq:spin1dualseom} declares that the field excitations are massless in nature. The second is a by-product from the definition of dual-vector and does not arise from the Lagrangian. In 2+1 dimensions, the first equation gives 3 equations for each component of the dual-vector. The second can be used to kill one component, leaving a system of equations with two components. Moreover, this system possesses a gauge symmetry:
\begin{equationsplit}
	\delta \phi_\mu &= \partial_{\mu} \xi
	\\
	\text{implies: } \delta \tensor[^*]{F}{_\mu} &= 0
\end{equationsplit}
Thus, using a scalar gauge parameter another component can be fixed, leaving a single independent dof in the system.

Overall the CS spin-1 lagrangian used in \eqnref{eq:csspin1modifiedlagrangian} describes a massless field with one dof, or a massless scalar field. This will be seen to be equivalent to electrodynamics in 2+1 in \fullref{sec:3delectrodynamics}.

The other issue, mentioned at the beginning of this subsection, regarding massive excitations leads to the theory of Topologically Massive Electrodynamics (TME) which will be dealt in \fullref{sec:3dTME}.

\subsection{Spin-2}\label{subsec:spin2csonly}

It is now desired to study the CS term for higher spins defined in eq\eqref{eq:highspinCSterm} for the case of spin-2. 
\begin{equationsplit}\label{eq:lagcsspin2}
	\mathcal{L}_{\text{cs spin-2}} = \sigma_2\ \phi^{\munu}\ \partial^\alpha \epsilon\indices{_\alpha_\beta_{(\mu}}\overline{F}_{\nu)}{^\beta}
\end{equationsplit}
Objects of relevance are:
\begin{equationsplit}
	&F_\munu = \Box \phi_\munu - 2\ \partial_{(\mu} \partial^\rho \phi_{\nu)\rho} + \del_\mu \del_\nu \phi'
	\\
	&F' = 2 (\Box \phi' - \del\cdot\del\cdot\phi)
	\\
	&\overline{F}_\munu = \Box \phi_\munu - 2\ \partial_{(\mu} \partial^\rho \phi_{\nu)\rho} + \del_\mu \del_\nu \phi' - \eta_\munu (\Box \phi' - \del\cdot\del\cdot\phi)
\end{equationsplit}
Plugging these back into the lagrangian gives:
\begin{equation}
	\mathcal{L}_{\text{cs spin-2}} = \frac{\sigma_2}{2}\ \phi^{\munu}\ \Bigg(\del_\alpha \epsilon\indices{^\alpha^\beta_{\mu}} \big( \Box \phi_{\nu\beta} - \del_{\nu} \del^\rho \phi_{\beta\rho}\big) + \mu\leftrightarrow\nu \Bigg)
\end{equation}

The lagrangian above is the linearized version of the well-know CS term from TMG. The above lagrangian can be solved for an equation of motion. That equation of motion can be further analyzed. This is given below:
\begin{equationsplit}
	\text{eom: }\Bigg(\del_\alpha \epsilon\indices{^\alpha^\beta_{\mu}} \big( \Box \phi_{\nu\beta} - \del_{\nu} \del^\rho \phi_{\beta\rho}\big) + \mu\leftrightarrow\nu \Bigg)  &= 0
	\\
	\text{trace: }\eta^\munu\ \Bigg(\del_\alpha \epsilon\indices{^\alpha^\beta_{\mu}} \big( \Box \phi_{\nu\beta} - \del_{\nu} \del^\rho \phi_{\beta\rho}\big) + \mu\leftrightarrow\nu \Bigg) &= 0
	\\
	\text{divergence: }\del^\mu  \Bigg(\del_\alpha \epsilon\indices{^\alpha^\beta_{\mu}} \big( \Box \phi_{\nu\beta} - \del_{\nu} \del^\rho \phi_{\beta\rho}\big) + \mu\leftrightarrow\nu \Bigg) &= 0
	\\
	\text{gauge invariance: } eom(\phi'_\munu) = eom(\phi_\munu + \delta\phi_\munu) - eom(\phi_\munu) &= 0
	\\
	\text{with: } \delta\phi_\munu = \del_{(\mu} \xi_{\nu)}&
\end{equationsplit}

The first equation is symmetric in free indices $\mu$ and $\nu$. In d-dimensions, this equation would have $\tfrac{d(d+1)}{2}$ independent components (put s=2 in $c_1$ \eqnref{eq:c1}). The trace condition presented above can be used to remove one of these components. The divergence condition has one free index, and therefore are actually $d$ equations, fixing $d$ components. The equation of motion also satisfies gauge symmetry for an arbitrary gauge parameter $\xi_\mu$. This freedom can be used to fix another $d$ components. Thus, the remaining truly independent components are the true dof described by the spin-2 CS lagrangian. It is: \footnote{Although the presence of $\epsilon$ symbol makes the CS term specific only to 2+1 dimensions, counting in general d-dimensions and seeing the result is more gratifying to the author}
\begin{equation}
	\text{CS spin-2 dof: } \frac{d(d+1)}{2} -1 -d -d = \frac{d(d-3)}{2}
\end{equation}

Clearly, the lagrangian given in eq\eqref{eq:lagcsspin2}, does not allow any propagating degrees of freedom in 2+1 dimensions. This situation is physically equivalent to GR in 2+1 dimensions, where the identical vanishing of the Weyl tensor ensures that there are no propagating gravitons (as will be seen in \fullref{sec:gravity2+1}).

The other aspect for the CS-term is its ability to give massive excitations. For spin-2, this leads to a theory called Topologically Massive Gravity (TMG) and will be studied in \fullref{sec:TMG}

To conclude, in this chapter pure HS gauge field theories were discussed. Some general group-theoretic arguments was enough to conclude that there are no high-spin excitations in 2+1 dimensions. It was readily seen that pure HS gauge field theories in 2+1 dimensions are equivalent in physical content to pure CS term based theories. Even more curious, are theories in which these two are simultaneously present. It will be seen that these theories, although independently lack any physical dof, together give rise to a massive physical excitation. Such theories are interesting due to their special properties, as will be studied in the following chapters.

\chapter{Fields of Spin-1}\label{ch:spin1}
\epigraph{``There are some questions in Astronomy, to which we are attracted rather on account of their peculiarity, as the possible illustration of some unknown principle, than from any direct advantage which their solution would afford to mankind."}{J. C. Maxwell\cite{maxwell}}

Spin-1 fields exist in nature. In fact four fundamental spin-1 fields have been observed in nature. They come in two kinds: (a) Massive fields such as the $Z$-bosons, and $W^{\pm}$-bosons, and (b) Massless fields with the archetypical example $\gamma$-bosons or the Photons. Since physical phenomenon of electromagnetism and electromagnetic waves are ubiquitous, well-understood and their well-established theoretical description has been around for a very long time, they form a rigid body-of-knowledge to compare other theories of spin-1 fields. Although this thesis is devoted to the study of \emph{massive gravity}, studying theories of massive spin-1 field will serve quite useful. For one, due to its simplicity the difference between massive and massless fields is quite transparent for spin-1 fields. This may provide some experience for the reader to discern and distinguish between massive and massless theories, when the computations get involving at spin-2, as it did for the author. Moreover, a study on spin-1 will certainly allow some instinctive expectations for the imminent spin-2 theories to be developed. As it always is with expectations, it will be exciting to see what becomes of such expectations.    

This thesis being dedicated to physics in 2+1 dimensions, the author finds it important to briefly include some tenets of the planar electromagnetic theory.

\section{Perspective: Planar Electromagnetism}\label{sec:3delectrodynamics}

With only 2 dimensions of space and 1 time-keeping dimension available, it is certain that light (or photons) in 2+1 electromagnetism will be quite different from its usual 3+1-dimensional avatar. As a starter: Consider the two polarization vectors of electromagnetic waves, which are characteristically transverse to the direction of propagation. In 2+1 dimensions, there are simply not enough space dimensions to sustain this behaviour. For any chosen direction, at most there can only be one transversal direction in the plane. Now, recalling that a vector field has a single scalar dof in 2+1 dimensions (\fullref{sec:irrep}), one may deduce that electromagnetic waves in 2+1 dimension will have only one transverse polarization vector. This is indeed true, as will be seen shortly. 

The lagrangian describing electromagnetic phenomena of electric fields $E$ and magnetic field $B$ is expressed, covariantly, through the electromagnetic field strength tensor $F_\munu$. These are related to the gauge field $A_\mu$ as follows:
\begin{equationsplit}\label{eq:EBfields}
	&F_\munu = \del_\mu A_\nu - \del_\nu A_\mu
	\\
	&E_i = F_{0i} = \del_0 A_i - \del_i A_0
	\\
	&\epsilon_{ijk} B_k = - F_{ij} = - \del_i A_j + \del_j A_i 
\end{equationsplit}
These definitions are clearly invariant under gauge transformation: $A_\mu \rightarrow A_\mu + \del_\mu \xi$, for an arbitrary gauge parameter $\xi$. In $d$-dimensions, the Maxwell Lagrangian description with conserved matter source $J_\mu$ is:
\begin{equationsplit}\label{eq:lagrangianmaxwell}
	\mathcal{L}_{maxwell} &= \frac{-1}{4} F^\munu F_\munu  - \half A^\mu J_\mu
	\\
	&=  \half A^\mu \big( \Box \eta_\munu - \del_\mu \del_\nu \big) A^\nu - \half A^\mu J_\mu
\end{equationsplit}
Putting $\mu=\nu=0$ in the above lagrangian, one quickly notes that there is no kinetic term for the $A_0$ component of the spin-1 field. Further, these definitions are independent of the number of dimensions. However, there is a major difference when physics from 3+1-dimensions is compared to physics in 2+1-dimensions. This can be seen by expressing $F_\munu$ in terms of fields $E$ and $B$. For 2+1 dimensions, this tensor can be obtained crudely by cutting of one-dimension as follows:

\begin{equation}
	F_\munu = 
	\underbrace{
		\begin{pmatrix}
		 0		&E_x		&E_y		&E_z\\
		-E_x	&0			&-B_z		&B_y\\
		-E_y	&B_z		&0			&-B_x\\
		-E_z	&-B_y		&B_x		&0
		\end{pmatrix}
	}_{\text{3+1 dimensions}}
	\rightarrow
		\underbrace{
		\begin{pmatrix}
		0		&E_x		&E_y		\\
		-E_x	&0			&-B		\\
		-E_y	&B		&0			\\
		\end{pmatrix}
	}_{\text{2+1 dimensions}}
\end{equation}

An immediate difference to note is that the B-field, having lost two components, is now only a (pseudo-)scalar\footnote{It may seem weird that the $x$,$y$ component of the B-field were lost but not so for the E-field. After all, $E_x$ and $B_x$ must have been pointing in the same $x$-direction? The answer lies in the use of the word `crudely'. The cutting off of an entire dimensions in the matrix-representation should only be read symbolically. That the B-field loses its components in 2+1 dimensions, can be seen more elegantly from the last equation in eq\eqref{eq:EBfields}.}. Solving the lagrangian in eq\eqref{eq:lagrangianmaxwell} for the equations of motion (ignoring the source term), one obtains:
\begin{equationsplit}\label{eq:eomspin1massless}
	\Box A_\mu &= 0
	\\
	\text{Lorenz gauge: }	\del \cdot A &= 0
\end{equationsplit}

In deriving these equations a Lorenz gauge choice has already been made. The first equation describes the massless nature of the excitations. The first equation is actually three separate equations for each component; the second equation can be used to remove one component from the $A_\mu$ gauge field. Next, noting that the $A_0$ component does not have a kinetic term in the lagrangian eq\eqref{eq:lagrangianmaxwell}, there is only one independent component left in the $A_\mu$ field. Hence, it is concluded that this theory describes a scalar massless excitation. The above equations of motion are quite reminiscent of eq\eqref{eq:spin1dualseom}, where a similar structure emerged for the electromagnetic dual-vector when using a pure CS lagrangian. Indeed, that theory had the same physical content as the present theory. The polarizations associated with perturbations of the scalar dof, or equivalently, electromagnetic waves in 2+1 dimensions, are transverse to the direction of momenta. This will be firmly established in \fullref{subsec:propagatorspin1massless}.

Finally, it is interesting to note that the $E$-field and $B$-field find a natural place in the electromagnetic dual-vector, $\tensor[^*]{F}{_\mu}$, introduced in eq\eqref{eq:dualvector}:
\begin{equation}\label{eq:dualvector2}
\tensor[^*]{F}{_\mu} = \half \epsilon_{\alpha\beta\mu} F^{\alpha\beta} = 
	\begin{pmatrix}
	-B \\
	E_y \\
	-E_x
	\end{pmatrix}
\end{equation}  

As promised in \fullref{subsec:fronsdalaction}, here it is checked whether the pure HS gauge field formulation due to Fronsdal does indeed reduces to the theory described above. This can be easily checked from the Fronsdal action eq\eqref{eq:lagrangianfronsdal}. 

\begin{equationsplit}\label{eq:lagrangianspin1}
	\mathcal{L}_{\text{Fronsdal spin-1}} &= \sigma_1 \ \phi^\mu \overline{F}_\mu
	\\
	&= \sigma_1\ \phi^\mu\ \big( \Box \eta_\munu - \del_\mu \del_\nu \big)\ \phi^\nu
\end{equationsplit}

Note, that this is the same as source-less Maxwell Lagrangian in eq\eqref{eq:lagrangianmaxwell} ($\sigma_1 = \half$). Thus, the physics that shall follow from these two lagrangians will necessarily be the same. 
\subsection{Propagator: Massless spin-$1$}\label{subsec:propagatorspin1massless}

Additionally, for future reference the propagator for massless spin-1 fields is derived here. Propagators are important objects because of their fundamental role. Abusing words, these objects transfer information between two sources. Consider the equation of motion from the above lagrangian in the presence of a source-term.
\begin{equation}\label{eq:dmunuoperatormassless}
	D_\munu \phi^\nu = \sigma_1 \big( \Box \eta_\munu - \del_\mu \del_\nu \big)\ \phi^\nu = J^\nu
\end{equation}
Classically, propagators are simply Green's function for an equation of motion, thereby allowing solutions for such inhomogeneous differential equations. They describe how effects are propagated between multiple sources $J_1, J_2, \dots$, each coupled to the same field $\phi_\mu$. Additionally in QFT, the boundary conditions imposed on propagators can be solved in multiple ways leading to different causal behaviours. For this thesis, only Feynman propagators will be considered, without any further qualification. 

To actually calculate the propagator, firstly it is noted that the differential operator $D_\munu$ sandwiched in the bi-linear kinetic term in the lagrangian eq\eqref{eq:lagrangianspin1} is degenerate due to gauge symmetry. Hence, to this lagrangian a gauge-fixing term is added. The gauge-fixed lagrangian becomes:
\begin{equationsplit}
	\mathcal{L} &= \mathcal{L}_{\text{Fronsdal spin-1}} + \mathcal{L}_{\xi-\text{gauge}}
	\\
	&= \sigma_1\ \phi^\mu\ \Big( \Box \eta_\munu - (1 - \frac{1}{\xi}) \del_\mu \del_\nu \Big)\ \phi^\nu
\end{equationsplit}

This lagrangian is now in a gauge called the \emph{$R_\xi$-gauge}, and keeping the $\xi$ term explicit has several benefits: (a) track the behaviour of gauge terms through a calculation, (b) allows one to use many popular gauge-choices in a neat manner, among others. Obtaining the propagator is now straight-forward. The $D_\munu$ operator in this gauge can be readily inverted by first going to momentum space. The propagator, in momentum space, is given as\footnote{Note - the Feynman $+i\epsilon$ prescription is inexplicitly understood. Also, usually these objects come dressed with $i$'s and wave factors such as $e^{-ikx}$. When performing further calculations, these aesthetics will be given due care.}:
\begin{equation}\label{eq:propagatorspin1massless}
	G^\munu = \frac{-1}{\sigma_1} \Bigg(\frac{\eta^\munu + (\xi-1) \frac{k^\mu k^\nu}{k^2}}{k^2}\Bigg)
\end{equation}

Firstly, that the excitations are massless is confirmed from the poles of the above object. Secondly, for the physical components in the gauge field $A_\mu$ which are actually propagated, the sign of the residue is positive (reminder metric signature is mostly minus).  Additionally, that the $A_0$ component comes with a wrong sign is not threatening since the lagrangian has no kinetic-term for this component. Finally, note that the propagating modes are transverse to the direction of propagation or momenta $k_\mu$. One can check this by taking an inner product of this propagator with momenta $k_\mu$. Only a $\xi$ dependent pure gauge term will remain, which being unphysical, can be removed with a gauge choice.

Thus, in this section Electromagnetism in 2+1 dimensions was studied. It was shown to be equivalent to the theory obtained from the spin-1 case in the HS Fronsdal formulation. Also, the propagator for this field was calculated, confirming transverse unitary propagation of massless scalar excitations. 
\section{Massive Electrodynamics: Proca Theory }\label{sec:proca}
A simple and insightful modification of the theory of massless spin-1 fields presented above is to give the massless field a mass term, called the Proca term. There are some important differences between the massive and the massless theory. The proca mass term breaks gauge symmetry inherently present in the massless theory. Thus, complications from gauge symmetry are absent in this theory. Another major difference between the two theories comes from the dof's excited by the two theories. As was shown in \fullref{sec:countinghighspindof} and explicitly confirmed in the section above, a  massless spin-1 field in 2+1 is equivalent to a single scalar dof. Whereas, a massive field in 2+1 carries 2 dof. As will be seen shortly, the massive theory excites an additional dof which is longitudinal to the momentum. 

The proca lagrangian is:
\begin{equationsplit}\label{eq:lagrangianproca}
	\mathcal{L}_{\text{massive spin-1}} &= \mathcal{L}_{\text{Fronsdal spin-1}} + \mathcal{L}_{\text{proca}}
	\\
	\text{with: } \mathcal{L}_{proca} &= \half m^2 \phi^\mu\phi_\mu
	\\
	\implies \mathcal{L}_{\text{massive spin-1}}	&= \sigma_1\ \phi^\mu\ \Big( \big(\Box + m^2 \big) \eta_\munu - \del_\mu \del_\nu \Big)\ \phi^\nu
\end{equationsplit}
And its equations of motion are:
\begin{equationsplit}\label{eq:eomproca}
	\big(\Box + m^2 \big)\ \phi^\nu &=0
	\\
	\del\cdot\phi =0
\end{equationsplit}

Contrasting the first equation with eq\eqref{eq:eomspin1massless}, it is clear that the field is now massive. The first equation describes the massive kinetics of each component of the proca-field. The second equation which has been derived from the lagrangian, can be used to get rid of one component. This leaves only 2 components in the proca-field. Thus, it is concluded that this theory describes the propagation of two massive dof, which is inline with earlier arguments. 
\subsection{Propagator: Massive spin-1}\label{subsec:propagatorspin1proca}
Since, there is no gauge symmetry the propagator can be constructed readily. First, the kinetic operator in the lagrangian is written in momentum space as:
\begin{equation}
	\tilde{D}_\munu = \sigma_1\ \Bigg(\big(-k^2 + m^2 \big)\ \eta_\munu + k_\mu k_\nu \Bigg)
\end{equation}
Demanding that the propagator $G^\munu$ is the inverse of the above object leads to:
\begin{equation}\label{eq:propagatorspin1massive}
	G^\munu = \frac{-1}{\sigma_1} \Bigg(\frac{\eta^\munu - \frac{k^\mu k^\nu}{m^2}}{k^2 - m^2}\Bigg)
\end{equation}

Solving for the poles of $k_0$ in the above expression clearly demonstrates that the excitations in the proca-theory are massive. Since, there is no kinetic term for the $A_0$ mode, the propagation described by the above propagator has correct sign for the remaining components and hence describes unitary propagation. Considering the $m \rightarrow 0$ limit of the above propagator, it is noted that this object blows up. Specifically, the $\frac{k^\mu k^\nu}{m^2}$ term. This is not an issue of major concern, neither does it point to any discontinuity in the parameters of the theory. This is an artefact of the longitudinal dof which was absent in the massless theory. This can be understood in two ways: (a) through explicit construction of the polarization vectors, and (b) by performing a St\"uckelberg analysis. Both of these methods lead to useful insights and are presented respectively.

The numerator of the pole in a propagator is the sum of the polarizations that are excited by the theory. In this case, there are two polarizations ($\lambda^i_\mu(k_\mu), i=(1,2)$ for 2-dofs) which can be explicitly constructed from the eom's as follows. First, a momentum state is chosen for the field's propagation:
$$ k_\mu = (E,0,k_y)$$
Next, the conditions that the polarization vectors have to satisfy are derived from the eom eq\eqref{eq:eomproca}:
$$k^\mu \lambda^i_\mu = 0\  \&\  k^2 = m^2$$
Finally, the polarization vectors can be explicitly constructed as:
\begin{equationsplit}\label{eq:polarizationmassivespin1}
	 \lambda^1_\mu &= (0,1,0)\  \&\  \lambda^2_\mu = (\frac{k_y}{m},0,\frac{E}{m})
\end{equationsplit}

Note, there is one polarization which is transverse to the chosen momenta $\lambda^1_\mu$ and another mass dependent polarization $\lambda^2_\mu$ which is longitudinal to the direction of field's propagation. Taking the massless $m\rightarrow0$ limit in these vectors will clearly affect the latter polarization, and clarifies why the propagator was blowing up. The explanation is that when $m\rightarrow0$, then $E\approx k$ and $\lambda^2_\mu$ becomes parallel to the proca-photon's momentum. Thus, its contributions dominate and they cannot be ignored cheaply with a $m\rightarrow0$ limit. However, the theory is continuous in its mass parameter $m$, and the massless theory can be recovered by using an appropriate projection operator. Indeed, if a complete calculation is performed using massive proca-photons, the massless limit of any observable's value will be the same, had the calculations been done for massless photons. This is in stark contrast with the vDVZ discontinuity observed in massive spin-2 Fierz-Pauli theory and massless spin-2 in GR  (discussed in \fullref{subsec:vdvzdiscontinuity}).
\subsection{Insights from St\"ckelberg Analysis - Proca Theory}\label{subsec:stuckelbergspin1proca}

It is very insightful to see that the theory is really continuous and propagates nothing more that two scalar dof through a \stuckelberg analysis. The continuity of the theory will be seen as an outcome of the decoupling of the two scalar dofs. This will not be the case in the \stuckelberg analysis for spin-2 fields, where one of the scalar mode does not decouple from the theory (\fullref{subsec:stuckelbergspin2}).  \stuckelberg analysis involves a chain of field redefinitions, and through such a trick many caveats in a theory can be unearthed. 

The formal meaning of the \stuckelberg formalism is easier to understand when the field in question has considerably high spin. Thus, the ideas of the formalism itself and their role is discussed in \fullref{subsec:stuckelbergspin2}, where this formalism is used for the theory of massive spin-2 fields. 

\paragraph{Caveat:} The theory described after a \stuckelberg field redefinitions will be equivalent (as in a derived theory), but not equal to the original proca theory. This is because, in essence, a \stuckelberg analysis diagonalizes the original lagrangian, thereby mixing all the dofs completely. \footnote{To muse, this is like rotating a vector. Even though the vector's description in terms of its components will change, the object in itself, obviously, remains the same.} This will become clear shortly.

The field redefinition is:
\begin{equation}\label{eq:stuckelbergspin1fieldredef}
	\phi_\mu = \psi^T_\mu + \del_{\mu} \chi
\end{equation}
Here, $\psi^T_\mu$ is demanded to be a transverse vector field, additionally satisfying \footnote{Note - some sloppiness has creeped in, and the reader is hereby alerted. A \stuckelberg analysis usually only involves the introduction of new fields, with no presupposed conditions on them \cite{reviewhinterbichler}. Imposing the transversality condition here has the effect of fixing a gauge. This allowed for a quicker route to the completely decoupled lagrangian. The name \stuckelberg is used here as an umbrella term for all such tricks. For a general analysis with no presupposed condition see \fullref{subsec:stuckelbergspin1PCS}. For the meaning and the role of the \stuckelberg formalism see \fullref{subsec:stuckelbergspin2}.},:
\begin{equation}
	\del^\mu \psi^T_\mu = 0
\end{equation}
And $\chi$ is a scalar field. Plugging this redefinition in the proca lagrangian in eq\eqref{eq:lagrangianproca}, gives the following new lagrangian. 
\begin{equationsplit}
	\mathcal{L}_{\text{massive spin-1}} &= \mathcal{L}_{\psi} + \mathcal{L}_{\chi}
	\\
	\text{where: } \quad & 
	\\
	\mathcal{L}_{\psi} &= \sigma_1\ \psi^T_\mu\ \big(\Box + m^2 \big)\  \psi^{T\mu}
	\\
	\mathcal{L}_{\chi} &= - \sigma_1\ m^2\ \chi\ \Box\ \chi
\end{equationsplit}

There are several key points to note: (i) Terms with both fields $\psi$ and $\chi$ (mixed terms) in the lagrangian, signalling a possible interaction between the two fields $\psi$ and $\chi$ get dropped out. This is, what was, meant by the de-coupling of the scalar mode. (ii) The importance of minus sign between the $\Box$ and $\del_\mu\del_\nu$ in eq\eqref{eq:dmunuoperatormassless} is understood\footnote{Constructing the propagator while doing a very general analysis by using coefficients like $a$ for $\Box$ and $b$ for $\del_\mu\del_\nu$ in the lagrangian, one clearly sees that there is an additional higher-derivate ghost mode riding in with the scalar field $\chi$. This is evaded by setting $a=-b$. Although, the author chose not to present such calculations, they are quite general to do and convince oneself upon the necessity of such particular signs.}.  (iii) Usually, a \stuckelberg field redefinition also introduces a gauge symmetry. This is not present in the lagrangian above, because the $\psi$ field was demanded to be transversal (see \fullref{subsec:stuckelbergspin1PCS} for a more general analysis with the gauge symmetry clearly present). 

Before proceeding, the $\chi$ field is rescaled as $\chi \rightarrow \frac{1}{m} \chi$. Thus, the equations of motion for the two \stuckelberg fields from their respective lagrangian are:
\begin{equationsplit}\label{eq:eomstuckelbergspin1}
	\text{For } \psi^T_\mu \text{ :}&
	\\
	&\big(\Box + m^2 \big)\  \psi^{T\mu} = 0
	\\
	&\del^\mu \psi^T_\mu = 0
	\\
	\text{For } \chi \text{ :}&
	\\
	&\Box \chi = 0
\end{equationsplit}

Neat. The dofs in the original proca-lagrangian have split into one massive dof $\psi$ and one massless dof $\chi$. While deriving these eoms it becomes clear that both the fields $\psi$ and $\chi$ \emph{are affected} by the presence of the proca-mass $m$ (in fact this is already seen at the lagrangian level). By putting $m\rightarrow0$ in the \stuckelberg lagrangian, it is concluded that the addition of the proca term has the following two affects:
\begin{itemize}
	\item It makes the original massless dof, from the massless spin-1 theory, massive. Since the $\psi$ field becomes massless on putting proca mass $m\rightarrow0$, the dof in $\psi$ field can be seen as the massive version of the massless dof from the massless theory. 

	\item The proca mass term also excites a new separate dof, which was not present in the massless theory. Since the lagrangian of the $\chi$ field completely drops away on putting $m\rightarrow0$, this confirms the said effect. This dof does not interact or couple with other dof/or external sources and this makes Proca theory, continuous in its mass $m$ parameter.
\end{itemize}

There is something unusual going on: The $\chi$ field which is a massless scalar, de-couples from the system when $m\rightarrow0$. Thus, strangely, it seems that the proca mass term has excited a massless scalar dof. Why would a mass term excite a massless dof? This also contradicts the original eom in eq\eqref{eq:eomproca}, which suggests that all components of the field $\phi$ are massive. The solution of this dilemma lies in realizing that the \stuckelberg field redefinitions have mixed all dofs. This confirms, that the effect of the \stuckelberg analysis is itself equivalent to that of diagonalizing the lagrangian. 

Thus, an interesting aspect has emerged. The original massless dof (from the massless theory of \fullref{sec:3delectrodynamics}) is still present in the proca theory, albeit as a massive dof. And upon diagonalizing this system, the newly excited dof emerges as a massless dof. This setup is reminiscent of the Higgs mechanism. Indeed, the \stuckelberg analysis corresponds to a special limit of the Higgs mechanism, in which the self-coupling $\lambda$ of Higgs is taken as $\lambda\rightarrow\infty$: thus a massive Higgs decouples and has disappeared, and what has remained is a massless goldstone boson $\chi$.

Before closing this section, the propagators for the two fields $\psi$ and $\chi$ are quickly discussed. 
\begin{equationsplit}
	G{^\munu_{\text{st\"uck-}\psi}} &=  \frac{-1}{\sigma_1} \Bigg(\frac{\eta^\munu }{k^2 - m^2}\Bigg)
	\\
	G_{\text{st\"uck-}\chi} &= \frac{1}{\sigma_1} \Bigg(\frac{1}{k^2}\Bigg)
\end{equationsplit}
It is seen that these propagators describe unitary propagation for both the dofs (Note - for the $\psi$ field: the time component was not excited right from the start, hence the sign is correct; and due to its transversal nature there is only one independent component in $\psi$). These propagators reaffirm the above mentioned observations. Additionally, the massless limit of these two propagators behaves as expected: the massive mode turns into a well-behaved massless mode and the massless field $\chi$ remains unaffected (though it gets dropped at the lagrangian level itself.).
 
With such an analysis, the affect of giving a mass-term to the massless theory is now crystal-clear. An executive summary follows: Giving a proca mass term to the spin-1 massless field resulted in a theory with two dofs (which is inline with the group theoretical arguments of \fullref{sec:irrep}). This theory does not possess gauge symmetry, but such a gauge symmetry can be re-introduced as done in the \stuckelberg analysis. To identify each dof in the theory, the propagator for massive spin-2 field was calculated, eq\eqref{eq:propagatorspin1massive}. This propagator confirmed unitary propagation but did not have a well-defined massless limit. Further insights were gained by explicitly constructing the polarization vectors and performing a \stuckelberg analysis. It can be concluded that the two excitations in the Proca theory are massive excitations. These could be further disentangled into a massless and a massive scalar mode. Their propagator has correct sign for unitary propagation and their massless limit is well-defined. 

\section{Topologically Massive Electrodynamics}\label{sec:3dTME}

There is yet another mass generating mechanism. This comes from the interplay of metric-based terms in the bulk and CS terms with their effects on the boundary of a manifold. Such a theory for spin-1 is called Topologically Massive Electrodynamics (TME). This theory provides a completely independent gauge-invariant mechanism for generating mass. Such a mechanism has important effects in fields like Condensed Matter Physics and may have many important consequences for fundamental physics in near future. 

The ingredients for constructing such a theory were already introduced in \fullref{ch:highspin} on high spin field theory. The strategy is to use the Fronsdal construction to provide a ``kinetic-term" for the spin-1 field and then to give it a mass by adding a CS ``mass" term. Throughout this thesis, the Fronsdal term and the CS term will be referred to as the kinetic term and mass term, respectively, yet these labels are not very accurate. It was seen in \fullref{sec:highspintopologicallymassive}, that a pure CS term can also lead to a kinetic term in the lagrangian\footnote{Note: that for the case of spin-1 a modified lagrangian was needed. See \fullref{subsec:spin1csonly} for details.}. Thus, these labels ``kinetic term" and ``mass term" in this context should only be seen as traditional notions for classifying and easy communication. With the strategy now laid out, the TME lagrangian follows ($\mu_{\phi}$ is used to indicate the relative coefficient of the two terms. This will turn out to be the mass of the field $\phi$.):

\begin{equationsplit}\label{eq:lagrangianTME}
	\mathcal{L}_{\text{TME}} &= \mathcal{L}_{\text{kinetic}} + \mu_{\phi}\ \mathcal{L}_{\text{cs}}
	\\
	\text{where: } \quad \quad &
	\\
	\mathcal{L}_{\text{kinetic}} &= \sigma_1 \ \phi^\mu \overline{F}_\mu
	\\
	\mathcal{L}_{\text{cs}} &= \sigma_1\ \phi^\mu \partial^\alpha \epsilon_{\alpha\nu\mu} \phi^\nu
	\\
	\text{together: } \quad \quad &	
	\\
	\mathcal{L}_{\text{TME}} &= \sigma_1\ \phi^\mu \Bigg( \Box \eta_\munu - \del_\mu \del_\nu + \mu_{\phi}\ \partial^\alpha \epsilon_{\alpha\nu\mu} \Bigg)\ \phi^\nu
\end{equationsplit}
The equations of motion coming from $\mathcal{L}_{\text{TME}}$ are:
\begin{equation}
	\Box \phi_\mu - \del_\mu\ \del\cdot \phi + 2 \mu_{\phi}\ \epsilon_{\alpha\nu\mu} \del^\alpha \phi^\nu = 0
\end{equation}
Note, due to the single derivative in the CS term it contributes twice in the eom. Next this eom can be written in terms of both the electromagnetic field strength tensor and its dual vector. This gives the following equations:
\begin{equationsplit}\label{eq:Kmunuoperatorspin1}
	\del^\sigma F_{\sigma\mu}  + \mu_{\phi}\ \epsilon_{\alpha\nu\mu} F^{\alpha\nu} &= 0
	\\
	\big(\del^\sigma \epsilon_{\sigma\mu\nu} + \mu_{\phi}\ \eta_{\mu\nu} \big) \tensor[^*]{F}{^\nu} &= 0
	\\
	K_\munu \tensor[^*]{F}{^\nu} &= 0
\end{equationsplit}

To obtain the massive nature of the excitation in this theory, the following is considered:
\begin{equation}
	K^{\mu\alpha} K_{\munu} \tensor[^*]{F}{^\nu} = 0
\end{equation}
Following this calculation through, one obtains:
\begin{equationsplit}
	\big(\Box + \mu{^2_\phi}\big) \tensor[^*]{F}{^\alpha} &=0
	\\
	\del_\alpha \tensor[^*]{F}{^\alpha} &=0
\end{equationsplit}

At last, the massive nature of the dof emerges. In deriving these equations the definition of the dual vector eq\eqref{eq:dualvector2} was used, which also implies the last equation above. Thus, it is noted that the parameter $\mu_{\phi}$ quantifies the mass of the dof described by the lagrangian $\mathcal{L}_{TME}$ in eq\eqref{eq:lagrangianTME}. One can notice, again, that there is no kinetic term for the $\phi_0$ component in the lagrnagian. Thus together these two equation suggest that this system describes one single massive dof. The derivation of this equation brings about some interesting points: (a) The number of dof and its massive nature does not agree with the group theoretical counting presented in the last chapter. Although intriguing, this is simply due to the fact that group theoretic arguments presented in the last chapter do not probe topological effects; (b) that the operator $K_\munu$ factorizes the operator $\big(\Box + \mu{^2_\phi}\big)$ in 2+1 dimensions; (c) unlike, the proca-massive theory, this theory does not add any dof but makes the massless dof massive. The last observation is not quite general, since the theory of TMG describes a spin-2 massive particle whereas GR in 2+1 has no dof, at all.

Under a gauge transformation of the field $\phi$ : $\delta \phi_\mu = \del_{\mu}\xi$, this lagrangian transforms as:
\begin{equationsplit}\label{eq:gaugeTME}
	&\mathcal{L} \rightarrow \mathcal{L} + \delta \mathcal{L}
	\\
	\text{where: } &\delta \mathcal{L} = \sigma_1\ \mu_{\phi}\ \del^\mu\ \big( \xi\ \epsilon_{\alpha\nu\mu} \del^\alpha \phi^\nu \big)
\end{equationsplit}
It is seen that the CS terms transforms by a total-derivative only. Although, transformations by a total derivative does not bear any effect on the equations of motion, such transformations bring out the topological nature of this theory. This will be addressed in \fullref{subsec:TMEpeculiar}, where other interesting features associated with this theory are collected.  

The peculiar massive dof excited in this theory will now be studied by constructing a propagator for this theory. Subsequently, a \stuckelberg analysis will be made.
\subsection{Propagator: TME}\label{subsec:propagatorTME}

To calculate the propagator in this theory, it is imperative to start by fixing the gauge. As in the massless case, this is done in \emph{$R_\xi$-gauge}. The lagrangian becomes:
\begin{equation}
	\mathcal{L}_{\text{TME }\xi\text{-gauge}} = \sigma_1\ \phi^\mu \Bigg( \Box \eta_\munu - (1- \frac{1}{\xi})\  \del_\mu \del_\nu + \mu_{\phi}\ \partial^\alpha \epsilon_{\alpha\nu\mu} \Bigg)\ \phi^\nu
\end{equation}
Inverting the kinetic operator, sandwiched above, follows the same steps as before. Going to momentum space this gives the following propagator:
\begin{equationsplit}\label{eq:propagatorTME}
	G^\munu &= \frac{-1}{\sigma_1}  \Bigg( \frac{\eta^\munu - \frac{k^\mu k^\nu}{ \mu{^2_\phi}} + \frac{i}{\mu_{\phi}}\epsilon^{\munu\rho}k_\rho }{k^2 - \mu{^2_\phi}} + \frac{+ \frac{k^\mu k^\nu}{ \mu{^2_\phi}} - \frac{i}{\mu_{\phi}}\epsilon^{\munu\rho}k_\rho}{k^2} + \xi \frac{k^\mu k^\nu}{k^4} \Bigg)
\end{equationsplit}

This propagator seems to have 3 poles, but not all of these are independent or physical. For, seeing the propagating dofs more clearly, this object can be simplified. The last pole clearly corresponds to pure gauge dof, since a gauge choice of $\xi = 0$ kills that piece. Such a gauge choice corresponds to the Lorenz gauge condition i.e. $\del\cdot\phi=0$\footnote{There are other important gauge choices such as the Feynman-'t Hooft gauge ($\xi=1$), Unitary gauge ($\xi\rightarrow\infty$) etc.}. In this gauge, the above propagator can be further simplified to the following form:
\begin{equationsplit}
	\xi=0; \quad\quad	G^\munu &= \frac{-1}{\sigma_1}  \Bigg( \frac{\eta^\munu - \frac{k^\mu k^\nu}{k^2} + \frac{i \mu_{\phi}}{k^2 }\epsilon^{\munu\rho}k_\rho }{k^2 - \mu{^2_\phi}}  \Bigg)
\end{equationsplit}
The poles from this simplified propagator confirm that there is only one massive dof. Contracting this object with a momentum state clarifies that this dof is transversal. Recall that in the massless theory there was only one dof which was also transversal. The effect of adding the CS term to the Maxwell lagrangian (eq\eqref{eq:lagrangianmaxwell}), can be seen explicitly by considering $\mu_{\phi}\rightarrow 0$ limit in the propagator in eq\eqref{eq:propagatorTME}. By carefully mixing all terms first, one obtains:
\begin{equation}\label{eq:propagatorTMElimit}
\lim_{\mu_{\phi}\rightarrow 0}  G^\munu = \frac{-1}{\sigma_1} \Bigg(\frac{\eta^\munu + (\xi-1) \frac{k^\mu k^\nu}{k^2}}{k^2}\Bigg)
\end{equation}
This is exactly the same as the propagator in eq\eqref{eq:propagatorspin1massless} obtained for the massless theory. This suggests that the addition of the CS-term has lead the massless dof to obtain a mass. This is also the reason why a CS term in such works, is usually referred to as the mass term. 

\subsection{Insights from \stuckelberg Analysis - TME Theory}\label{subsec:stuckelbergspin1TME}

The theory described before, can be further analyzed with a \stuckelberg analysis. The results shall not be surprising. To begin with, the same field redefinitions, from \fullref{subsec:stuckelbergspin1proca}, are used here. For convenience, they are reproduced:
\begin{equation}\label{eq:stuckelbergspin1fieldredef}
\phi_\mu = \psi^T_\mu + \del_{\mu} \chi
\end{equation}

As before, the $\phi$ field is thought of as a sum of a vectorial field $\psi$ and a scalar field $\chi$. The vector $\psi$ is also assumed to be transversal\footnote{Any analysis done without this assumption will obviously lead to the same physics (see \fullref{subsec:stuckelbergspin1PCS} for a more general analysis). This assumption simply allows for an easier quicker route.}. Plugging, these definitions back into the lagrangian $\mathcal{L}_{\text{TME}}$ produces the following new lagrangian:

\begin{equationsplit}
	\mathcal{L}_{\text{TME}} &= \mathcal{L}_{\psi} + \mathcal{L}_{\chi}
	\\
	\text{where: } \quad & 
	\\
	\mathcal{L}_{\psi} &= \sigma_1\ \psi^{T\mu}\ \big(\Box \eta_\munu + \mu_{\phi} \epsilon_{\alpha\nu\mu} \del^\alpha  \big)\  \psi^{T\nu}
	\\
	\mathcal{L}_{\chi} &= 0
\end{equationsplit}

Since, there is no proca mass term  to excite the scalar field $\chi$, its lagrangian is trivially zero. Also, due to the transversality condition imposed on field $\psi$, its lagrangian has turned into an easier expression. There is a price paid for this. This comes from noting that under these field redefinitions, the above lagrangian has lost its gauge symmetry. Under a gauge transformation of type: $\psi_\mu \rightarrow \psi'_\mu = \psi_\mu + \del_\mu \zeta$, a term proportional to the gauge parameter $\zeta$ remains (apart from the total derivatives coming from the CS term). This is not a problem, since the aim of this analysis is to look at the physical content only. 

Looking at the particle spectrum, the lagrangian above makes the dof count clear. The transversality condition, along with the continued absence of a kinetic term for the $0$-component already kills 2 components. Thus, the $\psi$ field propagates only one massive dof. That this dof is massive can be checked by looking at its propagator. This is:
\begin{equationsplit}
	G{^\munu_{\text{st\"uck-}\psi}} &= \frac{-1}{\sigma_1}  \Bigg( \frac{\eta^\munu + \frac{k^\mu k^\nu \mu{^2_\phi}}{k^4} - \frac{i \mu{_\phi}}{k^2}\epsilon^{\munu\rho}k_\rho }{k^2 - \mu{^2_\phi}}  \Bigg)
\end{equationsplit}
The poles clearly confirm the massive nature of this excitation. Interestingly, although the $k_\mu k_\nu$ piece is absent from the lagrangian of the $\psi$ field, the CS term excites such pieces in the propagator. This again serves as a gentle reminder that the role of the CS term is not restricted to a mass-generating term. This is quite different from other mass terms, such as a proca mass term. 

Finally, this \stuckelberg analysis along with the massless limit of the complete propagator in eq\eqref{eq:propagatorTMElimit} for the TME lagrangian makes it clear that the presence of the CS term in the Maxwell lagrangian has turned the massless dof into a massive one. This is not the only interesting feature of this theory. There are many others, of which some are discussed now. 

\subsection{Peculiarities of TME}\label{subsec:TMEpeculiar}

The following list is, empathetically, not comprehensive. It is, merely, suggestive:

\begin{enumerate}
	\item \textbf{Parity and Time-Reversal: }
	The CS term seen above violates both parity and time-reversal. Infact, this is not specific to TME and is true for the CS term of HS gauge fields as well as the mass term for fermions in 2+1 dimensions.
	\item \textbf{Topological Connection: }
	The gauge transformation of TME in eq\eqref{eq:gaugeTME}, shows that this lagrangian is gauge invariant upto total-derivatives. Moreover, when studying the gauge properties of the exponentiated action, TME is gauge invariant for small gauge transformations only. For large gauge transformations, the action changes by a finite interval. This corresponds to the topological quantity known as \emph{winding number}. This also suggests that the CS mass should be quantized.
	\item \textbf{Induced Mass: }
	As was mentioned in chapter-1, one need not start with a theory of topologically massive photons. A theory in which these gauge bosons are interacting with fermions will induce the CS mass. This will be explicitly calculated for a slightly different scenario in \fullref{ch:quantumloopcorrections}.
	\item \textbf{Anyons: }
	The irreducible representations studied in the last chapter also allow for massless particles with continuous spins. These particles, in stark contrast to Fermi-Boson statistics, allow fractional statistics. 
\end{enumerate}

\section{Massive Electrodynamics: Proca-Chern-Simons}\label{sec:massivespin1PCS}

As a finale to this chapter, one of the more interesting and, perhaps, natural theory to consider, is a theory of spin-1 fields where both these mass-generating mechanisms are simultaneously present. The analysis and outcome of this section can be seen as a preparatory step for analysing a similar theory for spin-2 fields in \fullref{sec:massivespin2FPCS} for \emph{massive gravity}. 

The setup for studying this Proca-Chern-Simons (PCS) massive electrodynamics, naturally follows the structure established in the preceding sections. A kinetic term is taken from the Fronsdal formulation. Two mass terms are taken from both proca theory as well as the pure CS theory. The resulting lagrangian is:
\begin{equationsplit}\label{eq:lagrangianPCS}
	\mathcal{L}_{\text{PCS}} &= \mathcal{L}_{\text{kinetic}} + \mathcal{L}_{\text{proca}} +\mu_{\phi}\ \mathcal{L}_{\text{cs}}
	\\
	\text{where: } \quad \quad &
	\\
	\mathcal{L}_{proca} &= \half m^2 \phi^\mu\phi_\mu
	\\
	\mathcal{L}_{\text{kinetic}} &= \sigma_1 \ \phi^\mu \overline{F}_\mu
	\\
	\mathcal{L}_{\text{cs}} &= \sigma_1\ \phi^\mu \partial^\alpha \epsilon_{\alpha\nu\mu} \phi^\nu
	\\
	\text{together: } \quad \quad &	
	\\
	\mathcal{L}_{\text{PCS}} &= \sigma_1\ \phi^\mu\ \Bigg( \big(\Box + m^2 \big) \eta_\munu - \del_\mu \del_\nu + \mu_{\phi}\ \partial^\alpha \epsilon_{\alpha\nu\mu} \Bigg)\ \phi^\nu
\end{equationsplit}

One immediately notices that the proca mass $m$ breaks gauge invariance of the lagrangian. Other than that, all the terms present in this lagrangian have already been discussed at length in this chapter. Proceeding on, the equations of motion are:
\begin{equationsplit}\label{eq:eomspin1PCS}
	\big(\Box + m^2 \big) \phi_\mu - \del_\mu\ \del\cdot \phi + 2 \mu_{\phi}\ \epsilon_{\alpha\nu\mu} \del^\alpha \phi^\nu = 0
\end{equationsplit}
It is not immediately clear, how many dof are present or what is the nature of the excitations in this theory. To address the first question, taking a divergence of the above equation is of help. It leads to the condition:
\begin{equation}
	\del\cdot\phi = 0
\end{equation}
The original eom were three equations for the three components in the field $\phi$. The above condition can be used to eliminate one of these components. Thus, it is now inferred that there are 2 dofs present in the system. To answer the second question above, regarding the massive/massless nature of these dofs, constructing a propagator seems necessary.

\subsection{Propagator: Massive spin-1 PCS}

While constructing the propagator, one meets a new feature of the field-theoretic language which was not discussed before. A mass-mixing polynomial. That is, when inverting the kinetic operator sandwiched in the above lagrangian, the determinant of this operator appears as a fourth order polynomial in momenta $k_\mu$ (only second order in $k^2$), with the masses $m$ and $\mu_\phi$ acting as coefficients. This polynomial is:
\begin{equation}\label{eq:massmixingspin1poly}
	(k^2 - m^2)^2 - k^2 \mu{^2_\phi}  = 0
\end{equation}
Factorizing this polynomial is necessary to express the propagator as a sum of poles. It is factorized by two functions $f_1$,$f_2$ of the masses $m$ and $\mu_\phi$ given by:
\begin{equationsplit}\label{eq:mixedmassspin1}
	&f_1 = \frac{1}{2} \big(\sqrt{\mu{^2_\phi} + 4 {m}^2} + \mu_\phi \big)
	\\
	&f_2 = \frac{1}{2} \big(\sqrt{\mu{^2_\phi} + 4 {m}^2} - \mu_\phi \big)
	\\
	\text{factorizing: } \quad \quad &
	\\
	&(k^2 - m^2)^2 - k^2 \mu{^2_\phi} = (k^2 - f_1^2)(k^2 - f_2^2) 
\end{equationsplit}

Although, not very complicated it is interesting to note how the two masses interact and mix in the mass-functions $f_1$ and $f_2$. Needless to guess, these two mass-mixing functions appear as the two masses in the poles of the propagator. The propagator can now be written as:

\begin{equationsplit}\label{eq:propagatorPCS}
	&G{^\munu_\text{PCS}} = \frac{-1}{\sigma_1} \frac{1}{f_1+f_2}\Bigg( \frac{Y_1^\munu}{k^2 - f_1^2} + \frac{Y_2^\munu}{k^2 - f_2^2}\Bigg)
	\\
	\text{where:} \quad\quad&
	\\
	&Y_1^\munu = f_1 \Bigg( \eta^\munu - \frac{k^\mu k^\nu}{f_1^2} + \frac{i}{f_1}\ \epsilon^{\munu\rho}k_\rho \Bigg)
	\\
	&Y_2^\munu = f_2 \Bigg( \eta^\munu - \frac{k^\mu k^\nu}{f_2^2} - \frac{i}{f_2}\ \epsilon^{\munu\rho}k_\rho \Bigg)	
\end{equationsplit}

There certainly is a pleasant symmetry in the above propagator with respect to the exchange of the mass functions $f_1\leftrightarrow f_2$. This is spoiled by the sign in front of the CS term. The functions $f_1$ and $f_2$ themselves differ by a sign for the mass $\mu_\phi$ in the same term. The above propagator clearly determines that in the PCS theory there are two massive dofs. The two mass-generating mechanisms are simultaneously present and they \emph{talk} with each other. At this stage, both the dof are massive and their masses are functions $f_1$ and $f_2$ of the original masses. 

It was shown that the proca mass $m$ excites two massive dof: which upon further analysis showed that there was one massless dof mixed with a massive scalar dof. In the PCS theory, one might expect that the two massive dofs $f_1$ and $f_2$ are simply a mixture of the massive dof coming from the proca term and massive version of the massless dof coming from the CS term. These expectations emerge directly from studying the two limiting cases for the mass functions $f_1$ and $f_2$. Consider these limits:
	\begin{equationsplit}
	m \rightarrow 0 &\implies f_1 = \mu_{\phi} \quad f_2 = 0
	\\
	\mu_{\phi} \rightarrow 0 &\implies f_1=f_2=m
	\end{equationsplit}
Thus, on setting the proca mass $m\rightarrow0$, the CS massive version of the original massless dof survives in the function $f_1$. Whereas, both functions $f_1$ and $f_2$ remain massive on setting the CS mass $\mu_\phi\rightarrow0$. This also confirms the conclusions drawn in \fullref{subsec:stuckelbergspin1proca}, upon the effect of adding a Proca term to the massless theory.

\textbf{Studying limits}

Since the $0$-component of the original field $\phi$ never had a kinetic term in the lagrangian, the propagator above has correct sign for other components, and this signals unitary propagation. It is interesting to consider how this propagator behaves under different limits. It maybe expected that this propagator would reduce to the propagator from the Proca-theory for $\mu_{\phi}\rightarrow 0$, and conversely for $m\rightarrow0$ to the TME propagator. Further putting both masses to zero, should result in the massless propagator. These limits can be implemented through their corresponding limits in terms of the mixed masses $f_1$ and $f_2$.

\begin{itemize}
	\item \textbf{Proca mass $m\rightarrow0$:}
	Plugging the limit, one sees that the $\frac{k^\mu k^\nu}{f_2^2}$ term in the $Y_2$ piece is blowing up. This is the same scenario as when the massless limit of the propagator in proca theory was considered in \fullref{subsec:propagatorspin1proca}. Ignoring this term, the rest of the propagator neatly reduces to:
\begin{equationsplit}\label{eq:propagatorTMExi0}
	\lim_{f_1=\mu_\phi,f_2=0}	G'^\munu &= \frac{-1}{\sigma_1}  \Bigg( \frac{\eta^\munu - \frac{k^\mu k^\nu}{k^2} + \frac{i \mu_{\phi}}{k^2 }\epsilon^{\munu\rho}k_\rho }{k^2 - \mu{^2_\phi}}  \Bigg)
\end{equationsplit}
	This is exactly the same as the Lorenz gauge propagator for TME in eq\eqref{eq:propagatorTMExi0}. There is a prime above to indicate that the blown-up term has been simply ignored. Although this is not the right-way, yet as the cause of blowing up is definitely understood there is no conceivable harm. A more elegant propagator will be made through a \stuckelberg analysis. Further note, that putting $\mu_{\phi}\rightarrow0$ in this object simply gives the massless propagator from eq\eqref{eq:propagatorspin1massless} (again in Lorenz gauge).
	
	\item \textbf{Chern-Simons mass $\mu_\phi \rightarrow0$:} This is a more elegant limit, since nothing blows up in this scenario. Putting in the limit the propagator simply becomes:
	\begin{equation}
	\lim_{f_1=f_2=m} G^\munu = \frac{-1}{\sigma_1} \Bigg(\frac{\eta^\munu - \frac{k^\mu k^\nu}{m^2}}{k^2 - m^2}\Bigg)
	\end{equation}
	This is exactly the same as the propagator for the massive proca theory in eq\eqref{eq:propagatorspin1massive}. Further, the $m\rightarrow0$ limit of this object was discussed in detail in \fullref{subsec:propagatorspin1proca}.
\end{itemize}

Having seen the propagator for the PCS theory in different limiting scenarios, all the expectations made from this study are confirmed. Indeed, the two massive dof propagated in the PCS theory are mixtures of the two original mass-generating mechanisms. The theory contains the two massive dof excited by the proca term mixed with the massive version of the massless dof due to the CS term. Thus, the two mechanism talk with each other and their discussion is encoded in eq\eqref{eq:mixedmassspin1}. Further study of this theory can be done through a \stuckelberg analysis. 

\subsection{Insights from \stuckelberg Analysis - PCS Theory}\label{subsec:stuckelbergspin1PCS}

Before presenting the actual details of the calculation, it is worthwhile to pause and develop an expectation for the results from this subsection. A \stuckelberg analysis has been done twice already: for proca theory in \fullref{subsec:stuckelbergspin1proca} and for TME theory in \fullref{subsec:stuckelbergspin1TME}. Based on the observations gathered earlier, one can guess that the \stuckelberg field redefinition should diagonalize the PCS lagrangian into a massive mode and a massless mode. Since the massless mode was absent in the \stuckelberg analysis for TME theory, it can be expected that if this mode has to reappear it should not be related to the CS mass $\mu_{\phi}$. Moreover, since both modes were present in the proca case, this suggests that both modes should be present in the current case as well and they should definitely depend on the proca mass $m$. Having such a general idea, the real calculations are now presented. 

The same field redefinitions are used here from the earlier \stuckelberg analysis. This is, for convenience, reproduced here from eq\eqref{eq:stuckelbergspin1fieldredef}:
\begin{equation}
\phi_\mu = \psi^T_\mu + \del_{\mu} \chi
\end{equation}
Under these definitions, the lagrangian becomes upto total derivatives:
\begin{equationsplit}\label{eq:lagrangianstuckelbergspin1A}
	\mathcal{L}_{\text{PCS}} &= \mathcal{L}_{\psi} + \mathcal{L}_{\chi} 
	\\
	\text{where: } \quad & 
	\\
	\mathcal{L}_{\psi} &= \sigma_1\ \psi^{T\mu}\ \Bigg( \big(\Box + m^2 \big) \eta_\munu  + \mu_{\phi}\ \partial^\alpha \epsilon_{\alpha\nu\mu} \Bigg)\ \psi^{T\nu}
	\\
	\mathcal{L}_{\chi} &= - \sigma_1\ \chi\ \Box\ \chi
\end{equationsplit}

No surprises. The \stuckelberg trick has again done a neat job of separating the two dofs into a massive and a massless mode. As a consequence, both the mass-generating mechanism are seen to be acting on the same single dof in the $\psi$ field, leaving the other completely untouched. However, since the $\chi$ field has already been rescaled as $\chi \rightarrow \frac{1}{m} \chi$, it shows the same behaviour as in the proca analysis, i.e. it will drop from the lagrangian in the limit $m\rightarrow 0$. The remarks made above, continue to stand. 

Having demanded transversality for the field $\psi$, this lagrangian does not possess any gauge symmetry. In order to restore this gauge symmetry, the transversality condition must be relaxed. To pursue this objective, a new field redefinition is necessary. 
\begin{equation}
\phi_\mu = \Psi_\mu + \del_{\mu} \chi'
\end{equation}
There are no presupposed conditions or demands on the new fields $\Psi$ and $\chi'$. Under this redefinition, the PCS lagrangian becomes:
\begin{equationsplit}\label{eq:lagrangianstuckelbergspin1B}
	\mathcal{L}_{\text{PCS}} &= \mathcal{L}_{\Psi} + \mathcal{L}_{\chi'} + \mathcal{L}_{\text{mix}}
	\\
	\text{where: } \quad & 
	\\
	\mathcal{L}_{\Psi} &= \sigma_1\ \Psi^{\mu}\ \Bigg( \big(\Box + m^2 \big) \eta_\munu  -\del_\mu \del_\nu + \mu_{\phi}\ \partial^\alpha \epsilon_{\alpha\nu\mu} \Bigg)\ \Psi^{\nu}
	\\
	\mathcal{L}_{\chi'} &= - \sigma_1\ \chi'\ \Box\ \chi' \quad \quad \quad \text{after: } \chi' \rightarrow \frac{1}{m} \chi'
	\\
	\mathcal{L}_{\text{mix}}  &= m \Psi^\mu \del_\mu \chi'
\end{equationsplit}
The above lagrangian has a new term $\mathcal{L}_{\text{mix}}$, where the scalar field $\chi'$ is coupled to the vector $\Psi$. If the the $\Psi$ field has been transversal, one immediately notices that this mixing term will be dropped (upon partial integration). This partly justifies the quicker route taken before. The other part, of this justification, will come from demonstrating that this lagrangian does indeed describe the same physics as the lagrangian in eq\eqref{eq:lagrangianstuckelbergspin1A}. Too this end, the first imperative step is to deal with $\mathcal{L}_{\text{mix}}$. Here, gauge freedom comes to rescue. Since, the $\Psi$ field is not transverse, the above lagrangian is gauge invariant! For an arbitrary gauge parameter $\zeta$, the gauge transformations are:
\begin{equationsplit}
	&\psi_\mu \rightarrow \psi'_\mu = \psi_\mu + \del_\mu \zeta
	\\
	&\chi' \rightarrow (\chi')' = - m \zeta
\end{equationsplit}

A gauge choice is made by demanding the following:
\begin{equationsplit}\label{eq:stuckelbergspin1Bgaugechoice}
	\del\cdot\Psi - m \chi' = 0
\end{equationsplit}
which forces the gauge parameter to satisfy
\begin{equationsplit}
	\big(\Box + m^2 \big) \zeta = 0
\end{equationsplit}
Upon adding the following gauge-fixing term:
\begin{equation}
	 \mathcal{L}_{gauge} = - \sigma_1 (\del\cdot\Psi - m \chi')^2
\end{equation}
The lagrangian in eq\eqref{eq:lagrangianstuckelbergspin1B} becomes:
\begin{equationsplit}\label{eq:lagrangianstuckelbergspin1B}
	\mathcal{L}_{\text{PCS}} &= \mathcal{L}_{\Psi} + \mathcal{L}_{\chi'}
	\\
	\text{where: } \quad & 
	\\
	\mathcal{L}_{\Psi} &= \sigma_1\ \Psi^{\mu}\ \Bigg( \big(\Box + m^2 \big) \eta_\munu  + \mu_{\phi}\ \partial^\alpha \epsilon_{\alpha\nu\mu} \Bigg)\ \Psi^{\nu}
	\\
	\mathcal{L}_{\chi'} &= - \sigma_1\ \chi'\ \big(\Box + m^2 \big)\ \chi '
\end{equationsplit}

The use of gauge choice to get rid of the mixing term was successful. The above lagrangian is clearly diagonalized. However, in this process the $\chi'$ field has picked up a mass term. If one is worried at this point, that a \stuckelberg analysis without the transversal demand has given the $\chi'$ field a mass-term whereas the $\chi$ was massless in the other scenario, attention is drawn to eq\eqref{eq:stuckelbergspin1Bgaugechoice}. Through the gauge choice that was made, one sees clearly that the transversal condition $\del\cdot\Psi$ has been exchanged for a massive $\chi'$. This continuous interplay between field-redefinitions, massless/massive nature of a field is a remarkable feature of the field-theoretic language. One can continue this chain of field-redefinitions until one sees the actual physics come-forth clearly. 

In the above form of the lagrangian, the \stuckelberg trick allows one to see the physical content of the PCS theory very explicitly. The dof count for the above lagrangian goes as follows: $\Psi$ and $\chi'$ field together require 4 components for their description. The gauge condition can be used to remove one of them. And since, the $0$-component of the $\Psi$ field never had a kinetic term, leaves only 2 independent components. Thus, there are two massive dof in the PCS theory. In addition to the proca mass, the $\Psi$ field also enjoys a CS mass. 

This chapter is now concluded with a short summary. In this chapter, the theories of massive fields of spin-1, and their interplay was studied. A brief discussion of planar physics confirmed the scalar nature of the massless excitation. The proca theory had two massive dofs, which were further investigated by the polarization vectors and a \stuckelberg analysis. The theory of topologically massive electrodynamics was introduced and its peculiarities looked into. This theory had one single massive excitation. Finally, a theory with both mass-generating mechanisms was considered. This theory had two massive dof, which were mixtures of the underlying proca mass $m$ and CS mass $\mu_{\phi}$. 

\emph{Onwards to gravity.}
\chapter{Fields of Spin-2}\label{ch:spin2}
\epigraph{``Now my own suspicion is that the Universe is not only queerer than we suppose, but queerer than we can suppose."}{J. B. S. Haldane\cite{haldane}}
The theory of General Relativity was briefly commented upon in \fullref{ch:intro}. The motivations for modifying gravity, especially for studying theories of massive gravity were also detailed in that chapter. In the last chapter, various theories of spin-1 fields, massive as well as massless were considered. This chapter can be seen as a logical extension of the same analysis to fields of spin-2. Multiple theories of spin-2 fields, massive and massless, will be studied in detail here. Although, the methods and, perhaps, the character of theories considered here is generally similar to the theories from the preceding chapter, in part due to the underlying field-theoretic framework, the analysis involved, however, is necessarily more complicated. 

Gravitational interactions are the results of an interplay between ``sources" (acting as gravitational charges) of such interactions and the gravitational field. Anything which possesses energy, momentum, and even pressure is coupled to the gravitational field and can act as a source for an interaction. The gravitational field is said to mediate these interactions. This field is the only known example of fundamental spin-2 field. That it is a field of spin-2, is concluded from the following facts\cite{qgVeltman:1975vx}:
\begin{itemize}
	\item The spin of the field must be even. This is because fields of odd spin allow positive and negative charges in their abelian gauge-group algebra leading to two different behaviours, namely attraction and repulsion (Example: electromagnetism with positive and negative electric charges)\footnote{Non-abelian groups allow for more than two kinds of charges. Example: the colour charge of quarks in QCD has three types labelled red, blue and green.}. An important characteristic feature of gravitational effects is their universality: particles and anti-particles have the same response to gravity\footnote{There are experiments underway to establish this more firmly (see \cite{antimatteralpha})}. This implies that the gravitational field must be of even spin.  
	
	\item The gravitational field is known to couple to the energy-momentum tensor $T_\munu$, which acts as a source. Since this tensor is not a scalar, that rules out spin-0 fields and clearly suggests spin-2 nature of the gravitational field. \footnote{Interestingly, the guiding principles used by Einstein (General Coordinate Invariance, and the Weak Equivalence Principle) to develop GR can also be satisfied by a spin-0 scalar field (the Einstein-Fokker theory \cite{reviewhinterbichler}). Remarkably, in spite of the lack of modern terminology for classification of fields through their mass and spin, Einstein's theory of GR is a nonlinear theory of massless spin-2 fields.}
	
	\item Additionally, polarization modes of a wave/perturbation in a field can also be used to deduce the spin of the field. If a polarization mode is invariant under a rotation of angle $\theta$, then the spin-$s$ of the field is:
	$$ s = \frac{\ang{360}}{\theta}$$
	The polarization modes for spin-1 fields were vectors, and they required a full rotation of \ang{360}, to return to their original configuration. The polarization modes of gravitational waves are invariant under a rotation by \ang{180}, suggesting that the gravitational field is spin-2.
\end{itemize}

Although, GR in 3+1 dimensions is a standard textbook material, GR in 2+1-dimension differs notably. The next section is devoted to planar gravity, bringing out its odd behaviour and fixing notation for further work.  
\section{Perspective: Gravity in 2+1 Dimensions}\label{sec:gravity2+1}
When the theory of Electromagnetism in 2+1 dimensions was discussed in \fullref{sec:3delectrodynamics}, it was seen that the magnetic field $B$ was a pseudo-scalar. When compared with the corresponding field in 3+1 dimensions, the nature of the electromagnetic field was, therefore, found to be contrastingly different in 2+1 dimensions. Namely, the massless spin-1 field behaved as a scalar and had only one transverse physical propagating dof. Given that GR is, inherently, dependent as well as constructed out of geometric properties of the spacetime manifold, it can be expected that planar GR will be drastically different in physical content than GR in 3+1 dimensions. 

An elegant starting point for GR, in any dimensions, is provided by the Einstein-Hilbert action. This action can be succinctly stated as (omitting possible cosmological constant and source terms):
\begin{equationsplit}\label{eq:lagrnagianEH}
	S_{\text{EH}} &= \int d{^d}x \mathcal{L}_{\text{EH}}
	\\
	\text{where:} \quad \quad &
	\\
	\mathcal{L}_{\text{EH}} &= -\half M_{pl}^{d-2} \sqrt{|g|}\ R
\end{equationsplit}
The minus sign comes from the chosen metric signature, the factor of half is in accordance with the usage of reduced Planck's mass $M_{pl}$, $\sqrt{|g|}$ is the square-root of the determinant of metric $g_\munu$ and $R$ is the curvature/Ricci scalar. 

Varying this action with respect to the inverse of the metric tensor, leads to the non-linear Einstein Field equations:
\begin{equation}\label{eq:eomeinsteinvacuum}
	G_\munu = R_\munu - \half R g_\munu = 0
\end{equation}
These, apparently innocuous looking equations, are the vacuum Einstein field equations. $G_\munu$ is called the Einstein tensor and $R_\munu$ is the Ricci tensor. This equation relates the curvature in the spacetime manifold encoded in the Einstein tensor with sources of curvature and a possible cosmological constant. The above equation (without matter sources) describes the behaviour of the gravitational field $g_\munu$ in vacuum. In 3+1 dimensions, the absence of sources does not necessarily imply uninteresting lifeless physics. Not surprisingly, there exist gravitational waves corresponding to local physical propagating dofs, which are perturbations in the gravitational field itself. 

However for 2+1 dimensions without a cosmological constant, the above equations imply that there is no curvature and all solutions are flat Minkowski solutions. Whether or not gravitational waves exist in 2+1 dimensions can be quickly established with a quick dof count. This is now done for an arbitrary $d$ dimensional manifold. Being symmetric in the two indices, the above eom is actually $\frac{d(d+1)}{2}$ separate equations for the $\frac{d(d+1)}{2}$ components in the symmetric tensor field $g_\munu$. GR is invariant under redefinition of coordinates (general coordinate invariance), thus this can be used to eliminate $d$ components. Additionally, the Einstein tensor is divergenceless: $\nabla^\mu G_\munu = 0$. This follows from the twice-contracted Bianchi identity (here $\nabla_\mu$ is the covariant derivative). This kills $d$ more components, leaving:
\begin{equationsplit}
	\text{dof GR } &= \frac{d(d+1)}{2} - d - d = \frac{d (d-3)}{2}
\end{equationsplit}

Thus, in 3+1 dimensions ($d=4$), there are 2 local physical propagating dof allowed in Einstein's theory of GR. These correspond to gravitational waves, which were directly observed in 2015 for the very first time\cite{gravwavemass}. This is also in agreement with the general dof count for massless fields from \fullref{sec:countinghighspindof} (see eq\eqref{eq:dofmasslessspins}). Plugging $d=3$ in the above formula, implies that the number of propagating dof for gravitational fields in 2+1 dimensions is 0! This is one of the most striking differences between GR in 2+1 and 3+1 dimensions. The vanishing of propagating dof in 2+1 dimensions is a generic feature of planar physics, true for all massless HS fields. For GR, this can also be expressed in terms of purely geometrical quantities which are used to describe a manifold's curvature. This would require a decomposition of the Riemann curvature tensor which is detailed in the next section. 
\subsection{Decomposing Curvature: The Riemann Tensor}\label{subsec:decomposeriemann}

The information regarding presence of curved spacetime, or equivalently curvature, in a manifold is encoded in an object called the Riemann tensor. The Riemann curvature tensor $R_{\mu\nu\alpha\beta}$ is an object with 4 indices and many internal symmetries. If this object is zero at any point on a manifold, then this is a definite indication of the absence of curvature at that point. A tensor of rank-4 without any special properties will require $d^4$ components, for its complete specification. Thankfully, the Riemann tensor has many symmetries and after accounting for all such symmetries, there are $\frac{d^2(d^2 -1)}{12}$ independent components in this tensor. This means that for $d=4$: 20 numbers and for $d=3$: 6 numbers, are required to completely describe the local curvature at any point on the manifold. 

Dealing with these components can be made manageable by decomposing the Riemann tensor. Depending on their properties, these decompositions can be grouped together into different tensorial objects. A general decomposition of the Riemann tensor into traceful and traceless part, in terms of arbitrary tensors would look like:
\begin{equationsplit}
	R_{\mu\nu\alpha\beta} &=   a_1 g_{\mu\nu} A_{\alpha\beta} + a_2 g_{\mu\alpha} A_{\nu\beta} + a_3 g_{\mu\beta} A_{\nu\alpha} + a_4 g_{\nu\alpha} A_{\mu\beta} +a_5 g_{\nu\beta} A_{\mu\alpha} + a_6 g_{\alpha\beta} A_{\mu\nu}
	\\
	& + B \big(b_1 g_{\mu\nu} g_{\alpha\beta} + b_2 g_{\mu\alpha} g_{\nu\beta} + b_3 g_{\mu\beta} g_{\nu\alpha}\big) 
	\\
	& +c_1 C_{\mu\nu\alpha\beta}
\end{equationsplit}
$A_{\mu\nu}$, $B$,and the traceless $C_{\mu\nu\alpha\beta}$ are 3 generic tensors, and $a_i$, $b_j$ and $c_1$ are 10 unknown coefficients. The idea is to find out how the information in the Riemann tensor can be distributed over the 3 generic tensors. Step one in finding this out is to force the symmetries of Riemann tensor on the RHS. This gives:
\begin{equationsplit}
	&a_1 = a_6 =0
	\\
	&a_2 = -a_3 = - a_4 = a_5
	\\
	&b_1= 0 \quad \quad b_2= -b_3
	\\
	&c_1 \quad \text{absorbed in} \quad C_{\mu\nu\alpha\beta}
\end{equationsplit}
Using these relations, the above decomposition simplifies to:
\begin{equationsplit}
	R_{\mu\nu\alpha\beta} &=   a_2 (g_{\mu\alpha} A_{\nu\beta} - g_{\mu\beta} A_{\nu\alpha} - g_{\nu\alpha} A_{\mu\beta} + g_{\nu\beta} A_{\mu\alpha})
	\\
	& + b_2 B \big(g_{\mu\alpha} g_{\nu\beta} - g_{\mu\beta} g_{\nu\alpha}\big) 
	\\
	& + C_{\mu\nu\alpha\beta}
\end{equationsplit}
Of the many possible decompositions, three will be presented here. These are given by three different choices of the tensors $A_{\mu\nu}$, $B$,and $C_{\mu\nu\alpha\beta}$. These are:
\begin{enumerate}
	\item \textbf{Choice A: } 
	This is \emph{the} conventional choice. Tensor $A_{\mu\nu}$ is identified with the Ricci tensor $R_{\mu\nu}$ and $B$ is identified with curvature scalar $R$. The decomposition is complete on determining coefficients $a_2$ and $b_2$ :
	\begin{equationsplit}
			&a_2 = \frac{1}{d-2}
			\\
			&b_2 = \frac{-1}{(d-1)(d-2)}
	\end{equationsplit}
	
	The Einstein field equations are expressed using this choice, and they describe the traceful part of the Riemann tensor. The Einstein tensor $G_\munu$, being symmetric in its two indices has $\frac{d(d+1)}{2}$ components, which for $d=4$ means 10 components. The Einstein field equations in eq\eqref{eq:eomeinsteinvacuum} fixes exactly this traceful part of $R_{\mu\nu\alpha\beta}$. Hence, in $d=4$, 10 components inside the Riemann tensor are completely determined by the Einstein equations. The remaining components are part of the conformal Weyl tensor: $C_{\mu\nu\alpha\beta}$. The number of components in Weyl tensor, for $d$-dimensions can be calculated: 
	\begin{equation} 
	\text{indepenent components} \quad C_{\mu\nu\alpha\beta} : \frac{d(d+1)(d+2)(d-3)}{12} 
	\end{equation}
	which for $d=4$ give 10 components. In a broader sense, the traceless part of the Riemann curvature, i.e. the Weyl tensor contains the dofs corresponding to gravitational waves. 
	
	In 2+1 dimensions, however, the Riemann tensor has 6 independent components, which is the same number of components as in the Ricci tensor. Thus, all the information in the Riemann tensor is present in the traceful part. In a contrast to 3+1 dimensions, the Einstein field equations completely fix the Riemann tensor in 2+1 dimensions. In fact, putting $d=3$ in the above formula shows that the traceless Weyl tensor vanishes identically in 2+1 dimensions\footnote{An amusing way to see this is to look at the `contradiction' itself. Start by labelling each component of the Weyl tensor and impose (a) all the symmetries of the Riemann tensor and (b) the tracelessness requirement. Now solve for each component in 3 dimensions. There are far too many conditions and only the trivial solution (all components being 0) can survive.}. This implies that there are no freely propagating physical dof associated with gravity in 2+1 dimensions. 
	
	\item \textbf{Choice B: }
	Another possibility is to demand that the tensor $A_{\mu\nu}$ is traceless. This means setting $A_{\mu\nu} = R_{\mu\nu} - \tfrac{1}{d} R g_{\mu\nu}$. Identifying $B$ with $R$, this decomposition gives:
	\begin{equationsplit}
		&a_2 = \frac{1}{d-2}
		\\
		&b_2 = \frac{1}{d(d-1)}
	\end{equationsplit}

	\item \textbf{Choice C: }
	Finally, another possibility could arise by demanding $B=0$. This leads to an interesting decomposition which will be used later on. Forcing $B=0$, is enough to determine the rest of the decomposition. This forces $A_{\mu\nu}$ to become a tensor, which is formally known as the Schouten tensor $S_{\mu\nu}$:
	\begin{equationsplit}\label{eq:schoutentensor}
		S_{\mu\nu} = \frac{1}{d-2} \big( R_{\mu\nu} - \frac{1}{2(d-1)} g_{\mu\nu} R \big)
	\end{equationsplit}
	In this decomposition, the Riemann tensor is simplified to:
	\begin{equationsplit}
	R_{\mu\nu\alpha\beta} &=  C_{\mu\nu\alpha\beta} + (g_{\mu\alpha} S_{\nu\beta} - g_{\mu\beta} S_{\nu\alpha} - g_{\nu\alpha} S_{\mu\beta} + g_{\nu\beta} S_{\mu\alpha})
	\end{equationsplit}
\end{enumerate}

Having seen multiple decompositions of the Riemann tensor, one obtains an insight into why there are no propagating dofs in 2+1-dimensional GR. The identical vanishing of the Weyl tensor can be traced back to the internal symmetries present in the Riemann tensor. Another aim of the above discussion was to introduce the Schouten Tensor as another alternative for expressing curvature. The Schouten tensor is used in \fullref{sec:TMG}, for defining the Cotton Tensor $C_{\mu\nu}$ which plays the role of Weyl Tensor in 2+1 dimensions. As will be seen, these objects are instrumental in the construction of Chern-Simons mass term for spin-2 fields.

To summarize: In 2+1 dimensions the vacuum Einstein field equations determine curvature completely. For $\Lambda=0$, all solutions in vacuum are flat, and otherwise have constant curvature. In planar physics, localized curvature, if present, is concentrated entirely at the source of matter. There are no local propagating dofs to be associated with gravitational waves\footnote{It may `feel' that 2+1 dimensional gravity is trivial. Indeed, Witten starts his classic paper on 2+1 dimensional gravity by addressing this looming sense of triviality \cite{WITTENexactly198846}.}. However, global dofs may exist for manifolds with non-trivial topology \cite{qgCarlipbook:1998uc}. 

The next section looks at the linearized version of the Einstein-Hilbert lagrangian in eq\eqref{eq:lagrnagianEH}. As before, the analysis of the next section is done keeping the dimensions $d$ of the manifold arbitrary. This will allow: (a) comparison between the massless theory in $d=3$ with the rest of the theories to be presented in this chapter, and (b) comparison between the linearized massless spin-2 field theory in different dimensions. Both of these comparisons shall bring out noteworthy features.

\section{Massless: Linearized General Relativity}\label{sec:masslessspin2}
Using the field theoretic language, one could forego all of the inspiring geometrical interpretations of GR and look upon it, cold-bloodedly perhaps, as a non-linear theory for spin-2 fields. Such a viewpoint is definitely encouraged by the successes seen when applying QFT to particle-physics. To the author's knowledge, this viewpoint originates with Rosen (1940)\cite{summinggravityrosen}, and comprised the ``Gupta Program" (1954)\cite{summinggravityGupta} (for an old review see \cite{summingravityfangfronsdal}). In fact, one can actually recover the entire non-linearities present in $\mathcal{L}_{\text{EH}}$, by summing-up higher order self-interactions of a linear spin-2 field. To this end, in this section the field theoretical view-point of gravity is presented using the linearized weak-field approximation. In this approximation, the metric tensor is expanded around a flat background (the minkowski metric $\eta_\munu$) as follows\footnote{Expansion about arbitrary backgrounds will be necessary in the next chapter, when discussing two independent metrics.}:
\begin{equation}
	g_{\mu\nu} = \eta_{\mu\nu} + \lambda\ h_{\mu\nu}
\end{equation}
Here, $\lambda$ is an infinitesimally small parameter making the perturbation field $|h_{\mu\nu}|\ll 1$ satisfy the weak-field criterion\footnote{If the discussion is only in the field theoretic language, this condition justifies using the weak-field approximation and hence perturbative methods. Geometrically, the freedom to chose coordinates mandates a more thorough definition for the weak-field approximation to be applicable.}. When discussing an interacting theory of spin-2 fields, $\lambda$ will serve as a coupling constant and, in anticipation, it is identified as:
\begin{equation}\label{eq:couplingspin2massless}
	\lambda = \frac{2}{(M_{pl})^\frac{d-2}{2}}
\end{equation}

In order to obtain a lagrangian for the linear perturbation field $h_{\munu}$, one needs to expand the following objects: (a) $\sqrt{|g|}$, (b) $\Gamma^\mu_{\nu\alpha}$, (c) $R_{\mu\nu\alpha\beta}$ (d) $R_{\mu\nu}$, and finally (e) $R$. In short, one expands $\mathcal{L}_{\text{EH}}$ and keeps terms which are at-most quadratic in the linear perturbations $h_\munu$\footnote{Higher order terms will be necessary when calculating vertex functions in \fullref{subsec:vertexfunctions}.}. Schematically, this is:
\begin{equationsplit}\label{eq:lagrangianspin2perturbation}
	\mathcal{L}_{\text{EH}} &= \lambda^0 (\mathcal{L}{^{(0)}_{\text{EH}}}) + \lambda^1 (\mathcal{L}{^{(1)}_{\text{EH}}}) + \frac{\lambda^2}{2!} (\mathcal{L}{^{(2)}_{\text{EH}}}) + O(\lambda^3)
\end{equationsplit}
At $0^{th}$-order, the lagrangian $\mathcal{L}{^{(0)}_{\text{EH }}}$ vanishes trivially due to the flat Minkowski background, while the $1^{st}$-order lagrangian $\mathcal{L}{^{(1)}_{\text{EH}}}$ gives rise to the eom. At second order, $\mathcal{L}{^{(2)}_{\text{EH}}}$ describes the dynamical behaviour of the perturbation field $h_\munu$ on the minkowski background. This is given by:
\begin{equationsplit}
	\frac{\lambda^2}{2!} (\mathcal{L}{^{(2)}_{\text{EH}}}) &= - \frac{1}{2}h^{\mu\nu} \Box h_{\mu\nu} + \frac{1}{2}  h' \Box h'
	\\
	&+ \half h^{\mu\nu} \partial_{\mu} \partial^\sigma  h_{\nu\sigma} + \half h^{\mu\nu} \partial_{\nu} \partial^\sigma  h_{\mu\sigma} 
	\\
	&- \frac{1}{2} h^{\mu\nu} \partial_\mu \partial_\nu h' - \frac{1}{2} h' \partial \cdot \partial \cdot h
\end{equationsplit}
 
Before proceeding to analyze these perturbations, the Fronsdal formulation for spin-2 fields from \fullref{ch:highspin} should be verified. Recalling the definition of the Fronsdal field given in  eq\eqref{eq:fronsdalfield} and the Fronsdal tensor from eq\eqref{eq:fronsdaltesnor}, the lagrangian for spin-2 fields from the Fronsdal formulation\footnote{$\sigma_2 = \tfrac{-1}{2}$, obtained by putting $s=2$ in eq\eqref{eq:factorsigmas}.} becomes:
\begin{equationsplit}\label{eq:lagrangianspin2}
	\mathcal{L}_{\text{Fronsdal spin-2}} &= \sigma_2 \ \phi^\munu \overline{F}_\munu
	\\
	&=\sigma_2 
	\begin{pmatrix*}[l]
	\phi^{\mu\nu} \Box \phi_{\mu\nu} -  \phi' \Box \phi' 
	\\
	- \phi^{\mu\nu} \partial_{\mu} \partial^\sigma  \phi_{\nu\sigma} - \phi^{\mu\nu} \partial_{\nu} \partial^\sigma  \phi_{\mu\sigma} 
	\\
	+ \phi^{\mu\nu} \partial_\mu \partial_\nu \phi' + \phi' \partial \cdot \partial \cdot \phi
	\end{pmatrix*}
\end{equationsplit}
Comparison between the two lagrangians clearly establishes that the Fronsdal formulation provides the same description for the linearized perturbations from GR. A careful reader might be bothered by seeing the trace of the Fronsdal field in the above equation. There is nothing to worry about; even though the Fronsdal field is made up of two traceless fields, it only satisfies the double-tracelessness condition eq\eqref{eq:doubletracelessness}. This can be seen by taking a trace over the Fronsdal field in eq\eqref{eq:fronsdalfield}. Further analysis of this massless spin-2 field theory, in this $d$-dimensional setup, is provided by several important aspects which are now looked into. These are gauge symmetry, eom and the propagator for this field. These will serve as important references for comparing this theory with other theories studied in this chapter. 
\subsection{Gauge Symmetry}
Consider a gauge transformation for the field $\phi_\munu$ with an arbitrary gauge parameter $\xi_\mu$:
\begin{equation}\label{eq:gaugetransformationspin2}
\begin{aligned}
\phi_{\mu\nu} &\to \phi'_{\mu\nu} = \phi_{\mu\nu} + \delta\phi_{\mu\nu}
\\
\text{where: }
\\
&\delta\phi_{\mu\nu} = \partial_{(\mu} \xi_{\nu) }
\end{aligned}
\end{equation}

Plugging, this transformation in the lagrangian in eq\eqref{eq:lagrangianspin2}, all $\xi$ dependent terms drop out. This confirms that the linear massless theory is gauge invariant. Geometrically, these gauge transformations arise from the diffeomorphism invariance in the full non-linear theory. 

\subsection{Equations of Motion}

There are two ways to look at the linear eom. One is to linearize the non-linear Einstein field equations eq\eqref{eq:eomeinsteinvacuum}. The other would be to us the massless spin-2 lagrangian in eq\eqref{eq:lagrangianspin2}. Ofcourse, the final eom, as it should be, is independent of the route taken. Following the latter method, the lagrangian is written in a bilinear form.
\begin{equationsplit}
	&\mathcal{L}_{\text{Fronsdal spin-2}} = \phi^\munu D_{\mu\nu\alpha\beta} \phi^{\alpha\beta}
\end{equationsplit}
This operator $D_{\mu\nu\alpha\beta}$ is called the Lichnerowicz operator. It is expressed as:
\begin{equation}
D_{\mu\nu\alpha\beta} = \sigma_2 \Bigg(\Box \Big( \frac{1}{2}(\eta_{\mu\alpha}  \eta_{\nu\beta}  + \eta_{\mu\beta}\eta_{\nu\alpha})  - \eta_{\mu\nu} \eta_{\alpha\beta} \Big) 
- \del_\mu \del_{(\alpha}\eta_{\beta)\nu}  - \del_\nu \del_{(\alpha}\eta_{\beta)\mu}
+ \del_\mu \del_\nu \eta_{\alpha\beta} + \del_\alpha \del_\beta \eta_{\mu\nu}\Bigg)
\end{equation}
The equations of motion take a particularly simple form when expressed in terms of the trace-reversed field $\overline{\phi}_\munu$. The eom are:
\begin{equationsplit}\label{eq:eomspin2massless}
	\Box \overline{\phi}_\munu &= 0
	\\
	\text{Lorenz gauge for }\overline{\phi}_\munu: \quad \del^\mu \cdot \overline{\phi}_\munu &= 0
\end{equationsplit}
The first equation is a Klein-Gordon type equation for massless fields. This confirms that there are massless dof in this theory. Since the last section confirms that this theory possesses gauge symmetry, a gauge choice was used to obtain the first equation. This gauge choice, the Lorenz gauge, forms the second equation. For $d\geq3$, a field subject to these two equations will have the following number of dofs (obtainable by either counting again or put s=2 in eq\eqref{eq:dofmasslessspins}.).
\begin{equation}
	\text{massless spin-2 dof:} \quad \quad \quad \frac{d(d-3)}{2}
\end{equation}
As already remarks, this clearly gives 0 dofs for 2+1 dimensional gravity.

\subsection{Propagator}
Since the kinetic operator $D_\munu$ in the bilinear lagrangian is gauge invariant, it cannot be inverted as such. Due to it's simplicity, the de-donder gauge is called upon for this task\footnote{Usage of the more general $R_\xi$ gauge will be seen in \fullref{sec:TMG}.}. The de-donder gauge choice is defined as:
\begin{equation}\label{eq:dedondergauge}
	\partial^\mu \phi_{\mu\nu}  - \frac{1}{2} \partial_\nu\phi' = 0
\end{equation}
Enforcing this condition fixes the gauge field $\xi_\mu$ only partially. It must satisfy:
\begin{equation}\label{eq:gaugeconditiononxi}
	\Box \xi_\nu = 0
\end{equation}
In accordance with de-donder gauge, a suitable gauge-fixing term can now be added to the original lagrangian. Quantum mechanically, this would require a Fadeev-Popov gauge fixing procedure. Here, the addition of this gauge-fixing term is justified by noting that the eom from both lagrangians are the same (however, the de-donder gauge condition needs to be enforced separately). 
\begin{equation}\label{eq:lagrangiangaugefixingterm}
\mathcal{L}_{\text{gauge-fixing term}} = -2\ \sigma_2 (\partial^\mu \phi_{\mu\nu}  - \frac{1}{2} \partial_\nu\phi')^2
\end{equation}
With the above term, the gauge fixed lagrangian gets simplified to:
\begin{equation}
\begin{aligned}
\mathcal{L}_{\text{gauge-fixed}}  &=  \sigma_2 (\phi^{\mu\nu} \Box \phi_{\mu\nu}  -  \frac{1}{2} \phi' \Box \phi')
\\&=  \phi^{\mu\nu} {D}_{\mu\nu\alpha\beta} \phi^{\alpha\beta}
\end{aligned}
\end{equation}
On inverting this operator in fourier space, the propagator for the gauge-fixed lagrangian describing spin-2 massless fields is obtained:
\begin{equation}\label{eq:propagatorspin2massless}
G^{\mu\nu\alpha\beta} = \frac{1}{k^2} \Bigg( (\eta^{\mu\alpha}  \eta^{\nu\beta}  + \eta^{\mu\beta}\eta^{\nu\alpha})  - \frac{2}{d - 2}  \eta^{\mu\nu} \eta^{\alpha\beta}  \Bigg)
\end{equation}

The poles of this object confirm the massless nature of the excitation. To check the sign of the residue, put $\mu=\nu=\alpha=\beta=1$, and obtain the amplitude for the propagation of the $\phi_{11}$ component as: $\frac{1}{k^2}\frac{2(d-3)}{d-2}$. Thus, for $d\geq4$, the residue is positive signalling unitary propagation. The propagator is smart enough to give 0 when $d=3$ is plugged in. 

For later convenience, it is noted that the coefficient of $\eta^{\mu\nu} \eta^{\alpha\beta}$ term is $\frac{-2}{d-2}$.

\section{Massive: Fierz-Pauli Theory}\label{sec:fierzpaulitheory}
As with the Proca theory of massive spin-1 fields, a first step towards modifying the linear behaviour of gravity would be to give the perturbations of gravitational field, from the last section, a mass term. This theory was first put forward by Fierz-Pauli in 1939 \cite{fierzpauli}. A seemingly natural setup that will be utilized here for studying this theory is to: take the Fronsdal lagrangian in eq\eqref{eq:lagrangianspin2} as a kinetic term and then to give the Fronsdal field $\phi_\munu$ a Fierz-Pauli (FP) mass term, with mass $m$\footnote{Note the Fronsdal field was meant to describe only massless fields. The Singh-Hagen lagrangian describing massive fields required many more auxiliary fields than present in the Fronsdal case. Nevertheless, the setup used here works.}. Note - since this chapter is devoted to the study of spin-2 fields, there should be no confusion about the reuse of the letter $m$ to denote the FP mass. When considering a theory of massive spin-1 fields along with massive spin-2 fields, then due care will be taken. 

The FP lagrangian is:
\begin{equationsplit}\label{eq:lagrangianfierzpauli}
	\mathcal{L}_{\text{massive spin-2}} &= \mathcal{L}_{\text{Fronsdal spin-2}} + \mathcal{L}_{\text{FP}}
	\\[0.5em]
	\text{with: } \mathcal{L}_{FP} &= \sigma_2 m^2 \big(\phi^\munu\phi_\munu - \phi'\phi'\big)
	\\[0.5em]
	\implies \mathcal{L}_{\text{massive spin-2}} &=\sigma_2 
\begin{pmatrix*}[l]
	\phi^{\mu\nu} \big(\Box + m^2 \big) \phi_{\mu\nu} -  \phi' \big(\Box + m^2 \big) \phi' 
	\\
	- \phi^{\mu\nu} \partial_{\mu} \partial^\sigma  \phi_{\nu\sigma} - \phi^{\mu\nu} \partial_{\nu} \partial^\sigma  \phi_{\mu\sigma} 
	\\
	+ \phi^{\mu\nu} \partial_\mu \partial_\nu \phi' + \phi' \partial \cdot \partial \cdot \phi
\end{pmatrix*}
\end{equationsplit}

In this lagrangian, plugging in the gauge transformations from eq\eqref{eq:gaugetransformationspin2}, it is observed that the FP mass $m$ breaks gauge symmetry. The equations of motion from this lagrangian can be worked out:
\begin{equationsplit}\label{eq:eomFPoriginal}
&\big(\Box + m^2 \big) \phi_{\mu\nu} -  \eta_\munu \big(\Box + m^2 \big) \phi' 
- \partial_{\mu} \partial^\sigma  \phi_{\nu\sigma} -  \partial_{\nu} \partial^\sigma  \phi_{\mu\sigma}
+ \partial_\mu \partial_\nu \phi' + \eta_\munu \partial \cdot \partial \cdot \phi = 0
\end{equationsplit}
Taking the divergence and the trace of the above equation gives:
\begin{equationsplit}
\text{Divergence:}  &\implies \del\cdot\phi_\mu - \del_\mu \phi' = 0
\\
\text{Trace:} &\implies \phi' = 0 
\end{equationsplit}

These two equations together imply the divergencelessness of the field $\phi_\munu$. Putting these two back into the original eom, the Fierz-Pauli eom takes a truly-transparent form.
\begin{equationsplit}\label{eq:eomFP}
	(\Box + m^2) \phi_{\mu\nu} &= 0
	\\
	\partial^\mu \phi_{\mu\nu} &=0
	\\
	\phi' &=0
\end{equationsplit}

There is some elegance to the simplicity of the above system of equations\footnote{And underlying this neatness is a rigid structure, which prohibits any careless tinkering. This will be seen shortly.}. The first equation confirms that the excitations in this theory are massive. The other two appear as additional conditions, derived from the eom. In fact, the first two equations constitute the Fierz-Pauli conditions for massive fields. The third equation is an odd trace condition. Being unique to the Fierz-Pauli theory, it will be discussed promptly in \fullref{subsec:boulwaredeserghost}. The simplicity of the above equations make dof counting easier. The first equation describes the behaviour of $\tfrac{d(d+1)}{2}$ components in the symmetric tensorial field $\phi_\munu$. The second equation can be used to eliminate $d$ components. Additionally, the peculiar trace-condition kills off one single component. Thus,
\begin{equation}\label{eq:dofmassivespin2}
	\text{massive spin-2 dof: }  = \frac{(d-2)(d+1)}{2}
\end{equation}

Plugging in $d=4$ gives 5 independent components for the massive field. This is the same as $2s+1$ components for a massive field with $s=2$, agreeing with the group-theoretic arguments of \fullref{sec:irrep}. For $d=3$, the massive theory propagates $2$ dof, which is in agreement with the general dof count for 2+1 dimensions. It is now confirmed that the FP mass term has excited 2 new dof's which were completely absent from the massless theory. In this regard, the behaviour of FP mass term is slightly different than the behaviour of the Proca mass term seen in \fullref{sec:proca}. As a matter of fact, there are two singularly distinguishing features arising out of the unconventional or `misbehaviour' of the FP theory. These are the vDVZ discontinuity and the Boulware-Deser ghost. These will be discussed shortly.
\subsection{Propagator}

To construct the propagator for this massive spin-2 field $\phi_\munu$, the lagrangian in eq\eqref{eq:lagrangianfierzpauli} is first written in the following form:
$$ \phi^{\mu\nu} {D}_{\mu\nu\alpha\beta} \phi^{\alpha\beta}$$
The kinetic operator ${D}_{\mu\nu\alpha\beta}$ is inverted by going to momentum space. This gives the following propagator.
\begin{equationsplit}\label{eq:propagatorspin2fierzpauli}
	G^{\mu\nu\alpha\beta} =  \frac{1}{(k^2-m^2)} &\Bigg\{ \Bigg((\eta^{\mu\alpha}  \eta^{\nu\beta}  + \eta^{\mu\beta}\eta^{\nu\alpha})  - \frac{2}{d-1} \eta^{\mu\nu} \eta^{\alpha\beta}\Bigg)
	\\
	& -  \Bigg(\frac{k^\mu k^{\alpha} \eta^{\beta\nu}}{m^2}  + \frac{k^\nu k^{\alpha}\eta^{\beta\mu}}{m^2}+ \frac{k^\mu k^{\beta} \eta^{\alpha\nu}}{m^2}  + \frac{k^\nu k^{\beta}\eta^{\alpha\mu}}{m^2}\Bigg)
	\\
	& +\frac{2}{d-1} \Bigg(\frac{k^\mu k^\nu \eta^{\alpha\beta}}{m^2} + \frac{k^\alpha k^\beta \eta^{\mu\nu}}{m^2} + (d-2)\frac{k^\mu k^\nu k^\alpha k^\beta}{m^4}\Bigg) \Bigg\}
\end{equationsplit}

This propagator's pole clarifies that the excitations being propagated have mass $m$. The residue of the propagator has correct sign for unitary propagation. Considering the massless limit $m\rightarrow0$, some terms in the propagator blow up. This is very reminiscent of a similar scenario seen for the propagator of the massive spin-1 fields in eq\eqref{eq:propagatorspin1massive}. There, the problem was identified as coming from the increasingly dominating contributions of the longitudinal polarizations in their corresponding massless limit. Looking at the structure of the problematic terms in the propagator above, it is observed that these terms correspond to a longitudinal dof. Since, in this setup the longitudinal dofs are mixed with other dof, the massless limit is blowing up. With a field decomposition, or a \stuckelberg analysis it will be seen that this difficulty goes away. 

More importantly, note the coefficient of $\eta^{\mu\nu} \eta^{\alpha\beta}$. It is equal to $\frac{-2}{d-1}$. The coefficient for the same term in the massless propagator in eq\eqref{eq:propagatorspin2massless} was $\frac{-2}{d-2}$. It is duly noted that there is a difference in the coefficients of this term in the two propagators. This is not merely a curious distinction between the two theories, but is an indication of a deep theoretical issue. As such this difference may be seen as a first gentle sign of a discontinuity in the theory. This will be further discussed in the \fullref{subsec:vdvzdiscontinuity}. 

For neatness, the propagator can also be expressed as:
\begin{equationsplit}
		G^{\mu\nu\alpha\beta} &=  \frac{1}{(k^2-m^2)}  \Bigg((P^{\mu\alpha}  P^{\nu\beta}  + P^{\mu\beta}P^{\nu\alpha})  + \frac{-2}{d-1} P^{\mu\nu} P^{\alpha\beta} \Bigg)
		\\
		\text{where: }
		\\
		P^{\mu\nu} &= \eta^{\mu\nu} - \frac{k^\mu k^\nu}{m^2}
\end{equationsplit}
The numerator of the pole in this propagator is a projector onto the sub-space of all symmetric, transverse and traceless tensors of rank-2. Thereby, this is the identity-operator for this sub-space. 

\subsection{vDVZ Discontinuity}\label{subsec:vdvzdiscontinuity}

From physical theories that hope to bear any chance for describing nature, there is a technical expectation: these theories must be continuous in their physical predictions for all values of the parameters that go into the theory. For example, all predictions made from the Proca theory of massive spin-1 fields always agree with corresponding predictions of the massless spin-1 fields, in the limit when the Proca mass is taken to 0. This leads one to the conclusion that the Proca theory \emph{is really} a massive version of the massless theory. However, if the predictions were to be off, then this would signal that the two theories are inherently different! 

For the case of spin-2 fields, the FP theory is an inherently different theory in comparison to the theory of linear massless spin-2 fields. This was first pointed out by van Dam, Veltman and separately by Zakharov in 1970, and is today referred to as the vDVZ discontinuity \cite{vdvz1, vdvz2}. A simple sign of this discontinuity has already appeared in the propagators for the two theories. This will now be seen more rigorously.

Consider an interaction between two conserved sources $A$ and $B$, with energy-momentum tensors $T{^a_\munu}$ and $T{^b_{\alpha\beta}}$ respectively. For some coupling constant $\lambda$, this interaction, at the linear level, is given by:
\begin{equation}
		\text{interaction} \quad \propto \quad \lambda\ T{^a_\munu}\ G^{\mu\nu\alpha\beta}\ T{^b_{\alpha\beta}}
\end{equation}
The coupling constant $\lambda$ for the massless case is given in eq\eqref{eq:couplingspin2massless}. 

A-priori the FP theory could be different from the massless version. To be general let the coupling constant for the FP theory be some $\lambda'$. Now calculating the interaction for the $-00-$ component of the energy-momentum tensor in these two theories leads to the following:
\begin{equationsplit}
		\text{GR massless spin-2: } \quad \quad & \lambda\ \frac{T{^a_{00}}T{^b_{00}}}{k^2}\ \frac{2(d-3)}{(d-2)} 
		\\
		\text{FP massive spin-2: } \quad \quad & \lambda'\ \frac{T{^a_{00}}T{^b_{00}}}{k^2-m^2}\ \frac{2(d-2)}{(d-1)} 
\end{equationsplit}

Apart from the obvious difference arising from the presence of the massive pole, one notices that the coefficients in the two results are also very different. This suggests that considering a massless limit of the FP theory will potentially lead to results which are different from those obtained from linearized GR. Doing this calculation for arbitrary $d$-dimensions, it is clear that this difference is not a particular feature of 2+1 or any other dimension. The two theories make differing predictions for the interaction considered in all dimensions! This difference is a spectacle in 2+1 dimensions: massless spin-2 fields can carry no interaction at all; whereas, there will be some finite interaction from massive spin-2 fields. 

There is still some hope. At this point, the coupling constant $\lambda'$ could be adjusted so that the interaction in the two theories match in the massless limit. This would force:
\begin{equation}
	\lambda' = \lambda \frac{(d-3)(d-1)}{(d-2)^2}
\end{equation}
This implies that the universal coupling strength of massive gravitons is indeed different than the coupling strength of massless gravitons. Surely, this can lead to testable predictions. Consider the scenario of light bending in the presence of a gravitational field. Since the energy-momentum tensor of massless photons is traceless, its interaction with gravitons is dictated by those terms in the propagator which have exactly the same coefficients for both theories. This implies that the difference in the prediction, for light-bending, between these two theories is proportional to the amount by which their coupling strengths differ. This can be quantified as:
\begin{equation}
	\text{Difference of light-bending: } \propto \frac{1}{(d-2)^2}
\end{equation}

This equation says that in 2+1 dimensions, there will be a 100\% difference between the prediction for light-bending in the two theories. For $d=4$, this difference is 25\%\footnote{Amusingly, if the Universe had infinite dimensions, then one could not use these arguments to distinguish between massive and massless gravitons. Thankfully, perhaps, this does not seem to be the case so far.}. This difference is large enough to be measured, and has been measured in experiments looking at light deflections from the Sun. Within the solar-system, the predictions from massless spin-2 fields or linearized GR are vindicated \cite{lightbending}. Experiments within the solar-system itself could, in principle, rule out the FP theory of massive gravity. Yet, this is only true for the linear-regime of GR. As was mentioned in \fullref{sec:gravityreview}, the linear-approximation works very well for the scales involved in the solar-system. A non-linear theory of massive gravity is still not ruled out. 

The origin of the vDVZ discontinuity lies in the extra dof present in the massive theory which were absent from the massless theory. Unlike the case of spin-1 fields, these dofs couple to the trace of the energy-momentum tensor and hence leave a distinguishable imprint. This will be seen more clearly with a \stuckelberg analysis. 
\subsection{Boulware-Deser Ghost}\label{subsec:boulwaredeserghost}

The last equation in eq\eqref{eq:eomFP} is a peculiar trace condition for FP theory. This originates from the Fierz-Pauli tuning of the mass term in $\mathcal{L}_{\text{FP}}$ which is reproduced here with a small modification.
\begin{equation}
	\mathcal{L}_{FP} = \sigma_2 m^2 \big(\phi^\munu\phi_\munu - (1 - \gamma) \phi'\phi'\big)
\end{equation}
Setting $\gamma = 0$ recovers the original mass term. Any violations to this coefficient, the Fierz-Pauli tuning, are heavily penalized. For $\gamma\neq0$, the action describes an additional massive scalar dof. This extra dof can be seen, most easily, by calculating the propagator again. This new propagator has some new terms which are an addition to the propagator in eq\eqref{eq:propagatorspin2fierzpauli}. These additional terms are:
\begin{equationsplit}
	\text{additional terms:}\quad G{_{\text{ghost}}^{\mu\nu\alpha\beta}} &= - \frac{2 \gamma (d-2)}{(d-1)} \Bigg( \frac{\eta^\munu \eta^{\alpha\beta} + \frac{k^\mu k^\nu \eta^{\alpha\beta}}{m^2} + \frac{k^\alpha k^\beta \eta^{\mu\nu}}{m^2} + \frac{k^\mu k^\nu k^\alpha k^\beta}{m^4} }{k^2 - m{^2_{\text{ghost}}}}\Bigg)
	\\
	m{^2_{\text{ghost}}} &= \frac{3 - 4 \gamma}{2 \gamma} m^2
\end{equationsplit}

There are a few things to notice about this new pole. First, it comes with a wrong sign. This signals non-unitary propagation and hence justifies calling this extra dof as a ghost. Ghosts are a nuisance in QFT and their presence in any theory, makes that theory strictly unsuitable for any application. Second, this is a massive ghost. On setting the tuning coefficient $\gamma=0$, the mass of this ghost diverges. Finally, the structure of indices in all of the terms present in the numerator, suggest that this ghost is mixed with the trace of the field $\phi'$. This is confirmed by noting that setting $\gamma=0$, sets this entire additional term to 0 and also provides the condition $\phi'=0$. 

It is clear that modifying the tuning in the FP mass term has introduced an additional ghost dof into the theory. The trace constraint in eq\eqref{eq:eomFPoriginal} is lost. The total number of propagating dof in FP theory without the FP tuning is:
$$ \gamma\neq0 \quad \text{massive spin-2 dof: } \quad  \frac{d(d-1)}{2} $$
This gives the theory \emph{one extra dof} than what is expected for a massive spin-2 field. This unwanted dynamical field is called as the Boulware-Deser ghost. This ghost mode complicates the construction of a nonlinear massive theory of gravity. This is due to the frustrating re-appearance of this ghost mode in non-linear theories even when they are explicitly removed from the linear version. 

In the past, this lead to the suggestion of a no-go theorem regarding the construction of non-linear massive theories of gravity \cite{reviewMay}. Nevertheless, a consistent non-linear theory of massive gravity, the dRGT theory has been found. In \fullref{ch:bimetric}, a more generalized version of this non-linear theory, the bimetric theory of gravitation, will be presented. As physics in 2+1 dimensions is special, consistent massive theory of gravity can be afforded without heeding to the bimetric model. This is the theory of Topologically Massive Gravity (TMG), presented in \fullref{sec:TMG}. A major outcome of this thesis has been to extend the bimetric theory with the topological mass-generating mechanism, available in 2+1 dimensions.

\section{The St\"uckelberg Analysis}\label{subsec:stuckelbergspin2}
\subsection{Massive \emph{vs} Massless}
The FP theory of massive spin-2 fields is an interesting variation to the theory of massless spin-2 fields. In 2+1 dimensions, it has been pointed out, through different routes (in \fullref{sec:irrep}, \fullref{sec:gravity2+1} and also \fullref{sec:masslessspin2}) that massless spin-2 fields have no propagating dof in 2+1 dimensions. This is in stark contrast with the massive theory which has excited 2 completely new dof in the theory. Where do these two new dofs come from?

A related question could be asked by comparing the dofs propagated by each theory in 3+1 dimensions. The massless theory propagates 2 dofs whereas the massive theory has 5. Infact, this feature can be extended to arbitrary dimensions $d$. Since the number of dof for a massive/massless field of spin-$s$ in arbitrary dimensions $d$ is known in eq\eqref{eq:dofmassivespins} and eq\eqref{eq:dofmasslessspins}, the extra dofs excited by the massive theory can be counted, in a more general way. This gives for $d$-dimensions:
\begin{equation}\label{eq:extradofspin2}
	\text{for spin-2: massive dof} - \text{massless dof } = d-1 
\end{equation}

Thus, irrespective of dimensions (for $d\geq2$) there are extra dofs excited by the massive theory. A legitimate question is: Where do these extra dofs come from? The \stuckelberg analysis in this section aims to address this inquiry.

As was seen in the \stuckelberg analysis from the preceding chapter, the general idea is to use field re-definitions. Unlike the case of spin-1 fields, however, as the spin of a tensorial fields increases the number of allowed field re-definitions also continues to increase. Consider for example, the case of spin-2 fields. A spin-2 field, such as the Fronsdal field $\phi_\munu$ (with $s=2$), can be further thought of as being mixed with a spin-1 field $\Pi_\mu$. This would be written as:
\begin{equation}
	\phi_\munu = \Psi_\munu + \del_\mu \Pi_\nu + \del_\nu \Pi_\mu
\end{equation} 
Here, the $\Psi_\munu$ field serves to store components from the original $\phi_\munu$ field which cannot be resolved into a vectorial field. However, there could also be a scalar hidden in the original $\phi_\munu$ field. Moreover, as seen in \fullref{subsec:stuckelbergspin1proca}, the vector field $\Pi_\mu$ could also have a scalar mode. Separating out these mixed modes constitutes a \stuckelberg chain of field re-definitions. For spin-2, this chain is:
\begin{equationsplit}
	\phi_{\mu\nu} & = \Psi_{\mu\nu} + \partial_\mu \Pi_\nu + \partial_\nu \Pi_\mu
	\\
	\Pi_\mu & = \pi_\mu + \partial_\mu \chi
	\\
	\Psi_{\mu\nu} & = \psi_{\mu\nu} + \chi' \eta_{\mu\nu}
\end{equationsplit}
Note: the last field re-definition is like a linearized conformal transformation. Both the spin-2 sector and the spin-1 sector contribute a scalar mode each. There is a-priori no reason that these scalar modes should be the same (although they will most certainly mix to form one unique scalar). Thus, finally the spin-2 Fronsdal field can be decomposed as:
\begin{equation}\label{eq:stuckelbergfieldredefspin2}
	\phi_{\mu\nu} =  \psi_\munu + \partial_\mu \pi_\nu + \partial_\nu \pi_\mu + \eta_\munu \chi' + 2 \del_\mu \del_\nu \chi
\end{equation}

The true effect of such field-redefinitions can only become clear by plugging these inside a good lagrangian. Using the FP lagrangian in eq\eqref{eq:lagrangianfierzpauli}, will do the job. A startling punchline is now waiting to be delivered. Schematically, the new lagrangian can be written as:
\begin{equationsplit}\label{eq:lagrangianstuckroleofmass}
	\mathcal{L}_{\text{massive spin-2}}(\phi_{\mu\nu}) = \mathcal{L}_{\psi_{\mu\nu}} + m \Bigg( \mathcal{L}_{\pi_\mu} + \mathcal{L}_\chi + \mathcal{L}_{\chi'} + \mathcal{L}_{mix} \Bigg)
\end{equationsplit}

The lagrangian can thus be seen to be composed of multiple lagrangians governing several independent fields $\psi_{\mu\nu}$, $\pi_\mu$, $\chi$, and $\chi'$ which may undergo interactions governed by a lagrangian containing the mixed terms. Usually, through proper choices, these mixed terms can be eliminated and the entire lagrangian is diagonalized. In its diagonalized form, a lagrangian will be composed of truly independent sectors of different fields \emph{which do not talk to each other}. The punchline is to note, what really happens when $m\rightarrow0$ is considered. It is clear, that the lagrangian for all the additional fields is dropped. Furthermore, any mass term present in the $\mathcal{L}_{{\psi_{\mu\nu}}}$ will also be dropped, turning $\psi_{\mu\nu}$ back into its massless version. As promised, the startling punchline is:

\emph{Adding a mass term has the effect of exciting new fields, which were absent in the corresponding massless theory.}

Since, this statement is true for all spins the number of extra dof excited by the mass term for spin-2 fields can now be understood. Suppose after a field decomposition of spin-2 fields, the vector mode is massive. Then, in accordance with the above principle, the mass term for the vector must have excited a scalar. Doing another field decomposition, the vector can be brought to its \emph{canonically massless} form. Thus, a massive spin-2 field can excite a massless vector (with $d-2$ dof for arbitrary dimensions) and a single scalar (only 1 dof). Together, these account for the new extra $d-1$ dof counted in eq\eqref{eq:extradofspin2}. 

It is now clear, for 2+1 dimensions, where the two dof in massive theory are coming from. They belong to a vector mode and a scalar mode. This will be demonstrated shortly, in a rather elegant manner. 

\subsection{Insights from St\"uckelberg Analysis - FP Theory}\label{subsec:stuckelbergFP}
Continuing with the spin-2 example, the $\chi'$ field is dynamically coupled to the spin-2 field with terms such as $\psi'\Box\chi'$. Demanding that such terms drop away, the $\chi'$ fields is set as:
\begin{equation}
	\chi' = \frac{-2 m^2}{d-2} \chi
\end{equation}
Finally, on rescaling the vector mode as $\pi_\mu \to \frac{1}{m} \pi_\mu$ and the scalar mode as $\chi \to \frac{1}{m^2} \chi$, the new lagrangian becomes:
\begin{equationsplit}\label{eq:lagrangianstuckelbergspin2}
&\mathcal{L}_{Stuck} = \mathcal{L}_{\psi_{\mu\nu}} + \mathcal{L}_{\pi_\mu} + \mathcal{L}_\chi + \mathcal{L}_{mix}
\\
\text{where: } \quad  &
\\
&\mathcal{L}_{\psi_{\mu\nu}}  = \mathcal{L}_{\text{kinetic}\ \psi_{\mu\nu}} + \mathcal{L}_{\text{FP}\ \psi_{\mu\nu}}
\\
&\mathcal{L}_{\pi_\mu}  = - \frac{1}{2} F_{\mu\nu} F^{\mu\nu} \quad\quad \Big(\text{with:} F_\munu = \del_\mu \pi_\nu - \del_\nu \pi_\mu\Big)
\\
&\mathcal{L}_{\chi}  = -2 \  \frac{d-1}{d-2} \  \chi \Bigg( \Box  - m^2 \frac{d}{d-2}\Bigg) \chi
\\
&\mathcal{L}_{mix}  =  -2  m \Bigg( \psi^{\mu\nu} \partial_{(\mu}  \pi_{\nu)}  - \psi'\  \partial \cdot \pi 
+ \  (\frac{d-1}{d-2}) \  ( 2 \chi \partial \cdot \pi + m \psi' \chi )			 \Bigg)
\end{equationsplit} 

Three important points are brought forth: 
\begin{enumerate}
\item Had the original lagrangian started with an energy momentum tensor $T_\munu$ for a source, then upon taking a massless limit in the above lagrangian one term which describes the coupling between scalar field $\chi$ and the source will survive. This term is:  $ \ \frac{2 \lambda}{d-2}\  \chi\  T'$. The existence of this term shows that the FP theory is discontinuous in its massless limit, since the extra scalar mode is still coupled to the trace of the energy-momentum tensor. This explains the root cause of the vDVZ discontinuity which was discussed in \fullref{subsec:vdvzdiscontinuity}. The contribution from this term exactly balances the differences in predictions calculated in that section, accounting for the discrepancy completely. 

\item If the original lagrangian was used with an arbitrary value for the FP tuning parameter $\gamma$, then their would have been a higher-derivative term for the scalar mode, corresponding to $\chi \Box^2 \chi$. These higher derivatives imply that there is an additional ghost dof riding with the scalar dof. This explains the origin of the Boulware-Deser ghost. Note that the $\chi$ field has a contribution coming from the $\chi'$ field which is coming directly from the trace of the original $\phi_\munu$ field. This explains why the Boulware-Deser ghost is linked with the trace of the original field, as discussed in \fullref{subsec:boulwaredeserghost}. 

\item In the original FP lagrangian eq\eqref{eq:lagrangianfierzpauli}, the mass term broke gauge invariance. This new lagrangian in eq\eqref{eq:lagrangianstuckelbergspin2} has an interesting gauge symmetry. This gauge symmetry can be divided into two parts. One of which is associated with an arbitrary vectorial gauge parameter $\xi_\mu$. The  gauge transformations are:
\begin{equationsplit}\label{eq:stuckelbergspin2gaugetransformations1}
	\delta 	\psi_{\mu\nu} & = \partial_\mu \xi_\nu + \partial_\nu \xi_\mu  
	\\
	\delta \pi_\mu & = - m \xi_\mu 
\end{equationsplit}
The second gauge transformation is associated with an arbitrary scalar gauge parameter $\Lambda$. The gauge transformations are:
\begin{equationsplit}\label{eq:stuckelbergspin2gaugetransformations2}
	\delta \pi_\mu & = -  \partial_\mu \Lambda
	\\
	\delta \chi & =  m \Lambda
\end{equationsplit}
These gauge transformations are kept separate, because the \stuckelberg lagrangian satisfies them separately. 
\end{enumerate}

To be sure that these field redefinitions and this \stuckelberg formalism does indeed describe the same physics as the original FP lagrangian, a dof count is necessary. In order to make the counting procedure easier, one must diagonalize the lagrangian completely. For doing this, the gauge freedom described above will be imperative.

A gauge choice inspired by the de-donder gauge in eq\eqref{eq:dedondergauge} is used. This gauge choice is given by:
\begin{equation}\label{eq:gaugechoice1}
(\partial_\mu \psi^{\mu\nu} - \frac{1}{2} \partial^\nu \psi'\  ) - m \pi^\nu = 0
\end{equation}
This gauge will work if the vectorial gauge parameter $\xi_\mu$ satisfies:
\begin{equation}\label{eq:gaugeresidual}
	(\Box + m^2) \xi_\mu = 0
\end{equation}
The gauge freedom coming from the scalar gauge parameter $\Lambda$ is used to force:
\begin{equation}\label{eq:gaugechoice2}
\partial \cdot \pi + \frac{m}{2} \psi' - 2 m\ \frac{d-1}{d-2}\ \chi = 0
\end{equation}
which is satisfied if the parameter $\Lambda$ satisfies:
\begin{equation}
(\Box + m^2) \Lambda = 0
\end{equation}

Inspired by these gauge-choices, a gauge fixing term can be designed to exactly cancel the terms in $\mathcal{L}_{mix}$. These gauge-fixing lagrangians are:
\begin{equationsplit}\label{eq:gaugefixingterms}
	\mathcal{L}_{GFI} & = \Bigg( (\partial_\mu \psi^{\mu\nu} - \frac{1}{2} \partial^\nu \psi'\  ) - m \pi^\nu \Bigg)^2
	\\
	\mathcal{L}_{GFII} & = -\Bigg( \partial \cdot \pi + \frac{m}{2} \psi' - 2 m \frac{d-1}{d-2} \chi   \Bigg)^2
\end{equationsplit}

Although, these gauge-fixing lagrangian do their job, they leave an imprint on the nature of the vectorial and the scalar modes. This is seen by adding the two lagrangian in eq\eqref{eq:gaugefixingterms}, to the lagrangian in eq\eqref{eq:lagrangianstuckelbergspin2}. This gives the gauge-fixed \stuckelberg lagrangian.
\begin{equationsplit}\label{eq:lagrangianstuckelbergspin2gaugefixed}
		&\mathcal{L}_{\text{gauge-fixed}}  = \mathcal{L}_{\psi_{\mu\nu}} + \mathcal{L}_{\pi_\mu} + \mathcal{L}_\chi
		\\
		\text{where: } \quad &
		\\
		&\mathcal{L}_{\psi_{\mu\nu}} =-\frac{1}{2}\psi^{\mu\nu} (\Box + m^2) \psi_{\mu\nu} 
		+  \frac{1}{4} \psi' (\Box + m^2) \psi'
		\\
		&\mathcal{L}_{\pi_{\mu}} = \pi_\mu \  (\Box + m^2) \  \pi^\mu
		\\
		&\mathcal{L}_{\chi} = -2 \  \frac{d-1}{d-2} \ \chi \  (\Box + m^2) \ \chi
\end{equationsplit}

Clearly, the lagrangian has now been completely diagonalized. During these steps, the vector mode $\pi_\mu$ has picked up a mass term. The equations of motion for each field can now be directly read from the lagrangian itself. Using these eoms, the dof counting is now presented. 
\begin{equationsplit}
	\begin{matrix*}[l]
	\text{dof: } &\psi_{\mu\nu} &= \frac{d(d+1)}{2}
	\\
	\text{dof: } &\pi_\mu &= d
	\\
	\text{dof: } &\chi &= 1
	\\
	\text{absence of kinetic term for } &\pi_0 &=-1
	\\
	\text{constraint from } &eq\eqref{eq:gaugechoice1} &= -d
	\\
	\text{residual gauge freedom} &eq\eqref{eq:gaugeresidual} &= -d
	\\
	\text{constraint from } &eq\eqref{eq:gaugechoice2} &= -1
	\\
	\text{final dof} &&= \frac{(d-2)(d+1)}{2}
	\end{matrix*}
\end{equationsplit}

This gives exactly the correct number of components required for a massive spin-2 field as in eq\eqref{eq:dofmassivespin2} (or by directly putting $s=2$ in eq\eqref{eq:dofmassivespins}). This verifies that the chain of \stuckelberg field definitions that have been employed in this section, and the lagrangian so obtained describes the same physical content. The propagators of these fields can also be straightforwardly calculated. These are:
\begin{equationsplit}\label{eq:propagatorstuckelbergspin2FP}
		G_\chi &= \frac{1}{2} \  (\frac{d-2}{d-1}) \ (\frac{1}{k^2 - m^2}) 
		\\
		G^{\mu\nu}_{\pi} &=   \frac{- \eta^{\mu\nu}}{k^2 - m^2}
		\\
		G^{\mu\nu\alpha\beta}_{\psi} &= \frac{1}{k^2 - m^2} \Bigg( (\eta^{\mu\alpha}  \eta^{\nu\beta}  + \eta^{\mu\beta}\eta^{\nu\alpha})  - \frac{2}{d - 2}  \eta^{\mu\nu} \eta^{\alpha\beta}  \Bigg)
\end{equationsplit}

The poles of all three propagators show massive excitations which can also be seen in the gauge-fixed lagrangian in eq\eqref{eq:lagrangianstuckelbergspin2gaugefixed}. The residue of all three propagator signal unitary propagation (Note - there is no kinetic term for $\pi_0$, thus the minus sign in that propagator cancels the minus sign coming from the metric). All the propagators have well-defined massless limit. This cures the problem of taking a massless limit in the propagator for the FP theory in eq\eqref{eq:propagatorspin2fierzpauli}. 

Additionally, the highlight of this procedure is the massless limit of the propagator for the spin-2 mode $G^{\mu\nu\alpha\beta}_{\psi}$. Not only is the coefficient of the term $\eta^{\mu\nu} \eta^{\alpha\beta}$ consistent with the massless propagator in eq\eqref{eq:propagatorspin2massless}, the entire propagator for the $\psi_\munu$ mode reduces to the massless propagator on setting $m\rightarrow0$! Thus, this spin-2 mode is now continuous in its mass-parameter unlike the FP massive theory. In a way, this cures the vDVZ discontinuity. 

Finally, going back to 2+1 dimensions, it is time to make full profit of $d$-dimensional calculations. The original query was: Where do the additional dof excited by the massive theory come from? The three propagators above are now smart enough to describe the propagating modes on their own. Just put $d=3$, and as an example set all open indices to 1. This gives:
 \begin{equationsplit}\label{eq:propmodesstuckspin-2}
 	&\text{for 2+1 dimensions only:}
 	\\
 	&G_\chi = \frac{1}{4}  \ (\frac{1}{k^2 - m^2}) 
 	\\
 	&G^{11}_{\pi} =   \frac{1}{k^2 - m^2}
 	\\
 	&G^{1111}_{\psi} = 0
 \end{equationsplit}

Speaking of elegance, these equations \emph{speak for themselves}. 
\section{Topologically Massive Gravity}\label{sec:TMG}
The theory of Topologically Massive Gravity was discovered in the early 1980s. It presents itself as an interesting possibility allowed only in an odd-dimensional spacetimes, such as 2+1 dimensions. It was shown earlier in \fullref{sec:masslessspin2}, that spin-2 fields which are massless do not carry any physically propagating dof. This situation can be changed by adding a Chern-Simons (CS) mass term. 

Most of the ingredient that go into this theory have already been introduced and dealt with. The only remaining ingredient is, obviously, the CS term itself. A starting point for this is the Schouten tensor $S_\munu$ which was introduced in \fullref{subsec:decomposeriemann}. The definition of this tensor is repeated from eq\eqref{eq:schoutentensor} (for the special case of $d=3$):
\begin{equationsplit}
	S_{\mu\nu} =  R_{\mu\nu} - \frac{1}{4} g_{\mu\nu} R
\end{equationsplit}
It was shown in the same section, that the Weyl tensor vanishes identically in 2+1 dimensions. The role of Weyl tensor in planar physics is played by the Cotton tensor. This tensor is defined as:
\begin{equation}
	C^\munu\; = \frac{1}{\detsqrt{g}}\; \epsilon^{\mu\alpha\beta}\; \nabla_\alpha\; S^\nu_{\beta}
\end{equation}
The action giving rise to TMG can now be stated as (with a relative coefficient of $\mu_g$):
\begin{equationsplit}\label{eq:lagrangianTMGnonlinear}
	\mathcal{L}_{\text{TMG non-linear}} &= m_g \big(\detsqrt{g} R + \frac{1}{2 \mu_g} \mathcal{L}_{cs-g} \big)
	\\
	\text{where: }\quad&
	\\
	m_g &= \half M_{pl}
	\\
	\mathcal{L}_{cs-g} &= \epsilon^{abc}\; \Gamma^e_{ad}\; \big( \partial_b \Gamma^d_{ce} + \frac{2}{3} \Gamma^d_{bf} \Gamma^f_{ce} \big)
\end{equationsplit}
The notation for Planck mass has been simplified from $\half M_{pl}$ to $m_g$in anticipation of the bimetric theory of gravity in \fullref{ch:bimetric}. As the name suggests, in that theory two independent metrics will be present and it will be important to distinguish between the Planck masses associated with each metric. This notation is justified since the subscript clearly indicates the metric for which the Planck mass stands. Next, note that the sign of the kinetic term in this lagrangian is opposite of the sign in $\mathcal{L}_{\text{EH}}$ eq\eqref{eq:lagrnagianEH}. This is necessary to ensure that the dof obtained from this theory is not a ghost. With these preliminary remarks, the study of the above lagrangian proceeds in the following manner. 

The variation of the TMG action with respect to the inverse metric gives the non-linear eoms. These are:
\begin{equation}\label{eq:eomnonlinearTMG}
	G_\munu\; + \frac{1}{\mu_g} C_\munu = 0
\end{equation}
In order to study the nature of excitations present in this theory, it is important to look at its linearized approximation. This approximation is obtained in a manner analogous to the case of linearized GR from \fullref{sec:masslessspin2}. The metric is expanded as:
\begin{equation}
g_{\mu\nu} = \eta_{\mu\nu} + \frac{1}{\sqrt{m_g}}\ \phi_{\mu\nu}
\end{equation}
Using this, all geometric objects that go into the TMG lagrangian need to be expanded. Performing this expansion, and keeping terms which are quadratic in the perturbation field gives the following lagrangian (showing only quadratic terms):
\begin{equationsplit}\label{eq:lagrangianTMGlinear}
	\mathcal{L}_{\text{TMG}} & = - \frac{1}{2}\phi^{\mu\nu} \Box \phi_{\mu\nu} + \frac{1}{2}  \phi' \Box \phi'
	\\
	&+ \phi^{\mu\nu} \big(\partial_{(\mu} \partial^\sigma  \phi_{\nu)\sigma} \big)
	- \frac{1}{2} \phi^{\mu\nu} \partial_\mu \partial_\nu \phi' - \frac{1}{2} \phi' \partial \cdot \partial \cdot \phi
	\\
	& -\frac{1}{4 \mu_g} \phi^{\mu\nu} \big(\epsilon{_\mu}^{\alpha\beta} \partial_\alpha (\Box \phi_{\beta\nu} - \partial_\nu \partial^\sigma \phi_{\beta \sigma} )  + (\mu \leftrightarrow \nu) \big)
\end{equationsplit}

At this point, it is indeed interesting to verify whether the HS CS term in eq\eqref{eq:highspinCSterm} conjectured in \fullref{sec:highspintopologicallymassive} gives the right lagrangian for TMG. A theory based on pure CS term for spin-2 was analyzed in \fullref{subsec:spin2csonly}. The recipe to obtain a TMG lagrangian from the HS theory is to use the Fronsdal lagrangian for the kinetic term and add the said CS term. This gives:
\begin{equationsplit}\label{eq:lagrangianTMG}
	\mathcal{L}_{\text{TMG}} &= \mathcal{L}_{\text{kinetic}} + \frac{1}{\mu_g}\ \mathcal{L}_{\text{cs spin-2}}
	\\
	\text{where: } \quad \quad &
	\\
	\mathcal{L}_{\text{kinetic}} &= \sigma_2 \ \phi^\munu \overline{F}_\munu
	\\
	\mathcal{L}_{\text{cs spin-2}} &= \sigma_2\ \phi^{\munu}\ \partial^\alpha \epsilon\indices{_\alpha_\beta_{(\mu}}\overline{F}_{\nu)}{^\beta}
	\\[0.5em]
	\text{together: } \quad \quad &	
	\\
	\mathcal{L}_{\text{TMG}} &= \sigma_2 
	\begin{pmatrix*}[l]
		\phi^{\mu\nu} \Box \phi_{\mu\nu} -  \phi' \Box \phi' 
		\\
		- \phi^{\mu\nu} \partial_{\mu} \partial^\sigma  \phi_{\nu\sigma} - \phi^{\mu\nu} \partial_{\nu} \partial^\sigma  \phi_{\mu\sigma} 
		\\
		+ \phi^{\mu\nu} \partial_\mu \partial_\nu \phi' + \phi' \partial \cdot \partial \cdot \phi
	\end{pmatrix*}
	\\
	&+\frac{\sigma_2}{2 \mu_g}\ \phi^{\munu}\ \Bigg(\del_\alpha \epsilon\indices{^\alpha^\beta_{\mu}} \big( \Box \phi_{\nu\beta} - \del_{\nu} \del^\rho \phi_{\beta\rho}\big) + \mu\leftrightarrow\nu \Bigg)
\end{equationsplit}

Noting that $\sigma_2 = \frac{-1}{2}$, it is clear that the above lagrangian is exactly the same as the lagrangian in eq\eqref{eq:lagrangianTMGlinear} which was obtained upon inearizing the non-linear TMG lagrangian. This demonstrates that the physics described by the TMG lagrangian above, inspired by the HS language will necessarily govern the same physics. This also provides a check for the HS CS term proposed earlier.  

To further analyze this theory, three aspects will be studied. These are the gauge-symmetries, eoms and the propagator for this theory.
\subsection{Gauge Symmetry}\label{subsec:TMGgauge}
This lagrangian enjoys gauge symmetry upto total derivatives. The gauge transformation for the field: $\phi_{\mu\nu} \to \phi'_{\mu\nu} = \phi_{\mu\nu} + \delta\phi_{\mu\nu}$ with an arbitrary gauge parameter $\xi_\mu$ is:
\begin{equationsplit}
\delta\phi_{\mu\nu} = \partial_\mu \xi_\nu + \partial_\nu \xi_\mu
\end{equationsplit}

When this transformation is plugged into the lagrangian eq\eqref{eq:lagrangianTMG}, terms with the gauge parameter remain. However, these terms are total-derivatives, making the entire theory gauge invariant upto total derivatives. Total derivative terms do not have any effect on the dynamics of the field. As was described in \fullref{subsec:TMEpeculiar}, these total derivative display the connection between CS terms and the topology of the manifold.
\subsection{Equations of Motion}\label{subsec:TMGeom}

The eom arising from the TMG lagrangian for the field $\phi_{\mu\nu}$ is:
\begin{equationsplit}
	0&=\Box (\phi_{\mu\nu} - \eta_{\mu\nu} \phi') - 2 \partial_{(\mu}\partial^\sigma\phi_{\nu)\sigma} + \partial_\mu \partial_\nu \phi' + \eta_{\mu\nu} \partial \cdot \partial \cdot \phi  
	\\
	&+ \frac{1}{2 \mu_g}    \big(\epsilon{_\mu}^{\sigma \alpha}\partial_\sigma (\Box \phi_{\alpha\nu} - \partial_\nu \partial^\beta \phi_{\alpha \beta} )  + (\mu \leftrightarrow \nu)\big)
\end{equationsplit}

To really see the nature of the excitation in this eom, it needs to be distilled to get rid of gauge dofs. To this end, a Lorenz gauge is called upon for service. Using this gauge choice, the above equation simplifies to:
\begin{equationsplit}
	0&=\Box (\phi_{\mu\nu} - \eta_{\mu\nu} \phi') + \partial_\mu \partial_\nu \phi' + \frac{1}{2 \mu_g}    \big(\epsilon{_\mu}^{\sigma \alpha}\partial_\sigma \Box \phi_{\alpha\nu}  + (\mu \leftrightarrow \nu)\big)
	\\
	&\text{Lorenz gauge for }\phi_\munu \quad \quad \del^\mu \phi_\munu =0
\end{equationsplit}
The Lorenz gauge choice will be satisfied if the gauge parameter $\xi_\mu$ satisfies:
\begin{equation}\label{eq:partialgaugelorenz}
	\Box \xi_\mu + \del_\mu \del\cdot\xi = 0
\end{equation}
Now taking a trace of the eom gives $\Box \phi' = 0 $, plugging this back into the eom and then taking a divergence implies $ \del_\mu \phi' = 0$, the eom can brought to the following form:
\begin{equationsplit}
	 \mu_g \phi_\munu  + \partial_\sigma \epsilon{_\mu}^{\sigma \alpha}  \phi_{\alpha\nu}  &= 0
	 \\
	 \text{rearranged to:} \quad \quad &
	 \\
	 \big(\del^\sigma \epsilon_{\sigma\mu\nu} + \mu_{\phi}\ \eta_{\mu\nu} \big) \phi{^\mu_\beta} &=0
	 \\
	 K_\munu \phi{^\mu_\beta} &=0
\end{equationsplit}
This structure is quite reminiscent of eq\eqref{eq:Kmunuoperatorspin1}. As was done in the case for spin-1 fields, consider:
\begin{equationsplit}
	K^{\mu\alpha} K_\munu \phi{^\nu_\beta} &= 0
	\\
	\implies \quad \big( \Box + \mu^2_g\big) \phi{^\alpha_\beta} &= 0
\end{equationsplit}

Thus, this equation confirms that the excitation of the field $\phi_\munu$ is massive with a mass given by $\mu_g$. The dof count goes as follows: The above equation is for a symmetric tensor field; thus these are 6 separate equations. The Lorenz gauge is used to eliminate 3 components. The trace of the field is not dynamical, hence it removes 1 more component. Finally, the Lorenz gauge has only fixed the gauge parameter partially (upto eq\eqref{eq:partialgaugelorenz}). The residual gauge-freedom can be used to remove one more component. This means that there is only one dof left.

The eoms confirmed that TMG has excited one single massive excitation. This single dof has helicity $\frac{|\mu_g|}{\mu_g}$. There is also a much more cleaner way of confirming the massive nature of this excitation. For this, the propagator is calculated.
\subsection{Propagator}\label{subsec:TMGpropagator}

The TMG lagrangian is gauge invariant. Thus, as done in the case of massless theory for spin-2 fields a suitable gauge-fixing term is required to fix the degeneracy in the kinetic operator and then obtain the propagator. In this case however, a more general approach is taken. The propagator is calculated in the \emph{$R_\xi$-gauge}. To this end, the gauge-fixing terms added to the TMG lagrangian is inspired by the de-donder gauge from eq\eqref{eq:dedondergauge}. The gauge fixing term, with an arbitrary gauge-keeping parameter $\xi$ is:
\begin{equation}\label{eq:lagrangiangaugefixingterm2}
\mathcal{L}_{\text{gauge term}} = \frac{-2}{\xi} (\partial^\mu \phi_{\mu\nu}  - \frac{1}{2} \partial_\nu\phi')^2
\end{equation}
Calculation of the propagator with such a term is certainly more involving. The benefit of keeping the gauge-parameter explicit, however, outweighs the troubles. With an arbitrary parameter different gauges can be selected, their effects studied, and it also highlights modes which have a contribution coming from the gauge dofs. The propagator in momentum space is given by:
\begin{equationsplit}\label{eq:propagatorTMG}
		&G^{\mu\nu\alpha\beta} = \frac{Y_1^{\mu\nu\alpha\beta}}{k^2} + \frac{Y_2^{\mu\nu\alpha\beta}}{k^4} +\frac{Y_3^{\mu\nu\alpha\beta}}{k^6}+ \frac{Y_4^{\mu\nu\alpha\beta}}{k^2 - \mu_g^2}
\end{equationsplit}
The tensors $Y_i^{\mu\nu\alpha\beta}$ are:
\begin{equationsplit}
		&Y_1^{\mu \nu \alpha \beta} = X_1^{\mu \nu \alpha \beta} + (-1 -  \frac{\xi}{8 - 11 \xi + 4 \xi^2}) X_2^{\mu \nu \alpha \beta} -  \frac{X_3^{\mu \nu \alpha \beta}}{\mu_g^2} + \frac{X_4^{\mu \nu \alpha \beta}}{\mu_g^2} + \frac{X_5^{\mu \nu \alpha \beta}}{\mu_g^4} -  \frac{i X_6^{\mu \nu \alpha \beta}}{2 \mu_g} + \frac{i X_7^{\mu \nu \alpha \beta}}{2 \mu_g^3}
		\\[0.5em]
		&Y_2^{\mu \nu \alpha \beta} =- \frac{2 (-1 + \xi) X_3^{\mu \nu \alpha \beta}}{-2 + \xi} + \frac{8 (-1 + \xi)^2 X_4^{\mu \nu \alpha \beta}}{8 - 11 \xi + 4 \xi^2} + \frac{X_5^{\mu \nu \alpha \beta}}{\mu_g^2} + \frac{i X_7^{\mu \nu \alpha \beta}}{2 \mu_g}
\end{equationsplit}
\begin{equationsplit}
		&Y_3^{\mu \nu \alpha \beta} =\frac{16 (-1 + \xi)^3 X_5^{\mu \nu \alpha \beta}}{(-2 + \xi) (8 - 11 \xi + 4 \xi^2)}
		\\[0.5em]
		&Y_4^{\mu \nu \alpha \beta} =- X_1^{\mu \nu \alpha \beta} + X_2^{\mu \nu \alpha \beta} + \frac{X_3^{\mu \nu \alpha \beta}}{\mu_g^2} -  \frac{X_4^{\mu \nu \alpha \beta}}{\mu_g^2} -  \frac{X_5^{\mu \nu \alpha \beta}}{\mu_g^4} + \frac{i X_6^{\mu \nu \alpha \beta}}{2 \mu_g} -  \frac{i X_7^{\mu \nu \alpha \beta}}{2 \mu_g^3}		
\end{equationsplit}
Here, the propagator is certainly more complicated than what has been seen before. First of all, the propagator consists of the sum of four poles. It is clear from the very structure of the theory, that not all of these poles correspond to a physically propagating dof. The numerators of these poles have been expressed in terms of tensors  $X_i^{\mu \nu \alpha \beta}$. These tensors are composed of the metric $\eta_{\mu\nu}$, momenta $k_\mu$ and the epsilon symbol $\epsilon_{\alpha\nu\mu}$ only. The complete forms of these 7 tensor is spelled out in \hyperref[ch:appendix1]{Appendix A}. 

One of the quickest ways to check whether this propagator is correct is to study it in its massless limit. From the lagrangian in eq\eqref{eq:lagrangianTMG}, it is seen that on sending $\mu_g\rightarrow\infty$, the lagrangian the theory of massless spin-2 fields. Hence, it is expected that this behaviour should also be reflected by this propagator. Indeed, setting $\xi=1$ and taking this limit in a careful manner leads to a propagator which is exactly the same as the propagator in eq\eqref{eq:propagatorTMG}.

The pole corresponding to tensor $Y_3^{\mu \nu \alpha \beta}$ is clearly completely unphysical, since setting $\xi=1$ gets rid of it completely. The rest of the propagator for this gauge choice becomes:
\begin{equationsplit}\label{eq:propagatorTMGxi=1}
	\text{for: }\xi&=1
	\\
	G^{\mu\nu\alpha\beta} &= \frac{1}{k^2 - \mu_g^2}
	\begin{pmatrix*}[l]
	- \frac{\mu_g^2}{k^2}X_1^{\mu\nu\alpha\beta} + (-1 + \frac{2 \mu_g^2}{k^2}) X_2^{\mu\nu\alpha\beta} 
	\\[0.5em]
	+ \frac{1}{k^2} (X_3^{\mu\nu\alpha\beta} -  X_4^{\mu\nu\alpha\beta} -  \frac{1}{k^2} X_5^{\mu\nu\alpha\beta})  
	\\[0.8em]
	+ \frac{i \mu_g}{2 k^2} (X_6^{\mu\nu\alpha\beta} -\frac{1}{k^2}X_7^{\mu\nu\alpha\beta})
	\end{pmatrix*}
\end{equationsplit}

This shows, that the excitation is massive. Also, contracting this object with a momentum vector such as $k_\mu$ automatically gives zero. This confirms that the massive excitation in this theory has transverse polarization. Finally, a \stuckelberg analysis of this theory does not seem to lead to any new insights. This can be seen by looking at all the possible field redefinitions which are given in eq\eqref{eq:stuckelbergfieldredefspin2} and reproduced here:
\begin{equation}
\phi_{\mu\nu} =  \psi_\munu + \partial_\mu \pi_\nu + \partial_\nu \pi_\mu + \eta_\munu \chi' + 2 \del_\mu \del_\nu \chi
\end{equation}
When these additional fields will be put in the TMG lagrangian in eq\eqref{eq:lagrangianTMG}, it is clear that none of the additional fields except $\psi_\munu$ can survive for the parity-odd part. This is because of the presence of the antisymmetric $\epsilon$ symbol, which will kill all the other fields. Additionally, since this theory has no FP mass term, none of the other fields should be excited in accordance with the arguments presented in \fullref{subsec:stuckelbergspin2}. 

In conclusion, the TMG theory propagates one massive dof where none was present in the massless case. This massive dof is, truly, a new dof, as it cannot be accounted for by either group-theoretic arguments or a \stuckelberg analysis. This dof finds its origin in global effects of the manifold \cite{qgCarlipbook:1998uc}. 

\section{Fierz-Pauli-Chern-Simons Gravity}\label{sec:massivespin2FPCS}
An interesting possibility, which occurs only in 2+1 dimensions, is to consider a theory of linear spin-2 fields where the gravitons are massive due to two different mass-generating mechanisms. This is analogous to \ref{sec:massivespin1PCS}, where the spin-1 fields were given a mass from both Proca theory and a CS mass term. In this section, such a possibility will be considered for gravity. 

All ingredients for setting up this theory are already available. As usual, the Fronsdal formulation gives the kinetic term for the spin-2 fields. Both FP mass terms and the CS mass terms will be present. The lagrangian describing a linear theory of massive gravitons or spin-2 fields with both Fierz-Pauli and Chern-Simons(FPCS) mass terms is:
\begin{equationsplit}\label{eq:lagrangianFPCS}
	&\mathcal{L}_{\text{FPCS}} = \mathcal{L}_{\text{kinetic}} + \mathcal{L}_{\text{FP}} + \frac{1}{\mu_g}\ \mathcal{L}_{\text{cs spin-2}}
	\\
	\text{where: } \quad& \quad 
	\\
	&\mathcal{L}_{\text{kinetic}} = \sigma_2 \ \phi^\munu \overline{F}_\munu
	\\
	&\mathcal{L}_{\text{cs spin-2}} = \sigma_2\ \phi^{\munu}\ \partial^\alpha \epsilon\indices{_\alpha_\beta_{(\mu}}\overline{F}_{\nu)}{^\beta}
	\\
	&\mathcal{L}_{\text{FP}} = \sigma_2 m^2 \big(\phi^\munu\phi_\munu - \phi'\phi'\big)	
\end{equationsplit}
Together, the lagrangian becomes:
\begin{equationsplit}
	\frac{\mathcal{L}_{\text{FPCS}}}{\sigma_2} &= \phi^{\mu\nu} (\Box + m^2) \phi_{\mu\nu}  -  \phi' (\Box + m^2) \phi'
	\\& - \phi^{\mu\nu} \Big(\partial_\mu \partial^\sigma  \phi_{\nu\sigma} +  \partial_\nu \partial^\sigma  \phi_{\mu\sigma}\Big)
	\\
	& + \phi^{\mu\nu} \partial_\mu \partial_\nu \phi' + \phi' \partial \cdot \partial \cdot \phi
	\\
	& +\frac{1}{2 \mu_g} \phi^{\mu\nu} \big(\epsilon{_\mu}^{\alpha\beta} \partial_\alpha (\Box \phi_{\beta\nu} - \partial_\nu \partial^\sigma \phi_{\beta \sigma} )  + (\mu \leftrightarrow \nu) \big)
\end{equationsplit}

Since there is the FP mass term, this lagrangian is not gauge-invariant. Thus, any complication arising from gauge modes is absent from this theory. This theory will now be studied and insights into the number of dofs, nature of excitations and their propagation will be looked into. Finally, a \stuckelberg analysis will reveal interesting insights into how the two mass-generating mechanism talk to each other, if at all.
\subsection{Equations of Motion}
To further pursue this lagrangian it is imperative to write it in a neat bilinear form with the kinetic operator $D_{\mu\nu\alpha\beta}$ sandwiched between the fields. This gives:
\begin{equationsplit}\label{eq:doperatorspin2FPCS}
	\frac{\mathcal{L}_{\text{FPCS}}}{\sigma_2} &= \phi^{\mu\nu}	D_{\mu\nu\alpha\beta} \phi^{\alpha\beta}
	\\
	\text{where: } \quad \quad &
	\\
	D_{\mu\nu\alpha\beta} &= \sigma_2 
	\begin{pmatrix*}[l]
	(\Box + m^2) \Big( \frac{1}{2}(\eta_{\mu\alpha}  \eta_{\nu\beta}  + \eta_{\mu\beta}\eta_{\nu\alpha})  - \eta_{\mu\nu} \eta_{\alpha\beta} \Big) 
	\\[0.5em]
	- \del_\mu \del_{(\alpha}\eta_{\beta)\nu}  - \del_\nu \del_{(\alpha}\eta_{\beta)\mu}
	\\[0.5em]
	+ \del_\mu \del_\nu \eta_{\alpha\beta} + \del_\alpha \del_\beta \eta_{\mu\nu}
	\\[0.5em]
	+\frac{1}{2 \mu_g} \epsilon_{\mu\sigma(\alpha} \del^\sigma  \big( \Box \eta_{\beta)\nu} - \del_{\beta)}\del_\nu \big)
	\\[0.5em]
	+\frac{1}{2 \mu_g} \epsilon_{\nu\sigma(\alpha} \del^\sigma  \big( \Box \eta_{\beta)\mu} - \del_{\beta)}\del_\mu \big)
	\end{pmatrix*}
\end{equationsplit}
The eoms are given by:
\begin{equationsplit}
	0&=D_{\mu\nu\alpha\beta} \phi^{\alpha\beta}
	\\
	\text{which gives:}&
	\\
	0&=	(\Box + m^2) \big(\phi_\munu - \eta_\munu \phi'\big) 
	\\
	&- \del_\mu \del\cdot\phi_\nu 	- \del_\nu \del\cdot\phi_\mu +\del_\mu\del_\nu\phi' + \eta_\munu \del\cdot\del\cdot\phi
	\\
	&+\frac{1}{\mu_g} \epsilon_{\sigma\alpha(\mu} \del^\sigma  \big( \Box \phi{{^\alpha}_{\nu)}} - \del_{\nu)} \del\cdot\phi^\alpha \big)
\end{equationsplit}

Inarguably, these eoms are not very transparent. Nevertheless some standard tricks can allow a dof count. To this end, the divergence and trace of the above equation yields:
\begin{equationsplit}
	\text{Divergence:}  &\implies \del\cdot\phi_\mu - \del_\mu \phi' = 0
	\\
	\text{Trace:} &\implies \phi' = 0 
\end{equationsplit}
Putting together the eom becomes:
\begin{equationsplit}\label{eq:eomFPCS}
	(\Box + m^2) \phi_{\mu\nu} +\frac{1}{\mu_g} \epsilon_{\sigma\alpha(\mu} \del^\sigma  \Box \phi{{^\alpha}_{\nu)}} &= 0
	\\
	\partial^\mu \phi_{\mu\nu} &=0
	\\
	\phi' &=0
\end{equationsplit}

The dof count is as follows. The first equation is for 6 symmetric components in the symmetric field $\phi_\munu$. The next equation is used to eliminate 3 components and the last kills one component. However, there is one extra dof due to the presence of third-order derivatives. Thus, together these equations have 3 dof. This is inline with what can be expected for a theory with 2 massive dofs coming from FP mass term and one coming from the CS mass term. To realize the nature of the propagating masses, constructing the propagator is necessary. 

\subsection{Propagator}

Since the kinetic operator $D_{\mu\nu\alpha\beta}$ is non-gauge invariant. It can be inverted without any gauge-fixing by going to momentum space. When inverting this operator one encounters a polynomial which needs to be factorized. This factorization is necessary from two points of view, which are not independent of each other. (a) This will allow an understanding of how exactly are the two mass mechanisms `communicating' with each other. (b) To express the propagator as sum of poles, which is necessary to see the propagating modes clearly. Without further ado, the mass-mixing polynomial is\footnote{When the author was doing this calculation, he was unaware that such a calculation for this theory was already done. The calculations in the literature have slightly differing routes (spin-projection operators in \cite{massmixingPinheiro1996}, and ADM-decomposition in \cite{massmixingdeser}). Thus, the calculations by the author provides an independent check for the calculations present in the literature.}:
\begin{equation}\label{eq:massmixingspin2}
k^6 - \mu_g^2 (k^2 - m^2)^2 = 0
\end{equation}

Notice that this polynomial is of sixth order in $k$ and cubic order in $k^2$. Comparing with the similar scenario for spin-1 fields, (eq\eqref{eq:massmixingspin1poly} in PCS theory) where the mass-mixing polynomial was quadratic in $k^2$, it is clear that the two scenarios are similar. Moreover, a cubic polynomial will necessarily imply three roots, which is exactly the propagating dofs in this theory, as per the counting from the last section. In PCS theory, with spin-1 fields the two roots of the mass-mixing polynomial were immediately identified as the masses of the propagating modes; and the same will be pursued here. 

To find the roots of this polynomial, two auxiliary functions $A$ and $B$ are defined. These will allow the roots of the polynomial to be expressed conveniently. These functions are:
\begin{equationsplit}
	A &= - \mu_g^2 (\mu_g^2 - 6 m^2)
	\\
	B &= \biggl(2 \mu_g^6 - 18 \mu_g^4 m^2 + 27 \mu_g^2 m^4 + 3 \sqrt{3} \sqrt{-4 \mu_g^6 m^6 + 27 \mu_g^4 m^8 }   \bigg)^{1/3}
\end{equationsplit}
In terms of these functions $A$ and $B$, the roots of the mass-mixing polynomial are:
\begin{equationsplit}\label{eq:massfunctionsf1f2f3spin2}
	&f_1  =  \tfrac{1}{3} \mu_g^2- \frac{2^{1/3}}{3} \frac{A}{B} + \frac{B}{3 (2^{1/3})} 
	\\
	&f_2  =  \tfrac{1}{3} \mu_g^2 + \frac{(1 + i \sqrt{3})}{3 (2^{2/3})} \frac{A}{B} -  \frac{(1 - i \sqrt{3})}{6 (2^{1/3})} B
	\\
	&f_3  =  \tfrac{1}{3} \mu_g^2 + \frac{(1 - i \sqrt{3})}{3 (2^{2/3})} \frac{A}{B}  -  \frac{(1 + i \sqrt{3}) }{6 (2^{1/3})} B
	\\
	\text{asserting:} &
	\\
	&k^6 - m_1^2 (k^2 - m_2^2)^2 = (k^2 - f_1)(k^2 - f_2)(k^2 - f_3)
\end{equationsplit}

The roots of the mass-mixing polynomial $f_1$,$f_2$, and $f_3$, by their very appearance suggest that this is not a simple mass-mixing. The square-root in $B$ enforces: $27 m^2 > 4 \mu{_g^2}$ to avoid complex roots. This theory is not a very healthy theory, since for a some parameters ghosts are present. So even though massive graviton modes are present, this theory cannot be considered physical \cite{fpcsghostsAccioly:2004qw}. 

In order to gain a notion on this mass-mixing, consider the extreme limits:
\begin{itemize}
	\item Taking the $\mu_g\rightarrow\infty$ limit corresponds to turning off the CS-mass term. This should revert the theory to the FP theory for massive spin-2 fields. Looking at the polynomial in eq\eqref{eq:massmixingspin2}, it is observed that the polynomial drastically changes to give $k^2 - m^2=0$. This is expected on the grounds of the theory turning into the FP theory. However, this limit is not well-defined for the mixed masses $f_1$,$f_2$, and $f_3$. They all diverge in this limit. This suggests that the mode containing contributions from the CS-term is mixed into all the mass functions $f_i$. 
	
	\item On the other hand, taking the limit $m\rightarrow0$, reduces the mass functions $f_1\rightarrow\mu_g^2$, and both $f_2=f_3=0$. This limit corresponds to removing the FP term from the lagrangian, reducing it to TMG. This directly suggests that two modes are switched off by switching off the FP-mass term. 
\end{itemize}
These vague notions on the distribution of masses in the dofs will be cleared up shortly. 

The propagator for this theory is: 
	\begin{equation}\label{eq:propagatorCS}
	G^{\mu\nu\alpha\beta} = \frac{Y_1^{\mu\nu\alpha\beta}}{k^2 - f_1} + \frac{Y_2^{\mu\nu\alpha\beta}}{k^2- f_2} + \frac{Y_3^{\mu\nu\alpha\beta}}{k^2 - f_3}
	\end{equation}
This clearly shows that there are 3 dofs in the theory. Since, the tensor $Y_i^{\mu\nu\alpha\beta}$ have very lengthy opaque expressions in terms of tensors $X_i^{\mu\nu\alpha\beta}$, they have been relegated to the \hyperref[ch:appendix1]{Appendix A}. These expressions even though of considerable length as they are, can be verified. A consistency check is provided by considering the propagator in the limits discussed above. 

Notice, that the lagrangian for FPCS theory break gauge symmetry due to the presence of the FP mass term. On removing the CS-term one gauge non-invariant theory turns into another non-gauge invariant theory. Since this could in principle happen without interfering with gauge-modes, this limit should work properly. Indeed this is reflected in the propagator as well. Taking the $\mu_g\rightarrow\infty$ limit, has one problem: the mass-functions $f_1$,$f_2$, and $f_3$ are all divergent in this limit. One way to side-step this issue is to first express all the $f_i$'s in terms of the original masses $m$ and $\mu_g$. Looking at the definition of the mass-functions $f_i$'s it is clear that such a step will give an extremely long-tedious expression. Putting $\mu_g\rightarrow\infty$ in this expression, however, reduces the entire propagator to the propagator from the FP theory in eq\eqref{eq:propagatorspin2fierzpauli}. 

Consider the opposite limit: Removing the FP-mass term from the FPCS lagrangian should reduce this theory to TMG. However, TMG enjoys gauge symmetry. Therefore, taking $m\rightarrow0$ or $f_1\rightarrow\mu_g^2$, along with $f_2=f_3=0$ should create problems, as one gauge non-invariant theory is turning into a gauge-invariant theory. Recall that theories with gauge symmetries, have redundant gauge dofs mixed with the physical dofs. Since the FP mass generally excites many more fields, than present in the massless theory, this limits should not work out smoothly. For 2+1 dimensions, FP theory had 2 dofs, while TMG had only 1 dof. Thus, putting $m=0$, when the dofs are still mixed should cause trouble. This is reflected in the propagator as well. Many of the terms in this case diverge. This also suggests, that decoupling these mixed dofs can cure this problem. Indeed, this will be seen to be true in the next section on \stuckelberg analysis.

\subsection{Insights from St\"uckelberg Analysis - FPCS Theory}

As has been remarked before, the mass-mixing in FPCS theory is complicated. The mass-functions $f_i$'s have complex expressions and are not always viable for a physically relevant theory. The aim of the analysis of this section is to resolve the uncertainties regarding the mass-mixing, resolve the massless limit of the propagators, and hopefully provide some clues for the cubic nature of the polynomial encountered in eq\eqref{eq:massmixingspin2}. 

Much of analysis in this section is closely related to the \stuckelberg analysis of the FP theory presented in \fullref{subsec:stuckelbergspin2}. Since, most of the important ideas were already presented in that section, here only the relevant details are discussed.

To begin with, the field re-definitions are (same as in eq\eqref{eq:stuckelbergfieldredefspin2}, reproduced here for convenience)
\begin{equationsplit}
	\phi_{\mu\nu} & = \Psi_{\mu\nu} + \partial_\mu \Pi_\nu + \partial_\nu \Pi_\mu
	\\
	\Pi_\mu & = \pi_\mu + \partial_\mu \chi
	\\
	\Psi_{\mu\nu} & = \psi_{\mu\nu} + \chi' \eta_{\mu\nu}
	\\
	\text{together:}&
	\\
	\phi_{\mu\nu} &=  \psi_\munu + \partial_\mu \pi_\nu + \partial_\nu \pi_\mu + \eta_\munu \chi' + 2 \del_\mu \del_\nu \chi	
\end{equationsplit}
Plugging these redefinitions, into the FPCS lagrangian in eq\eqref{eq:lagrangianFPCS} results in a lagrangian governing the dynamics of each of the additionally introduced fields. To disallow the scalar field $\chi'$ from being dynamically coupled to the trace of the spin-2 mode $\psi_\munu$, the scalar field is set as:
\begin{equation}
\chi' = \frac{-2 m^2}{d-2} \chi
\end{equation}
Now, the canonical vector field $\pi_\mu$ and the canonical scalar field are rescaled:
\begin{equationsplit}
	\pi_\mu &\to \frac{1}{m} \pi_\mu
	\\
	\chi &\to \frac{1}{m^2} \chi
\end{equationsplit}

Finally, the \stuckelberg lagrangian for FPCS theory becomes:
\begin{equationsplit}\label{eq:lagrangianstuckelbergspin2FPCS}
	&\mathcal{L}_{Stuck} = \mathcal{L}_{\psi_{\mu\nu}} + \mathcal{L}_{\pi_\mu} + \mathcal{L}_\chi + \mathcal{L}_{mix}
	\\
	\text{where: } \quad  &
	\\
	&\mathcal{L}_{\psi_{\mu\nu}}  = \mathcal{L}_{\text{kinetic}\ \psi_{\mu\nu}} + \mathcal{L}_{\text{FP}\ \psi_{\mu\nu}} + \frac{1}{\mu_g}\ \mathcal{L}_{\text{cs spin-2}}
	\\
	&\mathcal{L}_{\pi_\mu}  = - \frac{1}{2} F_{\mu\nu} F^{\mu\nu} \quad\quad \Big(\text{with:} F_\munu = \del_\mu \pi_\nu - \del_\nu \pi_\mu\Big)
	\\
	&\mathcal{L}_{\chi}  = -2 \  \frac{d-1}{d-2} \  \chi \Bigg( \Box  - m^2 \frac{d}{d-2}\Bigg) \chi
	\\
	&\mathcal{L}_{mix}  =  -2  m \Bigg( \psi^{\mu\nu} \partial_{(\mu}  \pi_{\nu)}  - \psi'\  \partial \cdot \pi 
	+ \  (\frac{d-1}{d-2}) \  ( 2 \chi \partial \cdot \pi + m \psi' \chi )			 \Bigg)
\end{equationsplit} 
The only difference between the \stuckelberg lagrangian of this section with that of \fullref{subsec:stuckelbergFP}, is the presence of the CS term for the field $\psi_\munu$. As was noted in the section on TMG \ref{subsec:TMGpropagator}, the CS term is not affected by these field re-definitions. Additionally this lagrangian also enjoys the same gauge symmetries as given in eq\eqref{eq:stuckelbergspin2gaugetransformations1} and eq\eqref{eq:stuckelbergspin2gaugetransformations2}. 

Get rid of the mixed terms in $\mathcal{L}_\text{mix}$, is possible thanks to the gauge symmetry mentioned above. Since this lagrangian along the mixed terms is exactly the same as was obtained before; the same steps are re-done. Using a de-donder like gauge, a suitable gauge fixing term is added to the lagrangian. These terms are given in eq\eqref{eq:gaugefixingterms}. Adding these terms to the above lagrangian leads to a completely diagonalized lagrangian. This lagrangian is:
\begin{equationsplit}\label{eq:lagrangianstuckelbergspin2FPCSgaugefixed}
	&\mathcal{L}_{\text{gauge-fixed}}  = \mathcal{L}_{\psi_{\mu\nu}} + \mathcal{L}_{\pi_\mu} + \mathcal{L}_\chi
	\\
	\text{where: } \quad &
	\\
	&\mathcal{L}_{\psi_{\mu\nu}} =-\frac{1}{2}\psi^{\mu\nu} (\Box + m^2) \psi_{\mu\nu} 
	+  \frac{1}{4} \psi' (\Box + m^2) \psi'
	\\
	& - \frac{1}{4 m_1} \psi^{\mu\nu} \big(\epsilon{_\mu}^{\alpha\beta} \partial_\alpha (\Box \psi_{\beta\nu} - \partial_\nu \partial^\sigma \psi_{\beta \sigma} )  + (\mu \leftrightarrow \nu)\big)
	\\
	&\mathcal{L}_{\pi_{\mu}} = \pi_\mu \  (\Box + m^2) \  \pi^\mu
	\\
	&\mathcal{L}_{\chi} = -2 \  \frac{d-1}{d-2} \ \chi \  (\Box + m^2) \ \chi
\end{equationsplit}

Note that the presence of the CS mass term in the $\psi_\munu$ field. The lagrangian for other fields is exactly the same as was obtained before. The propagators for these fields $\pi_\mu$ and $\chi$ are given in eq\eqref{eq:propagatorstuckelbergspin2FP}. Also, in that section a thorough dof count for this lagrangian without the CS term was presented. 

Now, on comparing these two lagrangians, it becomes clear that the mass-mixing in this theory is truly different. This is because each mass-generating mechanism is exciting a different mode. For 2+1 dimensions, the FP mass term excites a massive vector the $\pi_{\mu}$ field and a massive $\chi$ field. Contrarily, the CS term does not excite any of the auxiliary fields, and only gives a mass to the $\psi_\munu$ field. The $\psi_\munu$ field would be an unphysical non-propagating field without this CS term. This explains why the three mass functions $f_i$ were diverging when $\mu_g\rightarrow\infty$ limit was considered. Such a limit, would require the $\psi_\munu$ field to lose dofs. However, due to the mixing this limit could not be reached. 

In essence, it is now clear that the presence of two different mass-generating mechanisms has excited three different fields. All three have received a mass in this theory, and participate in the mass-mixing polynomial of eq\eqref{eq:massmixingspin2}. This also explains why the mass-mixing polynomial of the FPCS theory was of cubic order. In the spin-1 case, the proca mass excited a new longitudinal scalar which was mixed with another dof present in the theory already. In PCS theory, the CS term gave mass to the existing mode of the theory, and hence there were always two dofs to begin with. In the case of spin-2 fields, however, it is seen that the addition of CS term to the FP theory gives mass to modes which were absent from the theory. 

All of these comments suggest that the \stuckelberg lagrangian governing the dynamics of the decoupled mode should be continuous with both the respective theories. This is the case. The propagator for the fields $\pi_\mu$ and $\chi$ (see eq\eqref{eq:propagatorstuckelbergspin2FP})have well-defined massless limits. These modes decouple from the theory on setting $m\rightarrow0$, as can be seen from the eq\eqref{eq:lagrangianstuckroleofmass}. The propagator for the field $\psi_\munu$ will now be calculated. 

Although the different modes in the $\phi_\munu$ field have been decoupled, a non-trivial interesting mass-mixing still exists. The $\psi_\munu$ field obtains its mass from both the FP and the CS mechanisms, as can be seen from its lagrangian. For inverting the kinematic operator sandwiched in this lagrangian, one again meets the same mass-mixing polynomial eq\eqref{eq:massmixingspin2}. That the same polynomial appears is hardly a surprise, because of the presence of the 3rd-order derivatives in the CS term. Recall that a theory of pure CS term is equivalent in physical content to  a massless theory. Nevertheless, this object has an inverse and requires no-gauge fixing. The propagator for the $\psi_\munu$ field is:

\begin{equation}
	G^{\mu\nu\alpha\beta}_{\psi_{\mu\nu}} = \frac{Z_1^{\mu\nu\alpha\beta}}{k^2 - f_1} + \frac{Z_2^{\mu\nu\alpha\beta}}{k^2- f_2} + \frac{Z_3^{\mu\nu\alpha\beta}}{k^2 - f_3} + \frac{Z_4^{\mu\nu\alpha\beta}}{k^2 - \frac{2 f_1 f_2 f_3}{f_1 f_2 + f_2 f_3 + f_1 f_3 }}
\end{equation}

Alas, the expressions for the tensor $Z_i^{\mu\nu\alpha\beta}$ is too lengthy to put here. It can be found in \hyperref[ch:appendix1]{Appendix A}. The mass functions $f_i$ here are necessarily the same as those in eq\eqref{eq:massfunctionsf1f2f3spin2}, since they come from the same polynomial. The interesting thing about the above propagator is the addition of a new pole! It is this pole and its funny mass, which will allow this propagator to have well defined limits. The expression for this propagator, under the two extreme limits becomes:

\begin{equationsplit}
\lim_{\mu_g\to \infty}	G^{\mu\nu\alpha\beta}_{\psi} &= \frac{1}{k^2 - m^2} \Bigg( (\eta^{\mu\alpha}  \eta^{\nu\beta}  + \eta^{\mu\beta}\eta^{\nu\alpha})  - \frac{2}{d - 2}  \eta^{\mu\nu} \eta^{\alpha\beta}  \Bigg)
\\
\lim_{m\to0}			G^{\mu\nu\alpha\beta} &= \frac{1}{k^2 - \mu_g^2}
\begin{pmatrix*}[l]
	- \frac{\mu_g^2}{k^2}X_1^{\mu\nu\alpha\beta} + (-1 + \frac{2 \mu_g^2}{k^2}) X_2^{\mu\nu\alpha\beta} 
	\\[0.5em]
	+ \frac{1}{k^2} (X_3^{\mu\nu\alpha\beta} -  X_4^{\mu\nu\alpha\beta} -  \frac{1}{k^2} X_5^{\mu\nu\alpha\beta})  
	\\[0.8em]
	+ \frac{i \mu_g}{2 k^2} (X_6^{\mu\nu\alpha\beta} -\frac{1}{k^2}X_7^{\mu\nu\alpha\beta})
\end{pmatrix*}
\end{equationsplit} 

The first propagator is the same propagator as in eq\eqref{eq:propagatorstuckelbergspin2FP} that was obtained for the spin-2 mode in the after St\"ckelberging the FP theory. As such, it comes with the nice property of giving 0 (since in 2+1 dimensions, this mode cannot carry any dof). The second propagator is the same as the gauge fixed propagator for the TMG theory with $\xi=1$, as in eq\eqref{eq:propagatorTMGxi=1}. Thus, the \stuckelberg analysis has successfully disentangled all mixed dof's and has truly made the theoretical objects continuous. All propagating dofs in this theory are now accounted for. With this, the analysis of the spin-2 fields obtaining masses from both FP and CS mechanisms is concluded.

To conclude this chapter: multiple theories of massless and massive spin-2 fields corresponding to linearized gravity were studied. All of the theories studied can be constructed from many differing routes. In this chapter, the focus was kept on using the formulation of HS fields. This benefit of using this formulation is that it allows a clear insight into the ingredients of each theory. Gauge invariant massless theory of spin-2 fields corresponding to linearized GR was shown to have no propagating dofs. The FP theory was introduced, and it was shown to excite two dofs in 2+1 dimensions. The issues with this theory, namely vDVZ discontinuity and Boulware-Deser ghost were discussed. Using \stuckelberg analysis, the nature of these new dofs and their propagation was detailed. TMG theory was shown to have one parity-violating transverse dof. Finally, a theory with both mass-mechanisms FPCS was constructed and carefully looked into. The presented analysis details how this mass-mixing occurs and through a \stuckelberg analysis propagators for each dof with well-behaving massless limits were calculated.
\chapter{Topologically Massive Bimetric Gravity}\label{ch:bimetric}
\epigraph{To the beginning student mountains are mountains and water is water. To the advanced student mountains stop being mountains and water stops being water. To the master mountains are mountains again and water is water again.}{Reference Manual, FORM\cite{formmanual}}
\section{Perspective: Towards Non-Linear Massive Gravity}
In this thesis, so far, multiple theories of massive and massless fields in 2+1 as well as 3+1 dimensions have been considered. The Boulware-Deser ghost instability was discussed in \fullref{subsec:boulwaredeserghost}. It was seen that the FP-tuning of the coefficient $\gamma\rightarrow0$ was necessary to ensure that no higher-derivative ghost dof propagated in that theory. Since this additional mode comes with a wrong sign in the propagator (negative kinetic energy), this signalled a fatal blow to the theory of massive gravity: (a) Classically, giving rise to unbounded Hamiltonians, and (b) Quantum Mechanically, violating unitarity. For the linearized description of massive gravity, hence, the FP-tuning was sufficient to ensure a healthy theory. 

Generally, it was considered that a non-linear theory of massive gravity is necessarily sick because this ghost dof always seemed to reappear. The conclusions from the analysis of Boulware and Deser were so strongly accepted, that this field saw no considerable progress for a prolonged period of about 30-40 years\cite{reviewMay}. The twisted and amusing history of the development of a consistent non-linear theory of massive gravity is readily available in the literature (see \cite{reviewhinterbichler,reviewdeRham,reviewMay} for a review). The dRGT theory of massive gravity proposed in 2010, was the first consistent theory of non-linear massive gravity which did not suffer from the Boulware-Deser ghost instability. In 2012, Hassan and Rosen generalized the dRGT theory to the ghost free Bimetric theory of gravitation \cite{bimetricHassan2013, bimetrichassanrosenmay}. It is this generalized theory and its topological-extension which forms the subject of the present chapter. 

The first section is devoted to formulate and review the bimetric theory of gravity for arbitrary $d$-dimensions. Along the sides, special cases of 3+1 and 2+1 dimensions for this theory will be pointed. The presentation of both 3+1 and 2+1 dimensions will serve to bring for a comparison of the differences arising in the bimetric theory. After setting up the action, the non-linear eoms and the gauge symmetry in this theory is presented. The absence of the problematic Boulware-Deser ghost is qualitatively discussed through an ADM decomposition, which allows a dof count for the non-linear theory to be completed. To investigate the nature of excitations in this theory, it is important to study its linearized version. This, on the other hand, can only be done if flat spacetime is an allowed solution for this theory. This is explicitly verified and leads to the discussion of the parameter space in this theory. Perturbations in the bimetric theory will allow the complete lagrangian to be diagonalized into two separate excitations. These excitations correspond to gravitons of this theory, and are studied at the end of this review. 

Going back to 2+1 dimensions, it is interesting to note that the Boulware-Deser ghost problem was completely absent from the theory of TMG. The theory of TMG, hence, appears as a natural, healthy formulation of massive gravity in 2+1 dimensions. It will be, therefore, interesting to see what happens when the bimetric theory is deformed with a CS mass term. This novel extension of the bimetric theory will be studied in detail in the next section. 

Finally, a theory of minimally coupled photons to gravitons will be considered. Studying their interaction, the quantum loop corrections to gravitons arising from photons running in loops, will be calculated. Having, studied both CS-massive photons in TME and Proca-massive photons in Proca theory in \fullref{ch:spin1}, it will be interesting to calculate the interactions arising from photons which obtain their masses from both of these mechanisms. With these loop calculations this chapter will be concluded.
\section{The Theory of Bimetric Gravity}
The theory of bimetric gravity presents itself as the most general non-linear description of massive spin-2 fields which is free of the Boulware-Deser ghost. The theory posits an additional independent metric field $f_\munu$. This theory is different from Massive gravity (of the dRGT type) since it promotes the secondary reference metric field $f_\munu$ into a dynamical tensor with its own curvature and coupling to matter. Apart from being coupled to all matter sources, the other metric $f_\munu$ is also coupled to the original metric $g_\munu$ in a very special way. The construction of the interaction between the two metrics is such that it leads to an additional constraint which removes the Boulware-Deser ghost. The removal of this ghost from the complete non-linear theory can be considered as the completion of the massive gravity program initiated by Fierz-Pauli in 1939. 

Being a non-linear theory of massive gravity, the notion of mass, for arbitrary backgrounds, calls for attention. It was seen in \fullref{sec:irrep}, that the notion of mass was deeply connected with the symmetries of the background spacetime. Arising as a Casimir invariant, the concept of mass can be generalized to arbitrary background with the same amount of symmetries as the Minkowski background, i.e. de Sitter and Anti-de Sitter backgrounds. For other backgrounds, the notion of mass is difficult to define. Generally, the classification of fields as being massive or massless can still be made by looking at the number of dofs these fields propagate. As was seen from the \stuckelberg analysis of \fullref{subsec:stuckelbergspin2}, massive fields usually propagate many more dofs than their massless counterpart. To be precise, a parameter of the theory can earn the label of mass if on setting it to zero, the theory becomes gauge invariant, thereby reducing the number of physical dofs to that of a massless field. It is in this sense, that the words massive gravity are used for the non-linear theory.
\subsection{Action}\label{subsec:bimetricaction}

The setup of the bimetric theory, as the name suggests, will consist of two independent metrics $g_\munu$ and $f_\munu$, with their own dynamics. There is, also, an interaction between these the two metrics which will be mediated via a potential term. Following the same definitions used by GR, quantities describing curvature can be defined for each metric. Although mentioned earlier, the following remarks are repeated:
\begin{itemize}
	\item Geometrical objects with a tilde on top like, $\tilde{A}$, means that this object has been defined with respect to the metric $f_{\mu \nu}$. Conversely, $A$, without tilde is defined with respect to the metric $g_{\mu\nu}$.

	\item In anticipation of perturbations, objects with a bar on top such as, $\bar{g}_{\mu\nu}$ or $\bar{\nabla}$, denote objects which are defined with respect to the background. 
\end{itemize} 

The action for Bi-metric theory in $d$-dimensions is given by:
\begin{equationsplit}\label{eq:lagrangianbimetric}
	&\mathcal{S}_{bi} = \int d{^d}x \mathcal{L}_{bi}
	\\
	&\mathcal{L}_{bi} = m_g^{d-2} \detsqrt{g} R_g + m_f^{d-2} \detsqrt{f} \tilde{R}_f + 2 m^d \detsqrt{g} V(\mathbb{X};\beta_n)
\end{equationsplit}
Here, $R_g$ and $\tilde{R}_f$ denotes the Ricci Scalar defined with respect to metric $g$ and metric $f$, respectively and $m_g$ and $m_f$ denotes the Planck masses. The parameter $m$ is a mass scale, but is not an independent parameter of the theory. It is convenient to define ratios of the Planck masses and the mass scale. These definitions are: 
\begin{equationsplit}
	M^2 &= \frac{m^d}{m_g^{d-2}} 
	\\
	\alpha &= \frac{m_f}{m_g}
\end{equationsplit}
The first two terms in the bimetric lagrangian are the same as the terms coming from the Einstein-Hilbert action for each metric. The interesting part about this theory comes from the third term. The potential $V(\mathbb{X};\beta_n)$ is defined through a curious combination of the two metrics interacting via a square-root matrix $\mathbb{X}$ as follows:
\begin{equationsplit}\label{eq:Xmatrix}
	\mathbb{X} &= \sqrt{g^{-1}f}
	\\
	\text{or}
	\\
	X^\mu_\tau X^\tau_\nu & = g^{\mu\alpha}f_{\alpha\nu}
\end{equationsplit}
The potential is finally expressed via the invariants of this square-root matrix $\mathbb{X}$ through the usage of elementary symmetric polynomials $e_n(\mathbb{X})$. 
\begin{equationsplit} \label{eq:potentialbimetric}
	V(\mathbb{X};\beta_n) &= \sum_{n=0}^{d} \beta_n e_n(\mathbb{X})
	\\
	&\stackrel{4d}{=} \beta_0 e_0(\mathbb{X}) + \beta_1 e_1(\mathbb{X})  + \beta_2  e_2(\mathbb{X}) + \beta_3 e_3(\mathbb{X}) + \beta_4 e_4(\mathbb{X})
	\\
	&\stackrel{3d}{=} \beta_0 e_0(\mathbb{X}) + \beta_1 e_1(\mathbb{X})  + \beta_2  e_2(\mathbb{X}) + \beta_3 e_3(\mathbb{X})
\end{equationsplit}
From the definition of the potential term, it is clear that in a $d$-dimensional setup, the theory demands $d+1$ coefficients $\beta_n$. Not all of these coefficients are independent. Mathematicians have several ways of defining the elementary symmetric polynomials; in this thesis a comprehensive recursive relation is used. The first polynomial, for any matrix, is equal to 1. The rest are defined as: 
\begin{equationsplit}\label{eq:elementarysymmetricpolynomial}
&e_0 = 1	
\\
\text{and } \forall n \geq 1 \quad &
\\
&e_n(\mathbb{X}) = \frac{1}{n} \sum_{k=1}^{n} (-1)^{k+1} Tr(\mathbb{X}^k) e_{n-k}(\mathbb{X})
\end{equationsplit}
For any $d\times d$ matrix $\mathbb{X}$, $e_n(\mathbb{X})$ = 0, $\forall n > d$ and $e_d(\mathbb{X}) = \det \mathbb{X}$. This implies that these polynomials are finite in number. Note: $Tr(A)$ denotes the trace of matrix $A$.

In order to gain an impression on these polynomials, their explicit expressions for 2+1 dimensions is given:
\begin{equationsplit}
	e_0(\Xmatrix) &=1
	\\
	e_1(\Xmatrix) &=Tr(\Xmatrix)
	\\
	e_2(\Xmatrix) &= \half \Big( Tr(\Xmatrix)^2 - Tr(\Xmatrix^2) \Big)
	\\
	e_3(\Xmatrix) &= \frac{1}{6} \Big( Tr(\Xmatrix)^3  - 3\ Tr(\Xmatrix^2)\ Tr(\Xmatrix)  + 2\ Tr(\Xmatrix^3)\Big)
	\\
	&= det (\Xmatrix)
\end{equationsplit}
Due to the symmetric properties of the elementary symmetric polynomials, there is a symmetry relation for the potential term. This is expressed as:
\begin{equation}
	\sqrt{|g|}\ V(\sqrt{g^{-1} f; \beta_n}) = \sqrt{|f|}\ V(\sqrt{f^{-1} g}; \beta_{d-n})
\end{equation}

This suggests that the bimetric-theory of gravity is symmetric under the exchange of the two metrics (along with $\beta_n \leftrightarrow \beta_{d-n}$). This an elegant feature of the bimetric theory: both the metrics are placed on equal footing. The dRGT theory did not have this feature, wherein the other metric was only used to supply an additional constraint ensuring the absence of Boulware-Deser ghost. In this manner, the bimetric theory is a truly general theory of non-linear massive gravity. 
\subsection{Non-linear EOMs}\label{subsec:bimetricnonlineareom}

To obtain the non-linear equations of motion for both metrics $g_\munu$ and $f_\munu$, the variation of the  action $\mathcal{S}_{bi}$ with respect to each metric is considered. The elementary symmetric polynomials have various identities which are used to calculate their variations. These identities are listed in \hyperref[ch:appendix2]{Appendix B}. Using these identities the following can be proven, inductively:
\begin{equationsplit}
	\text{with: }\delta e_0(\Xmatrix) &= 0
	\\
	\text{for } n>1
	\\
	\delta e_n(\mathbb{X}) &= \sum_{k=1}^{n}\ (-1)^{k+1}\ Tr(\mathbb{X}^{k-1}\ \delta\mathbb{X}) e_{n-k}(\mathbb{X})
\end{equationsplit}
Using identities of trace such as 
$$Tr(\delta (\Xmatrix^m))= m Tr(\Xmatrix^{m-1}\ \delta \Xmatrix)$$
along with the variation of $\Xmatrix$, the variation of the potential term can be systematically evaluated. With these calculations, the variation of all the terms in the lagrangian is evaluated to be:
\begin{equationsplit}
	\delta (\detsqrt{g}R_g) & = \detsqrt{g} \ G_\munu  \ \delta g^\munu 
	\\
	\delta (\detsqrt{f}\tilde{R}_f) &= \detsqrt{f} \  \tilde{G}_\munu \  \delta f^\munu 
	\\
	\delta (\detsqrt{g} V ) &= \detsqrt{g} \ V^g_\munu \  \delta g^\munu  
	\\
	& = \detsqrt{f} \ V^f_\munu \  \delta f^\munu  
\end{equationsplit}
Using these variations, the non-linear eoms can be put in a rather readable form. The equations of motion for bimetric gravity are:
\begin{equationsplit}\label{eq:eombimetric}
	\text{for metric} \ & g_\munu :  \quad \quad
	G_\munu +  M^2 V^g_\munu = 0
	\\
	\text{for metric} \ & f_\munu :  \quad \quad
	\tilde{G}_\munu\: + \frac{M^2}{\alpha^{d-2}}\: V^f_\munu =0
\end{equationsplit}
Here, $G_\munu$ represents the Einstein tensor for metric $g_\munu$ and correspondingly $\tilde{G}_\munu$ is the Einstein tensor for the metric $f_\munu$. The contribution from the interaction term in the lagrangian is encoded in the potential terms $V^g_\munu$ and $V^f_\munu$. These, in turn, are related to the square-root $\Xmatrix$ by the following relations:
\begin{equationsplit}\label{eq:bimetricpotentials}
	V^g_\munu &= \sum_{n=0}^{d} (-1)^n\; \beta_n\; g_{\mu\rho}\; Y^\rho_{(n) \nu}(\mathbb{X})
	\\
	V^f_\munu &= \sum_{n=0}^{d}\: (-1)^n\: \beta_{d-n}\: f_{\mu \rho}\; Y^\rho_{(n) \nu}(\mathbb{X}^{-1})
\end{equationsplit}
These potentials involve a matrix $Y$ whose definition in component notation is:
\begin{equation}
	Y^\rho_{(n) \nu}(\mathbb{X})  = \sum_{r=0}^{n}\: (-1)^r\: (\mathbb{X}^{n-r})^\rho_\nu\: e_r(\mathbb{X})
\end{equation}
For clarity, in the above equation $\rho$ and $\nu$ are indices denoting the components of matrix $Y$, and $n$ is a running label. Finally, due to an overall covariance of the interaction term, there is a relation between the divergences of the potentials $V^g_\munu$ and $V^f_\munu$. This identity is:
\begin{equation}\label{eq:divergencepotentials}
	\sqrt{|g|}\ g^{\mu\alpha} \nabla_\alpha V^g_\munu = - \sqrt{|f|}\ f^{\mu\beta} \tilde{\nabla}_\beta V^f_\munu 
\end{equation}

\subsection{Gauge Symmetry}

If the interaction term in the bimetric lagrangian eq\eqref{eq:lagrangianbimetric} was absent, then the two metrics would enjoy separate diffeomorphism invariance. The presence of interactions, breaks this gauge symmetry into a diagonal subgroup. What this means is that the entire bimetric theory enjoys general coordinate invariance under those gauge transformations that transform both the metrics simultaneously. Consider a gauge transformation for an arbitrary vectorial gauge parameter $\xi_\mu$:
\begin{equationsplit}
	x_\mu \rightarrow x'_\mu &= x_\mu + \delta x_\mu
	\\
	\delta x_\mu = \xi_\mu
\end{equationsplit}
With this transformation, the bimetric theory is invariant for the following transformations of the two metrics:
\begin{equationsplit}
	\text{for metric} \ & g_\munu : \quad \quad
	\delta g_\munu = -2 g_{\alpha(\mu)} \nabla_{\nu)} \xi^\alpha
	\\
	\text{for metric} \ & f_\munu : \quad \quad 
	\delta f_\munu = -2 f_{\alpha(\mu)} \tilde{\nabla}_{\nu)} \xi^\alpha
\end{equationsplit}

There is gauge symmetry in bimetric gravity. As expected, a theory of two coupled metrics enjoys gauge symmetry under those transformations where the metrics transform simultaneously (each transformation cancelling the effect of the other). 
\subsection{Counting dof \& Absence of Boulware-Deser ghost}

A general dof count for the bimetric theory of gravity in arbitrary $d$ dimensions can now be done. This goes as follows: The two independent symmetric metrics satisfy their respective non-linear eom given in eq\eqref{eq:eombimetric}. Thus, there are $2\times \frac{d(d+1)}{2}$ components to begin with. The gauge-freedom described in the previous section can be used to eliminate $2d$ components ($d$ for each metric). The relation between the divergences of the potential terms in eq\eqref{eq:divergencepotentials} together with the non-linear eom in eq\eqref{eq:eombimetric} implies a Bianchi constraint ($\nabla^\mu V^g_\munu = 0$). This constraint eliminates $d$ components. The remaining components are:
$$ \text{remaining components: } \quad d(d-2)$$
Plugging $d=4$, implies that there are 8 components left. Of these 8 components, 5 belong to a massive spin-2 field, 2 belong to a massless spin-2 field, and the remaining dof is the Boulware-Deser ghost. Plugging $d=3$, gives 3 components. Noting that massless spin-2 fields do not carry any propagating dof in 2+1 dimensions, the Boulware-Deser ghost is seen again. It is imperative to keep in mind, that at the non-linear level the dofs have not been disentangled into massive and massless modes. The labels here are only meant to serve as an indicator. 

That the Boulware-Deser ghost instability is still present is not a surprise, since its absence has not been discussed yet. Although not discussed so far, the absences of this ghost-mode has been built into the bimetric theory. In fact the demand for the absence of this ghost mode, in the first place, is exactly the reason why a square-root matrix exists for the potential term in the lagrangian. A qualitative discussion for the absence of this ghost-mode is presented here (for a thorough proof see \cite{reviewMay}). The ADM decomposition of GR allows for a Hamiltonian formulation providing the theory with the possibility of a constraint analysis. In ADM decomposition, the metric $g_\munu$ is decomposed into a scalar called Lapse ($N$), a spatial-vector called Shift ($N_i$), and a spatial-metric $\gamma_{ij}$ (very reminiscent of \stuckelberg field re-definitions). For GR the Lapse and Shift do not correspond to physical dof and are only gauge dofs (there is no kinetic-term for these in the lagrangian). The spatial-metric contains propagating dofs and the gauge invariance of GR reduces the independent number of components in the spatial-metric to:
$$\text{GR: }\frac{d(d-1)}{2} - d = \frac{d(d-3)}{2}$$
These are, ofcourse, exactly the same number of components for a massless spin-2 field in $d$-dimensions. 

For massive gravity, the spatial-vector Shift ($N_i$) becomes dynamical and therefore adds $d-1$ components to massless theory. This gives:
$$\text{massive gravity: } \quad \frac{(d-2)(d+1)}{2}$$
components. This is exactly the same as that of a massive theory of spin-2 fields. However, the above is only true when the scalar Lapse ($N$) remains non-dynamical. This is the reason why the FP-tuning $\gamma=0$ was necessary in  the FP theory. Spoiling that tuning, almost always, makes the scalar Lapse become dynamical. This adds one more dof to the theory, which is the Boulware-Deser ghost!

For any non-linear theory of massive gravity, the absence of this ghost instability demands that the scalar Lapse always remains non-dynamical, or equivalently, the lagrangian is linear in $N$. In the ADM decomposition of metric $g_\munu$, the scalar Lapse ($N$) enters as a quadratic quantity. The square-root matrix $\Xmatrix = \sqrt{g^-1 f}$, is thus the primary reason why the lagrangian remains linear in Lapse $N$. This ensures that a square-root based interaction term in the lagrangian will always keep the extra scalar dof as non-dynamical. The dof count for bimetric theory can now be completed. 
\begin{equationsplit}
	\text{bimetric dof: } = d^2 - 2d -1 
\end{equationsplit}

Plugging $d=4$, gives 7 components where 5 will correspond to a massive spin-2 field and 2 belong to a massless field. Further for $d=3$, the bimetric theory offers 2 dofs all of which are in the massive component of a spin-2 field. Hence, the Boulware-Deser ghost is removed from the theory. 

The dof count in this section suggests that the dofs present in the bimetric theory have exactly the correct number of dof for a massive and a massless spin-2 field to be present simultaneously. This is not a coincidence, and indeed, the perturbations of bimetric theory can be resolved into one spin-2 field which is massive and another spin-2 field which is massless. Since, the only allowed theory for massive spin-2 fields (ofcourse excluding TMG which is special to 2+1 dimensions) is the FP-theory, it can be expected that on linearizing bimetric gravity one should obtain a theory which contains an FP theory plus linearized GR. Linearizing bimetric gravity involves an expansion of the two metrics $g_\munu$ and $f_\munu$ about the Minkowski flat background $\eta_\munu$. The possibility of such an expansion can only arise if the bimetric theory allows Minkowski background as a solution. This is verified in the next section for both 3+1 dimensions and for 2+1 dimensions. 

\subsection{Parameter Space for Minkowski Solutions}\label{subsec:bimetricparameterspace}

An important class of solutions allowed by the bimetric theory are derived by making an ansatz upon the nature of relation between the two metrics. This ansatz relates the two independent metric conformally and is called \textbf{Proportional Background Ansatz (PBA)}, given by\footnote{The letter $\rho$, instead of $c$ which is prevalent in the literature, is used here to denote the constant of proportionality.}: 
\begin{equation}\label{eq:PBAansatz}
f_\munu = \rho^2 g_\munu
\end{equation}
Here $\rho(x)$ is a spacetime dependent function. An additional benefit of this ansatz is the extreme simplifications it brings upon the analysis. These simplifications are brought forth by looking at the implications of this ansatz. These implications are presented below:

\subsubsection{Implications of Proportional Background Ansatz (PBA)}\label{subsubsec:bimetricPBA}
\begin{enumerate}
	\item \textbf{Elementary Symmetric Polynomial:}
	\newline
	Consider, the square root matrix $\mathbb{X}$:
	\begin{equationsplit}\label{eq:PBAsquarerootmatrix}
		X^\mu_\rho\; X^\rho_\nu\; & = g^{\mu \alpha}\; f_{\alpha \nu}\; = \rho^2\; \delta^\mu_\nu
		\\
		\implies\; \mathbb{X}\; &= \rho\; \mathbbm{1}
	\end{equationsplit}
	From this, the elementary symmetric polynomials simply become:
	\begin{equationsplit}\label{eq:PBAelementarysymmetricpolynomial}
		e_n(\mathbb{X}) &= \rho^n \binom{d}{n}
		\\
		e_n(\mathbb{X}^{-1}) &= \rho^{-n} \binom{d}{n}
	\end{equationsplit}
	
	\item \textbf{Bianchi on EOM \eqref{eq:eombimetric}:}
	\newline
	Since, the Einstein tensor is divergence-less, it implies
	\begin{equationsplit}
		\nabla^\mu\; V^g_\munu\; = \nabla^\mu\; V^f_\munu\; =\; 0
	\end{equationsplit}
	Following this through, leads to $\partial^\mu \rho = 0$, which implies that the $\rho$ parameter is a constant. 
		
	\item \textbf{Simplification of Potentials $V^g_\munu$ and $V^f_\munu$:}
	\newline
	Using \eqref{eq:PBAsquarerootmatrix} and \eqref{eq:PBAelementarysymmetricpolynomial} on the definition of $V^g_\munu$ and $V^f_\munu$ in \eqref{eq:bimetricpotentials}, one obtains:		
	\begin{equationsplit}\label{eq:PBAbimetricpotential}
		V^g_\munu\; &= g_\munu\; \sum_{n=0}^{d}\; \sum_{k=0}^{n}\; (-1)^{n+k}\; \beta_n\; \rho^n\; \binom{d}{k}
		\\
		V^f_\munu\; &= \frac{g_\munu}{\rho^{d-2}}\; \sum_{n=0}^{d}\; \sum_{k=0}^{n}\; (-1)^{n+k}\; \beta_{d-n}\; \rho^{d-n}\; \binom{d}{k}
	\end{equationsplit}
	Note, that apart from the explicit presence of the metric $g_\munu$, all the other terms are constants. As such, they now contribute as `new' \emph{cosmological constants} in the eom \eqref{eq:PBAbimetriceom}.
	
	\item \textbf{Einstein Tensors:}
	\newline
	Since, the Einstein tensor is scale invariant, there is another interesting simplification:
	\begin{equation}
	G_\munu = \tilde{G}_\munu
	\end{equation}
	
	\item \textbf{Equations of Motion:}
	\newline
	Finally, all of the above can be put together to write the equations of motion for the two metrics. These equations take the following simple form:
	\begin{equationsplit}\label{eq:PBAbimetriceom}
		\text{for metric} \ & g_\munu : 
		G_\munu\; +\;  \Lambda_g\; g_\munu = 0
		\\
		\text{for metric} \ & f_\munu : 
		G_\munu\; +\; \Lambda_f\; g_\munu =0
	\end{equationsplit}
	
	It is noted that now the difference between the dynamics of the two metrics comes only from the cosmological constants $\Lambda_g$ and $\Lambda_f$. These constants are:
	\begin{equationsplit}\label{eq:PBAbimetriccosmologicalconstant}
		\text{In 3+1 dimensions:} &
		\\
		\Lambda_g\; &=\; M^2\; (\beta_0\; + 3\beta_1\; \rho\; + 3\beta_2\; \rho^2\; + \beta_3\; \rho^3 )
		\\
		\Lambda_f\; &=\;  \frac{M^2}{\alpha^2 \rho^2}\; (\beta_1\; \rho\; + 3\beta_2\; \rho^2\; + 3 \beta_3\; \rho^3\; + \beta_4\; \rho^4 )
		\\
		\text{In 2+1 dimensions:} &
		\\
		\Lambda_g\; &=\; M^2\; (\beta_0\; + 2\beta_1\; \rho\; + \beta_2\; \rho^2\; )
		\\
		\Lambda_f\; &=\;  \frac{M^2}{\alpha \rho}\; (\beta_1\; \rho\; + 2\beta_2\; \rho^2\; +  \beta_3\; \rho^3\; )
	\end{equationsplit}	
\end{enumerate} 

From the new cosmological constants above, it is clear that the parameters $\beta_0$ and $\beta_d$ do not measure any interaction between the two metrics. They are present only as cosmological constants. This can be seen as follows: (a)For $\beta_0$ (cosmological constant for metric $g_\munu$), this is clear from eq\eqref{eq:PBAbimetriccosmologicalconstant}; (b) For $\beta_d$, it is recalled that if the eom in eq\eqref{eq:PBAbimetriceom} were written in terms of the metric $f_\munu$, then an additional factor of $\rho^{-2}$ from the PBA ansatz will be carried along. Plugging this in the cosmological constants, this implies that this $\rho^{-2}$ will exactly cancel the remaining factor of $\rho^2$ (standing in front of $\beta_d$ in $\Lambda_f$) and thereby, turn the parameter $\beta_d$ into a pure constant for metric $f_\munu$. 

\subsubsection{Parameter Space for Minkowski Solutions}
Finally, the parameter space which allows for Minkowski solutions can be looked into. This study has been divided into two parts; one for each of the dimensions of interest:
\begin{enumerate}
	\item \textbf{3+1 dimensions:}
	\newline
	The first constraints comes from the realization that for the eom of the two metrics in eq\eqref{eq:PBAbimetriceom} to be consistent, the following must hold:
	\begin{equation}\label{eq:PBAconstraint}
	\Lambda_g\; =\; \Lambda_f
	\end{equation}
	This condition can be used to eliminate one out of the 5 $\beta_n$ parameters available in 3+1 dimensions. As an example, it is used here to eliminate $\beta_4$. The parameter $\beta_4$ is now completely determined from consistency criterion alone. For completion, it is given by:
	\begin{equation}
	\beta_4 = - \frac{\beta_1 + 3 \beta_2 \rho -  \beta_0 \alpha^2 \rho + 3 \beta_3 \rho^2 - 3 \beta_1 \alpha^2 \rho^2 - 3 \beta_2 \alpha^2 \rho^3 -  \beta_3 \alpha^2 \rho^4}{\rho^3}
	\end{equation}
	
	Next, plugging in the minkowski metric $\eta_{\mu\nu}$ for both the eoms in eq\eqref{eq:PBAbimetriceom}, it is noted, that the Einstein tensor for minkowski metric vanishes trivially. This leads to:
	\begin{equation}\label{eq:bimetriconsistency}
	\Lambda_g\; =\; \Lambda_f\; =0
	\end{equation}
	This may seem to give rise to two conditions, however this is not the case. In essence, the two conditions are the same. Again, this freedom is used to eliminate one of the four remaining unknown coefficients. Here, for example, $\beta_0$ is chosen to enforce that both of the cosmological constants $\Lambda_g$ and $\Lambda_f$ are $0$, giving:
	\begin{equation}
	\beta_0 = -3 \beta_1 \rho - 3 \beta_2 \rho^2 -  \beta_3 \rho^3
	\end{equation}
	
	In conclusion, Minkowski backgrounds are an allowed solution for the bimetric theory under the PBA ansatz. There are 3 unknown remaining independent parameters which measure the strength of interaction between the two metrics\footnote{Further studies into the bimetric theory are done by first putting all but one $\beta_n$ parameters to 0. The theory that remains is called $\beta_i$ model of bimetric theory for the $i_{th}$ parameter which governs the interaction.}.

	\item \textbf{2+1 dimensions:}
	\newline
	In 2+1 dimensions, there are 4 $\beta_i$ parameters. Following the same procedure as above, $\beta_3$ is eliminated from consistency alone. Next $\beta_0$ is fixed to ensure that flat spacetime is a viable solution of the equations of motion in 2+1 dimensions. The eliminated parameters are:
	\begin{equationsplit}
		\beta_3 & = - \frac{\beta_1 -  \beta_0 \alpha^2 + 2 \beta_2 \rho - 2 \beta_1 \alpha^2 \rho -  \beta_2 \alpha^2 \rho^2}{\rho^2}
		\\
		\beta_0 &= -2 \beta_1 \rho -  \beta_2 \rho^2 
	\end{equationsplit}
	
	In contrast to 3+1 dimensions, there are only 2 remaining independent parameters, namely $\beta_1$ and $\beta_2$.
\end{enumerate}

\subsection{Perturbations in Bimetric Gravity}\label{subsec:bimetricperturbations}

Having confirmed that flat Minkowski spacetime is an allowed solution in the bimetric theory of gravity for both 3+1 and 2+1 dimensions, one could now venture to explore the dynamics of linear perturbations or `gravitons' in this theory. In order to do so, the following linear perturbations are defined:

\begin{equationsplit} \label{eq:metricperturbationdefinitions}
	f_\munu & = \bar{f}_\munu + \lambda \delta f_\munu 
	\\
	g_\munu & = \frac{1}{\rho^2} (\bar{g}_\munu + \lambda \delta g_\munu)
\end{equationsplit} 
PBA only implies the background metrics $\bar{f}_\munu$ and $\bar{g}_\munu$ to be related via eq\eqref{eq:PBAansatz}. The perturbations of the two metrics are independent of each other. Factorizing $\rho^{-2}$ from the perturbation $\delta g_\munu$ simplifies some of the calculations.

To obtain the lagrangian governing the dynamics of the perturbations, the above definitions are plugged back into the lagrangian for bimetric gravity. Some effort in this long and tedious calculation is saved by noting the following key points:
\begin{itemize}
	\item As the background for constructing propagators will be Minkowski, one can let the background metrics $\bar{f}$ and $\bar{g}$ go to $\eta_{\mu\nu}$ already. That is indices can be raised or lowered with respect to Minkoski metric. All covariant derivatives will reduce to partial derivatives. This implies that all of the curvature objects in the expansion of $\mathcal{L}_{bi}$ at $O(\lambda^0)$ will vanish completely.
	
	\item The expansion of $\mathcal{L}_f$ and $\mathcal{L}_g$ is not independent of each other. Noting that, $\detsqrt{g} R_g = \rho^{2-d} \detsqrt{f} \tilde{R}_f$, the expansion of geometric objects needs to be done once only. (This is the reason why $\rho^{-2}$ was explicitly factored out in eq\eqref{eq:metricperturbationdefinitions})
	
	\item Calculations should be done for a d-dimensional manifold. This allows the results for both 2+1 and 3+1 dimensions to be evaluated simultaneously.
\end{itemize}
This gives (showing only parts quadratic in perturbations):
\begin{equationsplit}\label{eq:perturbationbimetriclagrangian}
	\frac{\mathcal{L}_f}{\detsqrt{\eta}}\;  &=\;  m_f^{d-2}\; \Bigg( \lambda^{0} (0) + \lambda^{1} (\dots) + \frac{\lambda^2}{4}\; \mathcal{L}_{\text{kinetic spin-2}}\;(\delta f_\munu)\; \Bigg)
	\\
	\frac{\mathcal{L}_g}{\detsqrt{\eta}} &= (\frac{m_g}{\rho})^{d-2} \Bigg( \lambda^{0} (0) + \lambda^{1} (\dots) + \frac{\lambda^2}{4}\; \mathcal{L}_{\text{kinetic spin-2}}\;(\delta g_\munu)\; \Bigg)
	\\
	\frac{\mathcal{L}_{int}}{\detsqrt{\eta}} &=(\frac{m}{\rho})^{d} \lambda^2 \Bigg( v_1 (\delta g^\munu \delta g_\munu) + v_2 (\delta g' \delta g')
	+ v_3(\delta f^\munu \delta f_\munu) + v_4 (\delta f' \delta f')
	\\
	& \quad \quad	\quad \quad	\quad + v_5(\delta g^\munu \delta f_\munu) + v_6 (\delta g' \delta f') \Bigg)
\end{equationsplit}
Here, 6 new coefficients have been defined for simplifying the expression of the interaction lagrangian. The coefficients $v_i$ are composed of the $\beta_n$ parameters, $\rho$ and also have a dependence on the number of dimensions-$d$. Their explicit expressions are rather long, un-illuminating, thereby earning a place in \hyperref[ch:appendix2]{Appendix B}. On the other hand, imposing restrictions to enforce PBA and flat solutions gives them a rather simplified form. These can be expressed as:
\begin{equationsplit}\label{eq:ndparameter}
	&\text{On imposing PBA and Minkowski: } 
	\\
	v_1 &= - v_2 = v_3 = - v_4 = -2 v_5 =2 v_6 = n_d
	\\
	\text{and:} &
	\\
	n_4 &= \beta_1 \rho + 2 \beta_2 \rho^2 + \beta_3 \rho^3
	\\
	n_3 &= \beta_1 \rho + \beta_2 \rho^2
\end{equationsplit}

The presence of mixed terms in the lagrangian eq\eqref{eq:perturbationbimetriclagrangian} (corresponding to $v_5$ and $v_6$) clearly indicates that this lagrangian needs to be diagonalized. In order to do so, linear combinations of the perturbations are defined as follows:
\begin{equationsplit}\label{eq:perturbationlinearcombinationfordiagonalization}
	\delta g_\munu &= a_1 \phi_{\mu\nu} + a_2 \psi_\munu
	\\
	\delta f_\munu &= a_3 \phi_{\mu\nu} + a_4 \psi_\munu
\end{equationsplit}
The coefficients in the above equation now have to be carefully chosen such that the lagrangian in eq\eqref{eq:perturbationbimetriclagrangian} separates into two decoupled parts governing the dynamics of the two fields $\phi$ and $\psi$ separately. At this point a choice is present: de-coupling the mixing terms will lead to a mass term for one field  and make the other a massless field. Here, it is chosen to make the field $\phi$ massive with the following choice: 
\begin{equationsplit}
	&a_2=a_4
	\\
	&a_3=-\frac{a_1}{(\alpha \rho)^{d-2}}
\end{equationsplit}

Thus, the lagrangian governing the linear combination of perturbations has been completely diagonalized. This can be written very neatly as:
\begin{equationsplit}\label{eq:lagrangianbimetricperturbations}
	\mathcal{L}{^{(2)}_{bimetric}} &= \sigma_\phi  \Bigg(\mathcal{L}_{\text{kinetic spin-2}}(\phi_{\mu\nu}) + \mfp^2 (\phi^2 - \phi'^2) \Bigg) 
	+ \sigma_\psi \Bigg(\mathcal{L}_{\text{kinetic spin-2}}(\psi_\munu)\Bigg)
	\\
	\text{where:} \quad\quad&
	\\
	\sigma_\phi &= (\frac{m_g}{\rho})^{d-2} (1 + (\alpha \rho)^{2-d})
	\\
	\sigma_\psi &= (\frac{m_g}{\rho})^{d-2} (1 + (\alpha \rho)^{d-2})
\end{equationsplit}

The value of the Fierz-Pauli mass term for field $\phi_{\mu\nu}$ is given by: 
\begin{equation}\label{eq:bimetricmassfp}
\mfp^2 = \frac{M^2}{\rho^2} n_d (1 + (\alpha \rho)^{2-d}) 
\end{equation}
The above equation gives the expression of the FP mass for the field $\phi_{\mu\nu}$ in any dimension. Note: the expression of the parameter $n_d$ was given in eq\eqref{eq:ndparameter}. On comparing this value with literature (eq-3.10 in \cite{bimetricHassan2013}), it is noted that the value above (for d=4) has an extra factor of $\rho^{-2}$. This is due to the extra $\rho^{-2}$ which was factorized from the definition of the perturbations in eq\eqref{eq:metricperturbationdefinitions}. Thus, these calculations verify those in the literature. The Planck masses for each field are $\sqrt{\sigma_\psi}$ and $\sqrt{\sigma_\phi}$. For the massless field $\psi_\munu$ it becomes:
\begin{equation}
	\sqrt{\sigma_\psi} =  \sqrt{(\frac{m_g}{\rho})^{d-2} (1 + (\alpha \rho)^{d-2})}
\end{equation}
In d=4, this gives the same expression as (eq-3.11 in \cite{bimetricHassan2013}).(Ofcourse, differing by a factor of $\rho^{-2}$ for the same reasons as pointed before). Additionally, note during diagonalizing overall factors of $a_1$ and $a_4$ were ofcourse undetermined. They were conveniently set to 1.

\subsection{Propagators}\label{subsec:bimetricpropagators}
So far, the perturbations arising in bimetric theory have been used to diagonalize the bimetric lagrangian. This can be schematically written as:
$$ \mathcal{L}{^{(2)}_{bimetric}} =  \sigma_\phi \mathcal{L}_\phi + \sigma_\psi \mathcal{L}_\psi $$
The lagrangian for the perturbations in eq\eqref{eq:perturbationbimetriclagrangian} confirms explicitly that the linear perturbations under the PBA ansatz have resolved themselves into a massive spin-2 field governed by the FP lagrangian and a massless spin-2 field governed by the massless spin-2 lagrangian. That the FP lagrangian has reappeared should not come as a surprise, since it is the only consistent lagrangian for massive spin-2 fields. 

These lagrangians were analyzed rigorously in the preceding chapter. Their dofs and eoms were also studied. Most of the information can bee seen through the propagators. To calculate the propagator for the fields $\phi$ and $\psi$ arising in the bimetric theory, one only needs to put the correct expressions for the $\mfp$, and Planck masses at the right places (in the propagators calculated in the last chapter). Doing so, the following propagators for the two fields are obtained:
\begin{equationsplit}
	G_\psi^{\mu\nu\alpha\beta} & = \frac{1}{\sigma_\psi k^2} \Bigg(X_1^{\mu\nu\alpha\beta} - \frac{2}{d-2} X_2^{\mu\nu\alpha\beta} \Bigg)
	\\
	G_\phi^{\mu\nu\alpha\beta} & = \frac{1}{\sigma_\phi (k^2 - \mfp^2)} \Bigg( X_1^{\mu \nu \alpha \beta} - \frac{2}{d-1} X_2^{\mu \nu \alpha \beta} - \frac{X_3^{\mu \nu \alpha \beta}}{ \mfp^2} +  \frac{2}{(d-1) \mfp^2}  X_4^{\mu \nu \alpha \beta} +  \frac{2 (d-2) X_5^{\mu \nu \alpha \beta}}{(d-1) \mfp^4} \Bigg)
\end{equationsplit}

The first propagator confirms that the $\psi_\munu$ field is massless. In 2+1 dimensions, this propagator identically goes to 0! The second propagator shows the massive nature of the $\phi_{\mu\nu}$ field. These propagators have been written in terms of tensor $X{_i^{\mu \nu \alpha \beta}}$ for convenience. The expressions for these tensors is given in \hyperref[ch:appendix1]{Appendix A}.

\subsection{Important Limits of Bimetric Gravity}\label{subsec:limitsbimetric}

It is, ultimately, essential that GR is recovered from bimetric theory under appropriate limits. Also, the opposite limit that of obtaining a massive non-linear theory for one metric should exist. That was, in principle at least, the whole idea behind the program of massive gravity. Without any hesitations, these limits are now discussed\footnote{These limits are only discussed for the version of the theory presented here. A more realistic theory should include matter-gravity couplings for each metric. In the presence of such couplings, these limits need to be suitably modified \cite{reviewMay}.}:
\paragraph{Bimetric Theory $\rightarrow$ GR:}
The bimetric theory being completely symmetric in both metrics, allows for any of the two metric to become the metric used in GR. Choosing $g_\munu$ to take that role, the non-linear eom from eq\eqref{eq:eombimetric} are re-written here.
\begin{equationsplit}
	\text{for metric} \ & g_\munu :  \quad \quad
	G_\munu +  M^2 V^g_\munu = 0
	\\
	\text{for metric} \ & f_\munu :  \quad \quad
	\alpha^{d-2} \tilde{G}_\munu\: + M^2\: V^f_\munu =0
\end{equationsplit}

Thus, it is seen that on setting $\alpha\rightarrow0$, the dynamics of the $f_\munu$ metric drops away. This can also be seen at the level of the bimetric lagrangian in eq\eqref{eq:lagrangianbimetric}. The remaining equation $V^f_\munu =0$ is algebraic and its solutions are proportional background which fix the $\rho$ parameter of the theory. This turns the other potential $V^g_\munu$ into a purely cosmological constant. Thus, in this limit GR is recovered 

\paragraph{Bimetric Theory $\rightarrow$ Massive Gravity:}

Conversely, consider the limit $\alpha\rightarrow\infty$ in the EOMs:
\begin{equationsplit}
	\text{for metric} \ & g_\munu :  \quad \quad
	G_\munu +  M^2 V^g_\munu = 0
	\\
	\text{for metric} \ & f_\munu :  \quad \quad
	\tilde{G}_\munu\: + \frac{M^2}{\alpha^{d-2}}\: V^f_\munu =0
\end{equationsplit}

The effect of this limit on the metric $f_\munu$ is to turn it into a massless reference metric. This metric decouples from the theory and will have no dependence on the metric $g_\munu$. The other metric is now massive with a mass scalar governed by $M$. This is similar to the scenario in which non-linear massive gravity was first developed as in the dRGT theory. Note, a non-linear theory of massive gravity \emph{requires} a reference metric to get rid of the ghost-mode. This `strange' requirement is the reason why massive gravity took so long to be formulated consistently.

To conclude this section, it seems to the author, a summary is necessary to take stock of all that 
has happened so far. At first, the bimetric theory of gravitation was introduced. This theory involved two independent metrics, on equal footing, interacting with each other through a square-root potential. The interaction term of this theory was expressed in terms of the elementary symmetric polynomials for the matrix $\Xmatrix = \sqrt{g^-1f}$ as in eq\eqref{eq:potentialbimetric} or eq\eqref{eq:bimetricpotentials}. This square-root interaction term is a peculiar necessity in order to ensure the absence of the Boulware-Deser ghost from the full non-linear theory. To explicitly verify the mass-eigenstates in this theory flat background solutions were imperative. The PBA ansatz which means that although the two metric are flat; they could still in principle differ by some scale factor $\rho^2$ in eq\eqref{eq:PBAansatz} was put in force. Several implications of this ansatz were listed in \fullref{subsubsec:bimetricPBA}. Having confirmed that flat space is allowed by the theory, the nature of perturbations that arise in this theory was investigated. The lagrangian in eq\eqref{eq:lagrangianbimetricperturbations}, describes those perturbations. It is composed of a massless field $\psi_\munu$ as in standard linearized Einstein-GR along with a Fierz-Pauli massive field $\phi_{\mu\nu}$. 

That the complicated theory of bimetric gravity expressed in eq\eqref{eq:lagrangianbimetric} can be simplified enough to these two structures is most certainly a remarkable feature of bimetric gravity.

\section{Topologically Massive Bimetric Gravity}
So far, it has been seen that the bimetric theory of gravity is a very general and a significantly non-trivial extension to GR. This theory allowed massive gravity to obtain its own non-linear theory without the problematic appearance of the Boulware-Deser ghost. A major outcome of this thesis is to suggest an extension of the bimetric theory of gravity in 2+1 dimensions. This attractive extension involves giving a CS-term to each metric in the bimetric theory. Such an extension is interesting since: (a) GR by itself does not propagate any physical dof. The FP theory of linear spin-2 fields was successful in making this field massive, by exciting auxiliary vector and scalar modes. (b) pure CS theory of spin-2 fields in 2+1 dimensions is physically equivalent to GR, and TMG made the tensor mode of the spin-2 fields massive. (c) Thus, one immediately wonders what happens when the non-linear bimetric theory is deformed with a CS term? Does the theory reduce to two copies of TMG? Since CS terms find their origin in topology, will the presence of these terms affect the outcome of bimetric theory in 2+1 dimensions? 

On consistency grounds alone, the Topologically Massive Bimetric Gravity (TMBG) theory will turn out to be much more constrained than its corresponding bimetric version. It was noted in the last section, that under PBA, the background metrics were related by a constant factor of $\rho^2$. In TMBG, the presence of CS terms will completely fix the $\rho$ factor. The allowed values of this $\rho$ factor will turn out to be $0,\pm1$. This is an interesting outcome in the TMBG theory.

All the ingredients for the setup of this theory are completely in hand. The analysis will closely follow the presentation of the preceding section. In this extension, the geometric part of the action for each metric will be deformed by another term which is the Chern-Simons term. Since a term of this kind is only available in 3 dimensions, the following analysis is done explicitly in 2+1 dimensions.  
\subsection{Action}\label{subsec:topobimetricaction}

The bimetric lagrangian in 2+1 d was(set d=3 in eq\eqref{eq:lagrangianbimetric}):
\begin{equation}
	\mathcal{L}_{bi} = m_g \detsqrt{g} R_g + m_f \detsqrt{f} \tilde{R}_f + 2 m^3 \detsqrt{g} V(\mathbb{X};\beta_n)
\end{equation}
To this lagrangian a CS mass term is added. The lagrangian governing TMBG becomes:
\begin{equationsplit}\label{eq:lagrangianbimetricTMG}
	\mathcal{L}_{TMBG} &= \mathcal{L}_g + \mathcal{L}_f + \mathcal{L}_{int}
	\\
	\text{where:} \quad \quad&
	\\
	\mathcal{L}_g &= m_g \big(\detsqrt{g} R_g + \frac{1}{2 \mu_g} \mathcal{L}_{cs-g} \big)
	\\
	\mathcal{L}_f &= m_f \big(\detsqrt{f} \tilde{R}_f + \frac{1}{2 \mu_f} \mathcal{L}_{cs-f} \big)
	\\
	\mathcal{L}_{int} &= 2 m^3 \detsqrt{g} V(\mathbb{X};\beta_n)
\end{equationsplit}
The CS pieces are:
\begin{equationsplit}
	\mathcal{L}_{cs-g} &= \epsilon^{\alpha \beta \mu}\; \Gamma^\nu_{\alpha\delta}\; \big( \partial_\beta \Gamma^\delta_{\mu\nu} + \frac{2}{3} \Gamma^\delta_{\beta\kappa} \Gamma^\kappa_{\mu\nu} \big)
	\\
	\mathcal{L}_{cs-f} &= \epsilon^{\alpha \beta \mu}\; \tilde{\Gamma}^\nu_{\alpha \delta}\; \big( \partial_\beta \tilde{\Gamma}^\delta_{\mu \nu} + \frac{2}{3} \tilde{\Gamma}^\delta_{\beta\kappa} \tilde{\Gamma}^\kappa_{\mu \nu} \big)
\end{equationsplit}

Here, $\mu_g$ and $\mu_f$ appear as parameters or inverse-levels for the Chern-Simons terms. $\Gamma^\nu_{\alpha\delta}$ is the standard christoffel symbol, and the rest has already been explained. For further, convenience the ratios of Planck masses and the mass scale $m$ are defined:
\begin{equationsplit}
	M^2 &= \frac{m^3}{m_g} 
	\\
	\alpha &= \frac{m_f}{m_g}
\end{equationsplit}

To analyze the theory governed by this lagrangian: First, the complete non-linear eoms are derived. The parameter space for this theory to allow Minkowski background as solutions are derived by using Proportional Background Ansatz (PBA). This ansatz along with the flat space solutions will lead to some interesting conditions on the parameter space for this theory. Next perturbations in this non-linear theory are studied. This will allow the mass-eigenstates of the theory to be investigated. On perturbing the metric, and plugging it back into TMBG lagrangian, the action governing the dynamics of these perturbations is arrived at. Finally, on diagonalizing this system the nature of gravitons excited in this theory becomes clear. 

\subsection{Non-linear EOMs}\label{subsec:topobimetricnonlineareom}
The variation of the action with respect to the metrics $g_\munu$ and $f_\munu$ is given by:
\begin{equationsplit}
	\delta (\mathcal{L}_g) & = m_g \detsqrt{g} \big(\ G_\munu\; + \frac{1}{\mu_g} C_\munu \big)  \ \delta g^\munu 
	\\
	\delta (\mathcal{L}_f) &= m_f \detsqrt{f} \  \big(\tilde{G}_\munu\; + \frac{1}{\mu_f} \tilde{C}_\munu \big) \  \delta f^\munu 
	\\
	\delta (\detsqrt{g} V ) &= \detsqrt{g} \ V^g_\munu \  \delta g^\munu  
	\\
	& = \detsqrt{f} \ V^f_\munu \  \delta f^\munu  
\end{equationsplit}
Here, $C^\munu$ is the Cotton Tensor, introduced in \fullref{sec:TMG}. This tensor plays the role of Weyl tensor in 2+1 dimensions (since the latter vanishes identically). Cotton tensor is defined as: 
\begin{equation}
	C^\munu\; = \frac{1}{\detsqrt{g}}\; \epsilon^{\mu\alpha\beta}\; \nabla_\alpha\; S^\nu_{\beta}
\end{equation}
where $S_\munu$ is the Schouten tensor (a combination of Ricci Tensor and Ricci Scalar) given by (for 2+1 dimensions):
\begin{equation}
S_\munu =  R_\munu  - \frac{1}{4} g_\munu R
\end{equation}
The Cotton tensor $\tilde{C}_\munu$ and the Schouten tensor $\tilde{S}_\munu$ for the metric $f_\munu$ are defined in an obvious analogous manner. With these definitions, the non-linear eom are given by:
\begin{equationsplit}\label{eq:eomTMGbimetric}
	\text{for metric} \ & g_\munu : \quad \quad G_\munu\; +\; \frac{1}{u_g}\; C_\munu +\;  M^2 V^g_\munu = 0
	\\
	\text{for metric} \ & f_\munu : \quad \quad \tilde{G}_\munu\; +\; \frac{1}{u_f}\; \tilde{C}_\munu +\;  \frac{M^2}{\alpha} V^f_\munu = 0
\end{equationsplit}
On comparing these equations with the eom obtained for TMG theory (eq\eqref{eq:eomnonlinearTMG}), it is seen that for each of the metric: (a) the geometric part is modified by the presence of the CS term and  (b) there is an interaction given by the interaction potentials $V^g_\munu$ and $V^f_\munu$. Note: that the potentials are the same as in eq\eqref{eq:bimetricpotentials}. 
\subsection{Gauge Symmetry}

The TMBG lagrangian enjoys, essentially, the same gauge symmetry as in bimetric theory. This is because the CS-term upon gauge transformations are gauge invariant upto total-derivatives. The dynamics of the theory will not be affected by such surface terms. The dofs present in the theory can, therefore be counted by the same means as before. 

There are two copies of TMG to begin with, which implies two dof. The two metric also interact via the square-root matrix $\Xmatrix$ in the interaction term. The overall effect of this interaction in bimetric theory was to generate 2 dofs in 2+1 dimensions. These 2 dofs were seen through perturbations to be excited via a FP mass term. Thus, in total this theory should have 4 dofs. Following the same dof counting as in the bimetric theory, the number of dof being 4 can be seen as a direct increase coming from the presence of two CS terms. 
\subsection{Parameter Space for Minkowski Solutions}\label{subsec:topobimetricparameterspace}

To start discussing the parameter space for Minkowski solutions, the PBA ansatz is necessary. This ansatz is given in eq\eqref{eq:PBAansatz} and reproduced here. 
\begin{equation}
	f_\munu = \rho^2 g_\munu
\end{equation}
Most of its implications have already been discussed in detail in \fullref{subsubsec:bimetricPBA}. Hence, here only the key points are collected.
\begin{enumerate}
	\item Since the Cotton tensor is also divergence-less, Bianchi on the eom implies that the $\rho$ parameter is a constant in this theory. 

	\item Just as $\tilde{G}_\munu = G_\munu$, there is also a relation for the Schouten tensors: 
	$$ \tilde{S}_\munu = S_\munu$$ 
	Yet, due to the presence of $\detsqrt{g}$ in the definition of $C_\munu$, The Cotton tensors for each metric are related by:
	\begin{equation}
		\tilde{C}_\munu = \frac{1}{\rho^d}\; C_\munu
	\end{equation}
	
	\item Collecting the above implications, the EOM for topologically massive bimetric gravity with PBA becomes:
	\begin{equationsplit}\label{eq:eomTMGbimetricPBA}
		\text{for metric} \ & g_\munu : 
		G_\munu\; +\; \frac{1}{\mu_g} C_\munu +\;  \Lambda_g\; g_\munu = 0
		\\
		\text{for metric} \ & f_\munu : 
		G_\munu\; +\; \frac{1}{\mu_f \rho^3} C_\munu+\; \Lambda_f\; g_\munu =0
	\end{equationsplit}
	where:
	$$ \Lambda_g\; =\; M^2\; (\beta_0\; + 2\beta_1\; \rho\; + \beta_2\; \rho^2\; ) $$
	$$		\Lambda_f\; =\;  \frac{M^2}{\alpha \rho}\; (\beta_1\; \rho\; + 2\beta_2\; \rho^2\; +  \beta_3\; \rho^3\; )	$$
	
	As before, the contribution coming from the interaction between the two metrics has turned into a cosmological constant for both the metrics. The coefficients in this constant measure the strength of coupling between the two metrics. 
	\item For the consistency of the PBA ansatz, the following is required (from subtracting the two eoms):
	\begin{equation}
			0 = C_\munu \bigg( \frac{1}{\mu_g} - \frac{1}{\mu_f \rho^3}\bigg)\; +\; g_\munu \bigg( \Lambda_g - \Lambda_f \bigg)
	\end{equation}	
	This equation is drastically different (compare with eq\eqref{eq:bimetriconsistency}). This equation will in general hold only if both the expressions in the parentheses are identically zero. This gives the following conditions:
	\begin{enumerate}
		\item  
		\begin{equation}
			\mu_f = \frac{\mu_g}{\rho^3}
		\end{equation}
		This means that the two Chern-Simons terms for metric $g_\munu$ and $f_\munu$ are strongly related due to the PBA ansatz. If one tries to remove one of the C-S terms from the lagrangian in eq\eqref{eq:lagrangianbimetricTMG} by setting $\mu_f \to \infty$ (for finite $\rho$), then the other C-S term automatically gets dropped out as well.
		\item  
		\begin{equation}
			\Lambda_g = \Lambda_f
		\end{equation}
		This condition is the same as in eq\eqref{eq:PBAconstraint} arising as a constraint from applying PBA. This can be used to fix the parameter $\beta_3$ as:
		\begin{equation}
			\beta_3  = - \frac{\beta_1 -  \beta_0 \alpha^2 + 2 \beta_2 \rho - 2 \beta_1 \alpha^2 \rho -  \beta_2 \alpha^2 \rho^2}{\rho^2}
		\end{equation}
	\end{enumerate}
	\item 	Finally, the minkowski metric will be a solution if:
	 \begin{equation}
			\Lambda_g = \Lambda_f = 0
	 \end{equation}
	 This is used to eliminate $\beta_0$ as:
	 \begin{equation}
		\beta_0 = -2 \beta_1 \rho -  \beta_2 \rho^2 
	 \end{equation}
\end{enumerate}

In total, three conditions have emerged out of demanding that TMBG allows minkowski backgrounds under PBS. Two of them are exactly the same as in the case of bimetric theory. The third condition is new and unique to TMBG. For later convenience, these are summarized below:
\begin{equationsplit}\label{eq:parameterTMGbimetricPBA}
	\mu_f &= \frac{\mu_g}{\rho^3}
	\\
	\beta_3  &= - \frac{\beta_1 -  \beta_0 \alpha^2 + 2 \beta_2 \rho - 2 \beta_1 \alpha^2 \rho -  \beta_2 \alpha^2 \rho^2}{\rho^2}
	\\
	\beta_0 &= -2 \beta_1 \rho -  \beta_2 \rho^2 
\end{equationsplit}

\subsection{Perturbations in TMBG}\label{subsec:topobimetricperturbations}

It has now been confirmed that under a restricted parameter space, TMBG allows flat background solution. To understand the nature of excitations present in this theory, it is important to look at its linearized behaviour. For this, perturbations to both metrics are defined (same as eq\eqref{eq:metricperturbationdefinitions}, reproduced here for convenience):
\begin{equationsplit}
	f_\munu  &= \bar{f}_\munu + \lambda \delta f_\munu	
	\\
	g_\munu  &= \frac{1}{\rho^2} (\bar{g}_\munu + \lambda \delta g_\munu)
\end{equationsplit}
The result of expanding the TMBG lagrangian for these linear perturbations gives:
\begin{equationsplit}\label{eq:perturbationbimetricTMGlagrangian}
	\frac{\mathcal{L}_f}{\detsqrt{\eta}}\;  &=\;  m_f\; \Bigg( \lambda^{0} (0) + \lambda^{1} (\dots) + \frac{\lambda^2}{4}\; \big(\mathcal{L}_{\text{kinetic spin-2}}(\delta f_\munu) + \frac{1}{2 \mu_f} \mathcal{L}_{cs}\;(\delta f_\munu)\big)\; \Bigg)
	\\
	\frac{\mathcal{L}_g}{\detsqrt{\eta}} &= (\frac{m_g}{\rho}) \Bigg( \lambda^{0} (0) + \lambda^{1} (\dots) + \frac{\lambda^2}{4}\; \big(\mathcal{L}_{\text{kinetic spin-2}}(\delta g_\munu) + \frac{\rho^3}{2 \mu_f} \mathcal{L}_{cs}\;(\delta g_\munu)\big) \Bigg)
	\\
	\frac{\mathcal{L}_{int}}{\detsqrt{\eta}} &=(\frac{m}{\rho})^{3} \lambda^2 \Bigg( v_1 (\delta g^\munu \delta g_\munu) + v_2 (\delta g' \delta g')
	+ v_3(\delta f^\munu \delta f_\munu) + v_4 (\delta f' \delta f')
	\\
	& \quad \quad	\quad \quad	\quad + v_5(\delta g^\munu \delta f_\munu) + v_6 (\delta g' \delta f') \Bigg)
\end{equationsplit}
And the parameters $v_i$ satisfy:
\begin{equation}
	v_1 = - v_2 = v_3 = - v_4 = -2 v_5 =2 v_6 = n_3 = \beta_1 \rho + \beta_2 \rho^2 
\end{equation}

As before, the presence of mixed terms in the interaction lagrangian points towards a need to diagonalize this lagrangian. Linear combinations of the metric perturbations are defined.
\begin{equationsplit}
	\delta g_\munu &= a_1\; \phi_{\mu\nu} + a_2\; \psi_\munu
	\\
	\delta f_\munu &= a_3\; \phi_{\mu\nu} + a_4\; \psi_\munu
\end{equationsplit}
Plugging these field redefinitions into the lagrangian, and demanding that the dynamics of field $\phi$ and $\psi$ are completely decoupled leads to the following conditions. (A choice has been made to make the $\phi_{\alpha \beta}$ field gain the FP mass.)
\begin{equationsplit}\label{eq:diagonalizinglagrangian}
	a2 &= a4 
	\\
	a3 &= -\frac{a1}{\alpha \rho} 
	\\
	\mu_f &= \frac{\mu_g}{\rho}
\end{equationsplit}
The first two conditions are the same as for bimetric theory. The last of these conditions, is a new peculiar condition. It comes from demanding that the perturbations decouple from parity odd terms as well. Together with the constraints from PBA eq\eqref{eq:parameterTMGbimetricPBA}, this equation completely fixes the $\rho$ parameter as $-1,0,+1$. 

Finally, the completely diagonalized TMG-bimetric lagrangian becomes (keeping $\rho$ explicit):
\begin{equationsplit}\label{eq:lagrangianTMBGdecouple}
	\mathcal{L}{^{(2)}_{TMBG}}  &= \mathcal{L}_\phi + \mathcal{L}_\psi
	\\
	\mathcal{L}_\phi & = \sigma_\phi  \Bigg(\mathcal{L}_{\text{kinetic spin-2}}(\phi_{\mu\nu}) + \mfp^2 (\phi^2 - \phi'^2) + \frac{\rho}{2 \mu_g}  \mathcal{L}_{cs}(\phi_{\mu\nu})\Bigg)  
	\\
	\mathcal{L}_\psi &= \sigma_\psi \Bigg(\mathcal{L}_{\text{kinetic spin-2}}(\psi_\munu) + \frac{\rho}{2 \mu_g}  \mathcal{L}_{cs}(\psi_\munu)\Bigg)
	\\
	\text{where:} &
	\\
	\sigma_\phi &= \frac{m_g}{\rho} (1 + (\alpha \rho)^{-1})
	\\
	\sigma_\psi &= \frac{m_g}{\rho} (1 + (\alpha \rho))
\end{equationsplit}
The lagrangians for the two fields are now completely decoupled. It is seen that both fields have obtained a CS mass which is exactly the same. This happened when the two perturbations were being decoupled. Next, one of the field has obtained a mass term. This FP mass is given by:
\begin{equation}
		\mfp^2 = \frac{M^2}{\rho^2} n_3 (1 + (\alpha \rho)^{-1}) 
\end{equation}

\subsection{Mass Spectrum of TMBG}

The more startling issue with TMBG as opposed to any of the theories presented in this thesis so far, is that the two mass-generating mechanisms mixed the masses of each perturbation at the non-linear stage itself. This shows in the fact that when the perturbations are diagonalized the FP mass comes out to be dependent on $\rho$. This factor of $\rho$, in turn, was completely determined by the CS sector of the theory. Thus, the CS mass-generating mechanism, in a sense has governed the FP mass of the theory. The requirements of consistency, flat backgrounds and decoupled dynamics have forced the $\rho$ parameter to take only three possible values. This dictates the physics which is analyzed below:
\begin{enumerate}
	\item \textbf{For $\rho =+1$:} This is equivalent to setting all components of the two background-metrics as equal. Although the backgrounds are identified, there seem to be two separate fluctuations which have their own separate dynamics and are governed by two different Lagrangians:
	\begin{equationsplit}
		\sigma_\phi &= \frac{m_g (m_g + m_f)}{m_f}
		\\
		\sigma_\psi &= (m_g + m_f)
		\\
		\mfp^2 &=  \frac{m^3 (m_g + m_f) (\beta_1 + \beta_2)}{m_f m_g}
	\end{equationsplit}
	
	\item \textbf{For $\rho =-1$:} Since PBA relates the two backgrounds with $\rho^2$, this case is similar to the one before. The backgrounds for each metric are identified again. However, the parameters controlling the theory are different:
	\begin{equationsplit}
		\sigma_\phi &= \frac{m_g (m_g - m_f)}{m_f}
		\\
		\sigma_\psi &= (- m_g + m_f)
		\\
		\mfp^2 &=  \frac{m^3 (- m_g + m_f) (- \beta_1 + \beta_2)}{m_f m_g}
	\end{equationsplit}

	\item \textbf{For $\rho =0$:} This case is a rather different case than the ones above. It is equivalent to putting the background for the second metric to 0. Putting this condition in the expressions of the parameters $\sigma_\phi$, $\sigma_\psi$ and the FP mass $\mfp$, shows that they diverge. Since it is unclear how to make sense of the parameters this way, one alternative is to track the effect of setting $\rho\to0$ from early on. The ingredients that went into obtaining this particular value of $\rho$ are PBA, demand for flat solutions and consistency of background eoms. It can be noted that setting $\rho=0$, will seriously violate the basic input of PBA and all other equations derived from it. In the literature, it has been strongly suggested that away from PBA, fluctuations in the bimetric theory do not generally have an FP mass term with the correct tuning \cite{bimetricHassan2013}. The analysis of bimetric theories without PBA gets further complicated because the matrix $\Xmatrix$ does not have a simple expansion. Thus, it seems that the only viable choices are $\rho=\pm1$.
\end{enumerate}

Finally, the mass-spectrum excited by TMBG theory can be seen as follows: The lagrangian for the field $\phi$ is the same lagrangian for a FPCS theory. This theory was studied in the last chapter. It is therefore known that this perturbation will have two excitations. These excitations will carry a mixed mass. The mass-mixing and the propagators for this theory were already detailed when FPCS theory was studied. 

The other perturbative field $\psi$ is governed by a TMG lagrangian. From the studied of TMG lagrangian, it becomes clear that this field has one massive scalar dof. The propagator for this field was also calculated when TMG was studied. 

In total, the theory of TMBG seems to propagate 3 massive dof. A dof counting presented before suggested there should be 4 dof in this theory. It is seen that the ansatz of proportional backgrounds was too restrictive and it eliminated the fourth degree of freedom. Also, the proportionality constant was completely fixed due to the demand of decoupling the dynamics of the perturbation from the lagrangian. There may exist possibilities to diagonalize this lagrangian in other ways. Additionally, it remains to be seen what the effects of different signs ($\pm1$) for the EH part of each action will give rise to. One should also look into the possibility of deforming this theory with only one CS term. These possibilities are being looked into, by the author. 

In summary, the TMBG theory which can be viewed as an extension of the bimetric theory in 2+1 dimensions was studied. The setup included deforming the lagrangian of each metric with a CS term. The non-linear eom for this theory were derived. The contribution from the non-linear interaction appeared as interaction potentials in these eoms. For the version of the TMBG theory studied, the absence of Boulware-Deser ghost from the mass spectrum is confirmed. 

To understand the massive nature of the excitations in this theory, the linearized approximation was necessary. To this end, the parameter space for the theory which allows for Minkowski solutions under PBA was derived. At this point, in contrast to bimetric gravity in 2+1 dimensions, a new condition emerged for the consistency of the non-linear eoms. Next, the perturbations in this theory were defined and the lagrangian governing their dynamics was calculated. This lagrangian included mixed terms, and on diagonalizing this lagrangian a new condition emerged. This condition completely fixed the allowed values of the proportionality constant $\rho$ as $0,\pm1$. 

The theory was then considered under all the allowed values for $\rho$, and it was found that $\rho=0$ was not in conformity with the PBA ansatz. The Planck masses and the FP mass for the perturbations for both $\rho=\pm1$ theory was evaluated. Both of these theories give rise to 3 massive propagating dof. Using the insights gained from earlier studies, it is clear that there are 3 massive modes and they have been mixed in the TMBG theory in an interesting manner.

\chapter{Quantum Loop Corrections}\label{ch:quantumloopcorrections}
\epigraph{``Like the silicon chip of more recent years, the Feynman diagram was bringing computation to the masses."}{Julian Schwinger}
In this chapter, quantum loop corrections from minimally coupled photons to the graviton propagator in 2+1 dimensions is calculated. There are some related calculations on this topic in the literature. In early 1974, Capper et. al. calculated the corrections to the graviton propagator from photons\cite{loopCapper:1974ed}. After the discovery of gauge theories with topological masses in early 1980s many interesting calculations with the topological effects were carried out \cite{tmgSchonfeld:1980kb,tmgravityDeserJackiwTempleton:1981wh}. In 1986, van der Bij et. al. calculated the induced CS term for gravity from fermions and topologically massive photons\cite{tmgVANDERBIJ198687}. Lerda et. al. calculated the induced CS corrections from gravitons to Yang-Mills theory in 1987. Similar calculations were carried out by Ojima in 1988 using path-integral approach\cite{loopOjima:1988av}. Considering a self-dual model for photons, Yang calculated the induced CS terms for gravity in 1990 \cite{loopYang:1990yj}. In 1994, Pinheiro et. al. calculated the photon self-energy from topologically massive gravity\footnote{This is of course not a comprehensive historical review of such calculations.}.

Having studied the nature of photons which obtain their mass from both Proca mass term and a CS mass term, here a calculation of the graviton propagator correction from such photon is presented. Since, the primary interest for doing this calculation is to look into the induced CS masses of the theory, only the parity odd parts are evaluated here. In the section that follows, the theory describing interactions between minimally coupled photons and gravity is discussed. The propagators for such photons were already calculated, and are listed here. Then, the vertex functions necessary for this calculation are derived. Finally, the bubble loop is evaluated and the results discussed.

\section{Minimally Coupled Photons to Gravity}	

The theory describing the interaction of photons with gravitons is obtained by minimally coupling the spin-1 gauge field in the PCS lagrangian which was given in eq\eqref{eq:lagrangianPCS}. This lagrangian becomes:
\begin{equationsplit}
	\mathcal{L}_{\text{PCS}} &= \sigma_A \Bigg( \mathcal{L}_{\text{kinetic}} + \frac{m^2_{A}}{2}\;\mathcal{L}_{\text{proca}} +\frac{\mu_{A}}{2}\ \mathcal{L}_{\text{cs}} \Bigg)
	\\
	\text{where: } \quad \quad &
	\\
	\mathcal{L}_{\text{kinetic}} &= \frac{-1}{4}\; \sqrt{|g|}\;  F_{ab}\;F_{cd}\;g^{ac}\;g^{bd}
	\\
	\mathcal{L}_{\text{proca}} &= \sqrt{|g|}\; A_a\;A_b\;g^{ab}
	\\
	\mathcal{L}_{\text{cs}} &= \epsilon_{\mu\nu\alpha} A^\mu\; \del^\nu\; A^\alpha
\end{equationsplit}

Here, $A_a$ is a photon field which has both a Proca mass $m_A$ and a CS mass $\mu_A$. $F_{ab}$ denotes the electromagnetic field strength tensor, and $g^{ab}$ is the inverse of the metric\footnote{Letters of the alphabet are used to denote indices for convenience only.}. The metric is now expanded as follows:
\begin{equation}
	g_{ab} = \bar{g}_{ab} + \lambda\ \delta g_{ab}
\end{equation}

This allows the Lagrangian above to be expanded as follows (schematically):
\begin{equation}
		\mathcal{L}_{\text{PCS}} = 	\mathcal{L}^{(0)}_{\text{PCS}} + \lambda\; \mathcal{L}^{(1)}_{\text{PCS}} + \frac{\lambda^2}{4} \mathcal{L}^{(2)}_{\text{PCS}} + O(\lambda^4)
\end{equation}
The lagrangian at $0^{th}$-order will provide the propagation of the spin-1 field. The higher-order terms describe the interactions. These are dealt with separately.

\subsection{Propagators}

The propagator for the above lagrangian was already calculated in \fullref{sec:massivespin1PCS}. This propagator is reproduced here:
\begin{equationsplit}
	&G{^\munu_\text{PCS}} = \frac{-1}{\sigma_1} \frac{1}{f_1+f_2}\Bigg( \frac{Y_1^\munu}{k^2 - f_1^2} + \frac{Y_2^\munu}{k^2 - f_2^2}\Bigg)
	\\
	\text{where:} \quad\quad&
	\\
	&Y_1^\munu = f_1 \Bigg( \eta^\munu - \frac{k^\mu k^\nu}{f_1^2} + \frac{i}{f_1}\ \epsilon^{\munu\rho}k_\rho \Bigg)
	\\
	&Y_2^\munu = f_2 \Bigg( \eta^\munu - \frac{k^\mu k^\nu}{f_2^2} - \frac{i}{f_2}\ \epsilon^{\munu\rho}k_\rho \Bigg)	
\end{equationsplit}
Here, the mixed masses $f_1$ and $f_2$ are given by:
\begin{equationsplit}\label{eq:mixedmassspin1}
	&f_1 = \frac{1}{2} \big(\sqrt{\mu{^2_A} + 4 m^2_{A}} + \mu_A \big)
	\\
	&f_2 = \frac{1}{2} \big(\sqrt{\mu{^2_A} + 4 m^2_{A}} - \mu_A \big)
\end{equationsplit}

\subsection{Vertex Functions}\label{subsec:vertexfunctions}
The higher order terms from the Lagrangian contribute to the vertex functions. Since the CS mass term is independent of the metric it will not contribute to vertex functions. In total, there will be a contribution from both the kinetic term and the Proca mass term. These give (for all momenta flowing into the vertex):

1. From kinetic term at first order ($p$ and $r$ are photon momenta, and $k$ is graviton momenta)
\begin{equationsplit}
	V3^{\text{kin}}_{abcd}(k,p,r) = \frac{\sigma_A \lambda }{4}\Big(&- \eta_{cd}\; p_{b}\; r_{a}\; -  \eta_{cd}\; p_{a}\; r_{b}\; + \eta_{bd}\; p_{a}\; r_{c}\; + \eta_{ad}\; p_{b}\; r_{c}\; 
	\\
	&- \eta_{ab}\; p_{d}\; r_{c}\; + \eta_{bc}\; p_{a}\; r_{d}\; + \eta_{ac}\; p_{b}\; r_{d}\;  
	\\
	&+ p \cdot r (-  \eta_{ad}\; \eta_{bc}\;  -  \eta_{ac}\; \eta_{bd}\; + \eta_{ab}\; \eta_{cd}\;) \Big)
\end{equationsplit}

2. From Proca mass term at first order
\begin{equationsplit}
V3^{\text{Proca}}_{abcd}(k,p,r) = \frac{\sigma_A \lambda m^2_{A} }{4}\Big( - \eta^{ad} \eta^{bc} - \eta^{ac} \eta^{bd} + \eta^{ab} \eta^{cd} \Big)
\end{equationsplit}

The vertices coming from the Proca and kinetic term at second order have been calculated. Their expression is given in \hyperref[appendix3]{Appendix C}.
\section{The Bubble Loop}

Their are three diagrams which contribute to quantum loop corrections at the one-loop order. These three are: (a) Tadpole Diagram, (b) Bubble Diagram, and (c) Seagull Diagram. Since the results obtained from the Tadpole and the Seagull diagrams are still under being analysed, only the calculation of the bubble diagram is detailed.

The bubble diagrams can be depicted as:
\begin{center}
		\feynmandiagram[layered layout,horizontal=b to c] {
			a [particle =\(\delta g^{a_1 b_1}\)] --[plain, gluon	,	momentum	=\(p\)	] 
			b -- [	photon	,	half left	,	momentum	=\(k\)] c
			-- [	photon	,	half left	,	momentum	=\(k+p\)	] b,
			c -- [plain, gluon	,	momentum	=\(p\)	]  d [particle =\(\delta g^{a_2 b_2}\)],
		};
\end{center}

The amplitude is given by:
\begin{equation}
	i \mathcal{M}_{a_1 b_1 a_2 b_2} = N_s \int \frac{d^{3}k}{(2 \pi)^3} (iV1_{a_1 b_1 c_1 c_2})\;(iG^{{d_2}{c_2}})\;(iV2_{a_2 b_2 d_1 d_2})\;(iG^{{c_1}{d_1}})\;
\end{equation}

Since the calculation is rather tedious, only two key points are discussed. The first main issue is to reduce higher powers of the loop momenta in the numerator through algebraic reductions. The next main input comes from Veltman-Passarino reduction. These reduction identities relate the open indices present inside the loop integral to scalars which can then be reduced to standard integrals. The evaluation of the parity odd part gives the following form-factor:

\begin{equationsplit}
	i \mathcal{M}_{a_1 b_1 a_2 b_2} &= \frac{1}{16}\; f_1 \frac{A0(f_1)}{p^2} (\text{CS-term})
				\\
				& - \frac{1}{16}\; f_2 \frac{A0(f_2)}{p^2} (\text{CS-term})
				\\
				& + \frac{1}{8}\; f^3_1 \frac{B0(p, f_1, f_2)}{p^2} (\text{CS-term})
				\\
				& - \frac{1}{8}\; f^3_2 \frac{B0(p, f_1, f_2)}{p^2} (\text{CS-term})	
				\\
				& - \frac{1}{32}\; f_1 {B0(p, f_1, f_2)} (\text{CS-term})			
				\\
				& + \frac{1}{32}\; f_2 {B0(p, f_1, f_2)} (\text{CS-term})			
\end{equationsplit}

Here, the CS-term stands for the parity odd part. This is equal to:
\begin{equationsplit}
	\text{CS-term} = p^2 X^6_{a_1 b_1 a_2 b_2} - X^7_{a_1 b_1 a_2 b_2}
\end{equationsplit}
As before, the tensors $X^i_{a_1 b_1 a_2 b_2}$ are listed in \hyperref[ch:appendix1]{Appendix A}. This form factor clearly shows that photons in 2+1 dimensions are inducing a CS term for gravitons. The evaluation of the remaining integrals can be done through standard dimreg integrals. There have been some other calculations for different loops which have been done by the author, but since those results are still under analysis, they are not included here.

\chapter{Discussion and Outlook}\label{ch:final}

In this thesis, various theories of massive gravity were pursued. A short summary of the work presented in this thesis is now described:
\begin{itemize}
	\item First of all, the motivations for undertaking this study were discussed at length. Outstanding theoretical challenges in formulating Quantum Gravity, understanding the origin of Dark Matter and the Cosmological Constant Problem serve as some of the problems which invites researchers to look at IR modifications of gravity. Formulating a non-linear consistent theory of  Massive Gravity proved to be a difficult technical challenge. 

	\item The Higher Spin gauge field formulation was discussed. From group-theoretic arguments alone important details of 2+1 dimensional physics could be derived. Massless spin-1 fields were found to be equivalent to scalars and all massless fields of HS were seen to not have any physically propagating dof. This was verified by counting the dofs for both massive and massless fields of arbitrary spin-$s$ in arbitrary dimensions-$d$. A Chern-Simons term for HS fields was conjectured, and has been explicitly verified upto spin-2 in this thesis. 

	\item The analysis of spin-1 fields served as an important playground to look at topological effects. Massless photons and the Proca theory of massive photons has been known for a long time. The theory of Topologically Massive Electrodynamics was presented. Additionally, a theory of photons in which both mass mechanisms are present was studied. When both of these mass mechanisms are present, the theory propagates 2 massive dof. The masses of these dof had contributions from both Proca and CS mass. This was indeed confirmed by performing a \stuckelberg analysis. 

	\item The theory of spin-2 fields such as those coming from General Relativity and due to Fierz-Pauli were studied in detail. The massive spin-2 fields in FP theory suffered from the vDVZ discontinuity and required tuned coefficients to avoid Boulware-Deser ghost. Topologically Massive Gravity is a theory of gravity which allows gravitons to obtain mass from topological effects and is a unique possibility for 2+1 dimensions. A theory in which gravitons obtain mass from both FP mass terms and CS mass terms was setup. This theory had 3 propagating dof which on further \stuckelberg analysis were resolved into 2 dof excited by the FP mass term and 1 dof coming from the CS mass term. Since, the poles of this theory are not ghost-free for arbitrary parameters, this theory was deemed unsuitable. 
	
	\item The bimetric theory of gravitation was presented in detail. In hopes to obtain a healthy theory of gravity, in which gravitons were massive from both the FP and the CS mechanism - a Topologically Massive Bimetric theory of gravity was developed. This theory seemed to have many differences from the bimetric gravity in 2+1 dimensions. A \stuckelberg analysis was carried out and it was found that this theory excites 3 massive modes.
	
	\item The quantum loop corrections to graviton propagator from photons, which had a mass coming from both CS mass and Proca mass was calculated.
	
\end{itemize}

Although, 2+1 dimensions and massive gravity are both, seemingly, out of experimental reach the theoretical studies of this domain is well warranted. There are several important and necessary directions which emerge from this work. As it usually is with time, many of these could not be pursued and presented in this thesis. Some of the possibilities for future works are: (a) A CS term for HS fields was proposed. This term was verified for fields of spin-1 and spin-2. Preliminary calculations show that it seems to work for spin-2 fields as well (not presented here). This could motivate a study into the topological origins of this term. (b) The theory of TMBG is new and promises many unexplored chapters. Some of these explorations could look at the blackhole solutions allowed by this theory (in 2+1 dimensions), the mediated gravitational interactions, and loop corrections coming from massive gravitons. Some of these aspects are already being looked into by the author. 

For the author, working on this thesis has been a learning experience which was filled with excitement and various kinds of challenges. While many of the calculations that went in this thesis are done on paper by-hand, some others would not have been possible without the use of computer packages. The package xAct proved to be very helpful for doing routine tensor manipulations \cite{xpermMARTINGARCIA2008597}. The program FORM was extensively used to perform the loop calculations \cite{formmanual}. Both of these programs are gratefully acknowledged.
\label{mainmatterend}
\backmatter
\label{backmatterstart}
\appendix
\chapter{Appendix A}\label{ch:appendix1}
\setcounter{chapter}{1}
Tensors $X_i^{\mu\nu\alpha\beta}$ are defined as:
\begin{equationsplit}
	X_1^{\mu\nu\alpha\beta} & = \eta^{\alpha \nu} \eta^{\beta \mu} + \eta^{\alpha \mu} \eta^{\beta \nu}
	\\
	X_2^{\mu\nu\alpha\beta} & = \eta^{\alpha \beta} \eta^{\mu \nu}
	\\
	X_3^{\mu\nu\alpha\beta} & = k^{\beta} k^{\nu} \eta^{\alpha \mu} + k^{\beta} k^{\mu} \eta^{\alpha \nu} + k^{\alpha} k^{\nu} \eta^{\beta \mu} + k^{\alpha} k^{\mu} \eta^{\beta \nu}
	\\
	X_4^{\mu\nu\alpha\beta} & = k^{\mu} k^{\nu} \eta^{\alpha \beta} + k^{\alpha} k^{\beta} \eta^{\mu \nu}
	\\
	X_5^{\mu\nu\alpha\beta} & = k^{\alpha} k^{\beta} k^{\mu} k^{\nu}
	\\
	X_6^{\mu\nu\alpha\beta} & = k_\sigma (\epsilon^{\sigma\mu\alpha} \eta^{\nu\beta} + \epsilon^{\sigma\mu\beta} \eta^{\nu\alpha} + (\mu \leftrightarrow \nu) )
	\\
	X_7^{\mu\nu\alpha\beta} & = k_\sigma (\epsilon^{\sigma\mu\alpha} k^\nu k^\beta + \epsilon^{\sigma\mu\beta} k^\nu k^\alpha + (\mu \leftrightarrow \nu) )
\end{equationsplit}
\textbf{Propagator for $\phi_\munu$in FPCS theory}
The propagator is:
\begin{equation}
G^{\mu\nu\alpha\beta} = \frac{Y_1^{\mu\nu\alpha\beta}}{k^2 - f_1} + \frac{Y_2^{\mu\nu\alpha\beta}}{k^2- f_2} + \frac{Y_3^{\mu\nu\alpha\beta}}{k^2 - f_3}
\end{equation}
where:
\begin{equation}
\begin{split}
\begin{aligned}
Y_1^{\mu\nu\alpha\beta} 	&= 	\frac{\bigl(- 2 f_1^2 +  f_2 f_3 - f_1 (f_2 + f_3)\bigr)}{2 (f_1 -  f_2) (f_1 -  f_3)}  X_1^{\mu\nu\alpha\beta}
\\
&+ \frac{\bigl(2 f_1^2 - f_2 f_3 +  f_1 (f_2 + f_3)\bigr)}{2 (f_1 -  f_2) (f_1 -  f_3)}  X_2^{\mu\nu\alpha\beta}
\\ 
&+ \frac{\bigl(f_1^2 (f_2 + f_3) -  f_2 f_3 (f_2 + f_3) + f_1 (f_2^2 + f_2 f_3 + f_3^2)\bigr) }{(f_1 -  f_2) (f_1 -  f_3) \bigl(f_2 f_3 + f_1 (f_2 + f_3)\bigr)} X_3^{\mu\nu\alpha\beta}
\\ 
&+ \frac{\bigl(-f_1^2 (f_2 + f_3) +  f_2 f_3 (f_2 + f_3) - f_1 (f_2^2 + f_2 f_3 + f_3^2)\bigr)}{ (f_1 -  f_2) (f_1 -  f_3) \bigl(f_2 f_3 + f_1 (f_2 + f_3)\bigr)}  X_4^{\mu\nu\alpha\beta}
\\
&- 2 \frac{\bigl(- f_2 f_3 (f_2 + f_3)^2 + f_1^2 (f_2^2 + f_3^2) + f_1 (f_2^3 + f_3^3)\bigr) }{(f_1 -  f_2) (f_1 -  f_3) \bigl(f_2 f_3 + f_1 (f_2 + f_3)\bigr)^2} X_5^{\mu\nu\alpha\beta}
\\
&+  \frac{i  \sqrt{f_1 + f_2 + f_3} }{2 (f_1 -  f_2) (f_1 -  f_3)} (f_1 X_6^{\mu\nu\alpha\beta} - X_7^{\mu\nu\alpha\beta})
\end{aligned}
\end{split}
\end{equation}
and
\begin{equation}
\begin{split}
\begin{aligned}
Y_2^{\mu\nu\alpha\beta}	& =	\frac{\bigl(f_1 (f_2 - f_3) +  f_2 (2 f_2 + f_3)\bigr) }{2 (f_1 -  f_2) (f_2 -  f_3)} 						X_1^{\mu\nu\alpha\beta}
\\ 
&+ \frac{\bigl(-f_1 (f_2 -  f_3) - f_2 (2 f_2 + f_3)\bigr) }{2 (f_1 -  f_2) (f_2 -  f_3)} X_2^{\mu\nu\alpha\beta}
\\ 
&+ \frac{\bigl(-f_1^2 (f_2 -  f_3) - f_2 f_3 (f_2 + f_3) - f_1 (f_2^2 + f_2 f_3 -  f_3^2)\bigr) }{ (f_1 -  f_2) (f_2 -  f_3) \bigl(f_2 f_3 + f_1 (f_2 + f_3)\bigr)} X_3^{\mu\nu\alpha\beta}
\\
&+  \frac{\bigl(f_1^2 (f_2 -  f_3) + f_2 f_3 (f_2 + f_3) + f_1 (f_2^2 + f_2 f_3 -  f_3^2)\bigr) }{ (f_1 -  f_2) (f_2 -  f_3) \bigl(f_2 f_3 + f_1 (f_2 + f_3)\bigr)} X_4^{\mu\nu\alpha\beta}
\\
&+ 2 \frac{\bigl(- f_1 f_3^3 + f_1^3 ( f_2 - f_3) -  f_2 f_3^2 (f_2 + f_3) +  f_1^2 (f_2^2 - 2 f_3^2)\bigr) }{(f_1 -  f_2) (f_2 -  f_3) \bigl(f_2 f_3 + f_1 (f_2 + f_3)\bigr)^2} X_5^{\mu\nu\alpha\beta}
\\
&- \frac{i \sqrt{f_1 + f_2 + f_3} }{2 (f_1 -  f_2) (f_2 -  f_3)} (f_2 X_6^{\mu\nu\alpha\beta} - X_7^{\mu\nu\alpha\beta})	
\end{aligned}
\end{split}
\end{equation}
and, finally
\begin{equation}
\begin{split}
\begin{aligned}
Y_3^{\mu\nu\alpha\beta}	& = \frac{\bigl(f_1 (- f_2 + f_3) + f_3 (f_2 + 2 f_3)\bigr) }{2 (f_1 -  f_3) (- f_2 + f_3)} 					X_1^{\mu\nu\alpha\beta}
\\ 
&- \frac{\bigl(f_1 (- f_2 + f_3) + f_3 (f_2 + 2 f_3)\bigr) }{2 (f_1 -  f_3) (- f_2 + f_3)} X_2^{\mu\nu\alpha\beta}
\\
&+ \frac{\bigl(f_1^2 (f_2 - f_3) - f_2 f_3 (f_2 + f_3) + f_1 ( f_2^2 - f_2 f_3 - f_3^2)\bigr) }{ (f_1 -  f_3) (- f_2 + f_3) \bigl(f_2 f_3 + f_1 (f_2 + f_3)\bigr)} X_3^{\mu\nu\alpha\beta}
\\ 
&+ \frac{\bigl(f_1^2 (-f_2 +  f_3) +  f_2 f_3 (f_2 + f_3) - f_1 (f_2^2 -  f_2 f_3 -  f_3^2)\bigr)} { (f_1 -  f_3) (- f_2 + f_3) \bigl(f_2 f_3 + f_1 (f_2 + f_3)\bigr)} X_4^{\mu\nu\alpha\beta}
\\ 
&+ 2 \frac{\bigl(- f_1 f_2^3 + f_1^3 (- f_2 +  f_3) +  f_2^2 f_3 (f_2 + f_3) - f_1^2 (2 f_2^2 -  f_3^2)\bigr) }{(f_1 -  f_3) (- f_2 + f_3) \bigl(f_2 f_3 + f_1 (f_2 + f_3)\bigr)^2} X_5^{\mu\nu\alpha\beta}
\\					
&- \frac{i \sqrt{f_1 + f_2 + f_3} }{2 (f_1 -  f_3) (- f_2 + f_3)} (f_3 X_6^{\mu\nu\alpha\beta} - X_7^{\mu\nu\alpha\beta})	
\end{aligned}
\end{split}
\end{equation}
\newpage
\textbf{Propagator for $\psi_\munu$ in FPCS}
The propagator is given by:
\begin{equation}
G^{\mu\nu\alpha\beta}_{\psi_{\mu\nu}} = \frac{Z_1^{\mu\nu\alpha\beta}}{k^2 - f_1} + \frac{Z_2^{\mu\nu\alpha\beta}}{k^2- f_2} + \frac{Z_3^{\mu\nu\alpha\beta}}{k^2 - f_3} + \frac{Z_4^{\mu\nu\alpha\beta}}{k^2 - \frac{2 f_1 f_2 f_3}{f_1 f_2 + f_2 f_3 + f_1 f_3 }}
\end{equation}
where:
\begin{equationsplit}
	Z_1^{\mu \nu \alpha \beta} &= \frac{\bigl(-2 f_{1}^2 + f_2 f_3 -  f_1 (f_2 + f_3)\bigr) }{2 (f_1 -  f_2) (f_1 -  f_3)} X_1^{\mu \nu \alpha \beta}
	\\
	&+ \frac{\bigl(2 f_1^4 + f_1^2 (f_2 -  f_3)^2 + f_2^2 f_3^2 + 2 f_1^3 (f_2 + f_3) - 2 f_1 f_2 f_3 (f_2 + f_3)\bigr) }{(f_1 -  f_2) (f_1 -  f_3) \bigl(2 f_1^2 -  f_2 f_3 + f_1 (f_2 + f_3)\bigr)}X_2^{\mu \nu \alpha \beta}
	\\
	&+ \frac{2 f_1^2 (f_1 + f_2 + f_3)}{(f_1 -  f_2) (f_1 -  f_3) \bigl(2 f_1^2 -  f_2 f_3 + f_1 (f_2 + f_3)\bigr)}  X_3^{\mu \nu \alpha \beta}
	\\
	&-  \frac{2 f_1^2 (f_1 + f_2 + f_3) }{(f_1 -  f_2) (f_1 -  f_3) \bigl(2 f_1^2 -  f_2 f_3 + f_1 (f_2 + f_3)\bigr)}X_4^{\mu \nu \alpha \beta}
	\\
	&-  \frac{2 f_1 (f_1 + f_2 + f_3) }{(f_1 -  f_2) (f_1 -  f_3) \bigl(2 f_1^2 -  f_2 f_3 + f_1 (f_2 + f_3)\bigr)} X_5^{\mu \nu \alpha \beta}
	\\
	&+  \frac{i  \sqrt{f_1 + f_2 + f_3} }{2 (f_1 -  f_2) (f_1 -  f_3)} (f_1 X_6^{\mu\nu\alpha\beta} - X_7^{\mu\nu\alpha\beta})			
\end{equationsplit}
\begin{equationsplit}
	Z_2^{\mu \nu \alpha \beta} &= \frac{\bigl(f_1 (f_2 -  f_3) + f_2 (2 f_2 + f_3)\bigr) }{2 (f_1 -  f_2) (f_2 -  f_3)} X_1^{\mu \nu \alpha \beta}
	\\
	& -  \frac{\bigl(f_1^2 (f_2 -  f_3)^2 - 2 f_1 f_2 (- f_2^2 + f_2 f_3 + f_3^2) + f_2^2 (2 f_2^2 + 2 f_2 f_3 + f_3^2)\bigr) }{(f_1 -  f_2) (f_2 -  f_3) \bigl(f_1 (f_2 -  f_3) + f_2 (2 f_2 + f_3)\bigr)} X_2^{\mu \nu \alpha \beta}
	\\
	&-  \frac{2 f_2^2 (f_1 + f_2 + f_3) }{(f_1 -  f_2) (f_2 -  f_3) \bigl(f_1 (f_2 -  f_3) + f_2 (2 f_2 + f_3)\bigr)}X_3^{\mu \nu \alpha \beta}
	\\
	& + \frac{2 f_2^2 (f_1 + f_2 + f_3) }{(f_1 -  f_2) (f_2 -  f_3) \bigl(f_1 (f_2 -  f_3) + f_2 (2 f_2 + f_3)\bigr)} X_4^{\mu \nu \alpha \beta}
	\\
	&+ \frac{2 f_2 (f_1 + f_2 + f_3) }{(f_1 -  f_2) (f_2 -  f_3) \bigl(f_1 (f_2 -  f_3) + f_2 (2 f_2 + f_3)\bigr)} X_5^{\mu \nu \alpha \beta}
	\\
	\\
	&- \frac{i \sqrt{f_1 + f_2 + f_3} }{2 (f_1 -  f_2) (f_2 -  f_3)} (f_2 X_6^{\mu\nu\alpha\beta} - X_7^{\mu\nu\alpha\beta})
\end{equationsplit}
\begin{equationsplit}
	Z_3^{\mu \nu \alpha \beta} &=\frac{\bigl(f_1 (- f_2 + f_3) + f_3 (f_2 + 2 f_3)\bigr) }{2 (f_1 -  f_3) (- f_2 + f_3)} X_1^{\mu \nu \alpha \beta}
	\\
	&-  \frac{\bigl(f_1^2 (f_2 -  f_3)^2 - 2 f_1 f_3 (f_2^2 + f_2 f_3 -  f_3^2) + f_3^2 (f_2^2 + 2 f_2 f_3 + 2 f_3^2)\bigr) }{(f_1 -  f_3) (- f_2 + f_3) \bigl(f_1 (- f_2 + f_3) + f_3 (f_2 + 2 f_3)\bigr)} X_2^{\mu \nu \alpha \beta}
	\\
	&-  \frac{2 f3^2 (f_1 + f_2 + f_3) }{(f_1 -  f_3) (- f_2 + f_3) \bigl(f_1 (- f_2 + f_3) + f_3 (f_2 + 2 f_3)\bigr)}X_3^{\mu \nu \alpha \beta}
	\\
	& + \frac{2 f_3^2 (f_1 + f_2 + f_3) }{(f_1 -  f_3) (- f_2 + f_3) \bigl(f_1 (- f_2 + f_3) + f_3 (f_2 + 2 f_3)\bigr)} X_4^{\mu \nu \alpha \beta}
	\\
	&+ \frac{2 f_3 (f_1 + f_2 + f_3) }{(f_1 -  f_3) (- f_2 + f_3) \bigl(f_1 (- f_2 + f_3) + f_3 (f_2 + 2 f_3)\bigr)} X_5^{\mu \nu \alpha \beta} 
	\\
	&- \frac{i \sqrt{f_1 + f_2 + f_3} }{2 (f_1 -  f_3) (- f_2 + f_3)} (f_3 X_6^{\mu\nu\alpha\beta} - X_7^{\mu\nu\alpha\beta})
\end{equationsplit}
\begin{equationsplit}
	Z_4^{\mu \nu \alpha \beta} &=- \frac{\bigl(f_2 f_3 + f_1 (f_2 + f_3)\bigr)^3 X_2^{\mu \nu \alpha \beta}}{\bigl(2 f_1^2 -  f_2 f_3 + f_1 (f_2 + f_3)\bigr) \bigl(f_1 (f_2 -  f_3) + f_2 (2 f_2 + f_3)\bigr) \bigl(f_1 (f_2 -  f_3) -  f_3 (f_2 + 2 f_3)\bigr)} 
	\\
	&+ \frac{2 (f_1 + f_2 + f_3) \bigl(f_2 f_3 + f_1 (f_2 + f_3)\bigr)^2 X_3^{\mu \nu \alpha \beta}}{\bigl(2 f_1^2 -  f_2 f_3 + f_1 (f_2 + f_3)\bigr) \bigl(f_1 (f_2 -  f_3) + f_2 (2 f_2 + f_3)\bigr) \bigl(f_1 (f_2 -  f_3) -  f_3 (f_2 + 2 f_3)\bigr)} 
	\\
	&-  \frac{2 (f_1 + f_2 + f_3) \bigl(f_2 f_3 + f_1 (f_2 + f_3)\bigr)^2 X_4^{\mu \nu \alpha \beta}}{\bigl(2 f_1^2 -  f_2 f_3 + f_1 (f_2 + f_3)\bigr) \bigl(f_1 (f_2 -  f_3) + f_2 (2 f_2 + f_3)\bigr) \bigl(f_1 (f_2 -  f_3) -  f_3 (f_2 + 2 f_3)\bigr)} 
	\\
	&-  \frac{4 (f_1 + f_2 + f_3)^2 \bigl(f_2 f_3 + f_1 (f_2 + f_3)\bigr) X_5^{\mu \nu \alpha \beta}}{\bigl(2 f_1^2 -  f_2 f_3 + f_1 (f_2 + f_3)\bigr) \bigl(f_1 (f_2 -  f_3) + f_2 (2 f_2 + f_3)\bigr) \bigl(f_1 (f_2 -  f_3) -  f_3 (f_2 + 2 f_3)\bigr)}
\end{equationsplit}
\chapter{Appendix B}\label{ch:appendix2}
Some important identities used for the bimetric theory and the TMBG theory are:
\begin{equation}
	Tr (\delta \Xmatrix^n) = n Tr (\Xmatrix^{n-1} \delta \Xmatrix)
\end{equation}
\begin{equation}
Tr (\Xmatrix^{k-1}\; \delta \Xmatrix) = \half Tr (\Xmatrix^{k-2}\; \delta \Xmatrix^2)
\end{equation}

The coefficients $v_i$ in bimetric gravity obtained on expansion of the interaction term are:

\begin{equationsplit}
	v_1 &=- \tfrac{1}{2} \beta_0 + \tfrac{3}{4} \beta_1 \rho -  \tfrac{1}{2} \beta_1 d \rho -  \beta_2 \rho^2 + \beta_2 d \rho^2 
	\\
	&-  \tfrac{1}{4} \beta_2 d^2 \rho^2 + \tfrac{5}{4} \beta_3 \rho^3 -  \tfrac{37}{24} \beta_3 d \rho^3 + \tfrac{5}{8} \beta_3 d^2 \rho^3 -  \tfrac{1}{12} \beta_3 d^3 \rho^3 
	\\
	&-  \tfrac{3}{2} \beta_4 \rho^4 + \tfrac{17}{8} \beta_4 d \rho^4 -  \tfrac{53}{48} \beta_4 d^2 \rho^4 + \tfrac{1}{4} \beta_4 d^3 \rho^4 -  \tfrac{1}{48} \beta_4 d^4 \rho^4;
\end{equationsplit}
\begin{equationsplit}
	v_2 &= \tfrac{1}{4} \beta_0 -  \tfrac{1}{2} \beta_1 \rho + \tfrac{1}{4} \beta_1 d \rho + \tfrac{3}{4} \beta_2 \rho^2 -  \tfrac{5}{8} \beta_2 d \rho^2 
	\\
	&+ \tfrac{1}{8} \beta_2 d^2 \rho^2 -  \beta_3 \rho^3 + \tfrac{13}{12} \beta_3 d \rho^3 -  \tfrac{3}{8} \beta_3 d^2 \rho^3 + \tfrac{1}{24} \beta_3 d^3 \rho^3 
	\\
	&+ \tfrac{5}{4} \beta_4 \rho^4 -  \tfrac{77}{48} \beta_4 d \rho^4 + \tfrac{71}{96} \beta_4 d^2 \rho^4 -  \tfrac{7}{48} \beta_4 d^3 \rho^4 + \tfrac{1}{96} \beta_4 d^4 \rho^4
\end{equationsplit}
\begin{equationsplit}
	v_3&=- \tfrac{1}{4} \beta_1 \rho -  \tfrac{1}{4} \beta_2 d \rho^2 + \tfrac{1}{4} \beta_3 \rho^3 + \tfrac{1}{8} \beta_3 d \rho^3 -  \tfrac{1}{8} \beta_3 d^2 \rho^3 
	\\
	&-  \tfrac{1}{2} \beta_4 \rho^4 + \tfrac{1}{6} \beta_4 d \rho^4 + \tfrac{1}{8} \beta_4 d^2 \rho^4 -  \tfrac{1}{24} \beta_4 d^3 \rho^4
\end{equationsplit}
\begin{equationsplit}
	v_4 &= \tfrac{1}{4} \beta_2 \rho^2 -  \tfrac{1}{2} \beta_3 \rho^3 + \tfrac{1}{4} \beta_3 d \rho^3 
	\\
	&+ \tfrac{3}{4} \beta_4 \rho^4 -  \tfrac{5}{8} \beta_4 d \rho^4 + \tfrac{1}{8} \beta_4 d^2 \rho^4
\end{equationsplit}
\begin{equationsplit}
	v_5&=- \tfrac{1}{2} \beta_1 \rho + \beta_2 \rho^2 -  \tfrac{1}{2} \beta_2 d \rho^2 -  \tfrac{3}{2} \beta_3 \rho^3 + \tfrac{5}{4} \beta_3 d \rho^3 
	\\&-  \tfrac{1}{4} \beta_3 d^2 \rho^3 + 2 \beta_4 \rho^4 -  \tfrac{13}{6} \beta_4 d \rho^4 + \tfrac{3}{4} \beta_4 d^2 \rho^4 -  \tfrac{1}{12} \beta_4 d^3 \rho^4
\end{equationsplit}
\begin{equationsplit}
	v_6&=\tfrac{1}{2} \beta_1 \rho -  \beta_2 \rho^2 + \tfrac{1}{2} \beta_2 d \rho^2 + \tfrac{3}{2} \beta_3 \rho^3 -  \tfrac{5}{4} \beta_3 d \rho^3 
	\\
	&+ \tfrac{1}{4} \beta_3 d^2 \rho^3 - 2 \beta_4 \rho^4 + \tfrac{13}{6} \beta_4 d \rho^4 -  \tfrac{3}{4} \beta_4 d^2 \rho^4 + \tfrac{1}{12} \beta_4 d^3 \rho^4
\end{equationsplit}

\chapter{Appendix C}\label{ch:appendix3}

The vertex functions coming from second order expansion of the minimally coupled lagrangian is:

\begin{equationsplit}
	V4^{{\text{kin}}{\mu \nu \alpha \beta \rho \sigma}}(k,p,r) &=	- \eta^{\alpha \sigma} \eta^{\beta \rho} \eta^{\mu \nu} p_{a} r^{a} -  \eta^{\alpha \rho} \eta^{\beta \sigma} \eta^{\mu \nu} p_{a} r^{a} 
	\\
	&+ \eta^{\alpha \sigma} \eta^{\beta \nu} \eta^{\mu \rho} p_{a} r^{a} + \eta^{\alpha \nu} \eta^{\beta \sigma} \eta^{\mu \rho} p_{a} r^{a} + \eta^{\alpha \rho} \eta^{\beta \nu} \eta^{\mu \sigma} p_{a} r^{a} \\
	&+ \eta^{\alpha \nu} \eta^{\beta \rho} \eta^{\mu \sigma} p_{a} r^{a} + \eta^{\alpha \sigma} \eta^{\beta \mu} \eta^{\nu \rho} p_{a} r^{a} + \eta^{\alpha \mu} \eta^{\beta \sigma} \eta^{\nu \rho} p_{a} r^{a} -  \eta^{\alpha \beta} \eta^{\mu \sigma} \eta^{\nu \rho} p_{a} r^{a} 
	\\
	&+ \eta^{\alpha \rho} \eta^{\beta \mu} \eta^{\nu \sigma} p_{a} r^{a} + \eta^{\alpha \mu} \eta^{\beta \rho} \eta^{\nu \sigma} p_{a} r^{a} -  \eta^{\alpha \beta} \eta^{\mu \rho} \eta^{\nu \sigma} p_{a} r^{a} 
	\\
	&-  \eta^{\alpha \nu} \eta^{\beta \mu} \eta^{\rho \sigma} p_{a} r^{a} -  \eta^{\alpha \mu} \eta^{\beta \nu} \eta^{\rho \sigma} p_{a} r^{a} + \eta^{\alpha \beta} \eta^{\mu \nu} \eta^{\rho \sigma} p_{a} r^{a} \\
	&+ \eta^{\mu \sigma} \eta^{\nu \rho} p^{\beta} r^{\alpha} + \eta^{\mu \rho} \eta^{\nu \sigma} p^{\beta} r^{\alpha} -  \eta^{\mu \nu} \eta^{\rho \sigma} p^{\beta} r^{\alpha} -  \tfrac{1}{2} \eta^{\beta \sigma} \eta^{\nu \rho} p^{\mu} r^{\alpha} -  \tfrac{1}{2} \eta^{\beta \rho} \eta^{\nu \sigma} p^{\mu} r^{\alpha} 
	\\
	&+ \eta^{\beta \nu} \eta^{\rho \sigma} p^{\mu} r^{\alpha} -  \tfrac{1}{2} \eta^{\beta \sigma} \eta^{\mu \rho} p^{\nu} r^{\alpha} -  \tfrac{1}{2} \eta^{\beta \rho} \eta^{\mu \sigma} p^{\nu} r^{\alpha} + \eta^{\beta \mu} \eta^{\rho \sigma} p^{\nu} r^{\alpha} + \eta^{\mu \sigma} \eta^{\nu \rho} p^{\alpha} r^{\beta} 
	\\
	&+ \eta^{\mu \rho} \eta^{\nu \sigma} p^{\alpha} r^{\beta} -  \eta^{\mu \nu} \eta^{\rho \sigma} p^{\alpha} r^{\beta} -  \tfrac{1}{2} \eta^{\alpha \sigma} \eta^{\nu \rho} p^{\mu} r^{\beta} 
	\\
	&-  \tfrac{1}{2} \eta^{\alpha \rho} \eta^{\nu \sigma} p^{\mu} r^{\beta} + \eta^{\alpha \nu} \eta^{\rho \sigma} p^{\mu} r^{\beta} -  \tfrac{1}{2} \eta^{\alpha \sigma} \eta^{\mu \rho} p^{\nu} r^{\beta} -  \tfrac{1}{2} \eta^{\alpha \rho} \eta^{\mu \sigma} p^{\nu} r^{\beta} + \eta^{\alpha \mu} \eta^{\rho \sigma} p^{\nu} r^{\beta} 
	\\
	&-  \tfrac{1}{2} \eta^{\beta \sigma} \eta^{\nu \rho} p^{\alpha} r^{\mu} -  \tfrac{1}{2} \eta^{\beta \rho} \eta^{\nu \sigma} p^{\alpha} r^{\mu} + \eta^{\beta \nu} \eta^{\rho \sigma} p^{\alpha} r^{\mu} 
	\\
	&-  \tfrac{1}{2} \eta^{\alpha \sigma} \eta^{\nu \rho} p^{\beta} r^{\mu} -  \tfrac{1}{2} \eta^{\alpha \rho} \eta^{\nu \sigma} p^{\beta} r^{\mu} + \eta^{\alpha \nu} \eta^{\rho \sigma} p^{\beta} r^{\mu} 
	\\
	&+ \eta^{\alpha \sigma} \eta^{\beta \rho} p^{\nu} r^{\mu} + \eta^{\alpha \rho} \eta^{\beta \sigma} p^{\nu} r^{\mu} -  \eta^{\alpha \beta} \eta^{\rho \sigma} p^{\nu} r^{\mu} -  \tfrac{1}{2} \eta^{\beta \sigma} \eta^{\mu \rho} p^{\alpha} r^{\nu} 
	\\
	&-  \tfrac{1}{2} \eta^{\beta \rho} \eta^{\mu \sigma} p^{\alpha} r^{\nu} + \eta^{\beta \mu} \eta^{\rho \sigma} p^{\alpha} r^{\nu} -  \tfrac{1}{2} \eta^{\alpha \sigma} \eta^{\mu \rho} p^{\beta} r^{\nu} 
	\\
	&-  \tfrac{1}{2} \eta^{\alpha \rho} \eta^{\mu \sigma} p^{\beta} r^{\nu} + \eta^{\alpha \mu} \eta^{\rho \sigma} p^{\beta} r^{\nu} + \eta^{\alpha \sigma} \eta^{\beta \rho} p^{\mu} r^{\nu} + \eta^{\alpha \rho} \eta^{\beta \sigma} p^{\mu} r^{\nu} 
	\\
	&-  \eta^{\alpha \beta} \eta^{\rho \sigma} p^{\mu} r^{\nu} + \eta^{\beta \sigma} \eta^{\mu \nu} p^{\alpha} r^{\rho} -  \eta^{\beta \nu} \eta^{\mu \sigma} p^{\alpha} r^{\rho} -  \eta^{\beta \mu} \eta^{\nu \sigma} p^{\alpha} r^{\rho} + \eta^{\alpha \sigma} \eta^{\mu \nu} p^{\beta} r^{\rho} 
	\\
	&-  \eta^{\alpha \nu} \eta^{\mu \sigma} p^{\beta} r^{\rho} -  \eta^{\alpha \mu} \eta^{\nu \sigma} p^{\beta} r^{\rho} -  \eta^{\alpha \sigma} \eta^{\beta \nu} p^{\mu} r^{\rho} -  \eta^{\alpha \nu} \eta^{\beta \sigma} p^{\mu} r^{\rho} 
	\\
	&+ \eta^{\alpha \beta} \eta^{\nu \sigma} p^{\mu} r^{\rho} -  \eta^{\alpha \sigma} \eta^{\beta \mu} p^{\nu} r^{\rho} -  \eta^{\alpha \mu} \eta^{\beta \sigma} p^{\nu} r^{\rho} + \eta^{\alpha \beta} \eta^{\mu \sigma} p^{\nu} r^{\rho} + \tfrac{1}{2} \eta^{\alpha \nu} \eta^{\beta \mu} p^{\sigma} r^{\rho} 
	\\
	&+ \tfrac{1}{2} \eta^{\alpha \mu} \eta^{\beta \nu} p^{\sigma} r^{\rho} -  \tfrac{1}{2} \eta^{\alpha \beta} \eta^{\mu \nu} p^{\sigma} r^{\rho} + \eta^{\beta \rho} \eta^{\mu \nu} p^{\alpha} r^{\sigma} -  \eta^{\beta \nu} \eta^{\mu \rho} p^{\alpha} r^{\sigma} -  \eta^{\beta \mu} \eta^{\nu \rho} p^{\alpha} r^{\sigma} 
	\\
	&+ \eta^{\alpha \rho} \eta^{\mu \nu} p^{\beta} r^{\sigma} -  \eta^{\alpha \nu} \eta^{\mu \rho} p^{\beta} r^{\sigma} -  \eta^{\alpha \mu} \eta^{\nu \rho} p^{\beta} r^{\sigma} -  \eta^{\alpha \rho} \eta^{\beta \nu} p^{\mu} r^{\sigma} 
	\\
	&-  \eta^{\alpha \nu} \eta^{\beta \rho} p^{\mu} r^{\sigma} + \eta^{\alpha \beta} \eta^{\nu \rho} p^{\mu} r^{\sigma} -  \eta^{\alpha \rho} \eta^{\beta \mu} p^{\nu} r^{\sigma} -  \eta^{\alpha \mu} \eta^{\beta \rho} p^{\nu} r^{\sigma} 
	\\
	&+ \eta^{\alpha \beta} \eta^{\mu \rho} p^{\nu} r^{\sigma} + \tfrac{1}{2} \eta^{\alpha \nu} \eta^{\beta \mu} p^{\rho} r^{\sigma} + \tfrac{1}{2} \eta^{\alpha \mu} \eta^{\beta \nu} p^{\rho} r^{\sigma} -  \tfrac{1}{2} \eta^{\alpha \beta} \eta^{\mu \nu} p^{\rho} r^{\sigma}
\end{equationsplit}
\begin{equationsplit}
	V4^{{\text{Proca}}{\mu \nu \alpha \beta \rho \sigma}}(k,p,r) &= - m_A^2 \eta^{\alpha \sigma} \eta^{\beta \rho} \eta^{\mu \nu} -  m_A^2 \eta^{\alpha \rho} \eta^{\beta \sigma} \eta^{\mu \nu} + m_A^2 \eta^{\alpha \sigma} \eta^{\beta \nu} \eta^{\mu \rho} 
	\\
	&+ m_A^2 \eta^{\alpha \nu} \eta^{\beta \sigma} \eta^{\mu \rho} + m_A^2 \eta^{\alpha \rho} \eta^{\beta \nu} \eta^{\mu \sigma} + m_A^2 \eta^{\alpha \nu} \eta^{\beta \rho} \eta^{\mu \sigma} 
	\\
	&+ m_A^2 \eta^{\alpha \sigma} \eta^{\beta \mu} \eta^{\nu \rho} + m_A^2 \eta^{\alpha \mu} \eta^{\beta \sigma} \eta^{\nu \rho} -  m_A^2 \eta^{\alpha \beta} \eta^{\mu \sigma} \eta^{\nu \rho} 
	\\
	&+ m_A^2 \eta^{\alpha \rho} \eta^{\beta \mu} \eta^{\nu \sigma} + m_A^2 \eta^{\alpha \mu} \eta^{\beta \rho} \eta^{\nu \sigma} -  m_A^2 \eta^{\alpha \beta} \eta^{\mu \rho} \eta^{\nu \sigma} 
	\\
	&-  m_A^2 \eta^{\alpha \nu} \eta^{\beta \mu} \eta^{\rho \sigma} -  m_A^2 \eta^{\alpha \mu} \eta^{\beta \nu} \eta^{\rho \sigma} + m_A^2 \eta^{\alpha \beta} \eta^{\mu \nu} \eta^{\rho \sigma}
\end{equationsplit}
\bibliographystyle{myIEEEtran} 
\bibliography{references} 
\label{backmatterend}
\end{document}